\documentclass[12pt,a4paper]{article}

\usepackage[utf8]{inputenc}
\usepackage[dvipsnames,table]{xcolor}

\usepackage[left=2.3cm,right=2.3cm,top=2cm,bottom=2.5cm]{geometry}
\usepackage{cite}
\usepackage[bookmarks=true]{hyperref}
\usepackage{booktabs}
\usepackage{enumitem}

\newif{\ifarxiv}
\newif{\ifdraft}
\newif{\ifremarks}

\arxivtrue 
\drafttrue 
\remarkstrue 

\ifdraft\else\fi

\ifremarks\newcommand{\remarktb}[1]{%
  {\renewcommand{\bfdefault}{b}\color[RGB]{0,150,0}{\textbf{T:~#1}}}}\fi
\providecommand{\remarktb}[1]{\ignorespaces}
\ifremarks\newcommand{\remarkfc}[1]{%
  {\renewcommand{\bfdefault}{b}\color[RGB]{0,0,150}{\textbf{F:~#1}}}}\fi
\providecommand{\remarkfc}[1]{\ignorespaces}
\ifremarks\newcommand{\remarkcb}[1]{%
  {\renewcommand{\bfdefault}{b}\color[RGB]{150,0,150}{\textbf{C:~#1}}}}\fi
\providecommand{\remarkcb}[1]{\ignorespaces}
\ifremarks\newcommand{\remarkab}[1]{%
  {\renewcommand{\bfdefault}{b}\color[RGB]{221, 160, 221}{\textbf{A:~#1}}}}\fi
\providecommand{\remarkab}[1]{\ignorespaces}

\usepackage[T1]{fontenc}
\usepackage{lmodern}
\usepackage{microtype}

\usepackage{xspace}
\def\etal.{et\penalty50\ al.}
\newcommand*{\eg}{e.\,g.\@\xspace}
\newcommand*{\ie}{i.\,e.\@\xspace}
\newcommand*{\vs}{vs.\@\xspace}
\newcommand*{\aka}{a.\,k.\,a.\@\xspace}

\providecommand{\hypersetup}[1]{}
\providecommand{\hyperref}[2][]{#2}
\providecommand{\texorpdfstring}[2]{#1}
\providecommand{\pdfbookmark}[3][]{}
\newcommand{\email}[1]{\href{mailto:#1}{\nolinkurl{#1}}}

\hypersetup{plainpages=false}
\hypersetup{pdfpagemode=UseOutlines}
\hypersetup{bookmarksnumbered=true}
\hypersetup{bookmarksopen=true}
\hypersetup{pdfstartview=FitH}
\hypersetup{colorlinks=true}
\hypersetup{citecolor=[rgb]{0 .3 0}}
\hypersetup{urlcolor=[rgb]{.35 0 0}}
\hypersetup{linkcolor=[rgb]{0 0 .5}}
\hypersetup{citebordercolor={.6 .9 .6}}
\hypersetup{urlbordercolor={.7 .8 1}}
\hypersetup{linkbordercolor={1 .7 .7}}
\hypersetup{pdfborder={0 0 .5}}

\makeatletter
\let\@myabstract\@empty
\let\@keywords\@empty
\let\@subject\@empty
\providecommand{\affiliation}[1]{\gdef\@affiliation{#1}}
\providecommand{\myabstract}[1]{\gdef\@myabstract{#1}}
\providecommand{\keywords}[1]{\gdef\@keywords{#1}}
\providecommand{\subject}[1]{\gdef\@subject{#1}}
\def\thetitle{\@title}
\def\theauthor{\@author}
\def\theaffiliation{\@affiliation}
\def\theabstract{\@myabstract}
\def\thesubject{\@subject}
\def\thedate{\@date}
\def\thekeywords{\@keywords}
\makeatother
\AtBeginDocument{
\hypersetup{pdftitle={\thetitle}}%
\hypersetup{pdfauthor={\theauthor}}%
\hypersetup{pdfsubject={\thesubject}}%
\hypersetup{pdfkeywords={\thekeywords}}%
}

\usepackage{graphbox}

\usepackage[font=small,labelfont=bf,margin=0.05\textwidth]{caption}

\makeatletter
\newcommand\footnoterepeat[1]{\protected@xdef\@thefnmark{\ref{#1}}\@footnotemark}
\makeatother

\usepackage{amsmath,amssymb,mathtools}
\usepackage[mathbold,autobold,greekcaps,greeklower]{mathfixs}
\providecommand{\mathbold}{\mathbf}
\newcommand{\nn}{\nonumber}
\allowdisplaybreaks
\numberwithin{equation}{section}

\newcommand{\namedref}[2]{\hyperref[#2]{#1~\ref*{#2}}}
\newcommand{\secref}[1]{\namedref{Section}{#1}}
\newcommand{\appref}[1]{\namedref{Appendix}{#1}}
\newcommand{\tabref}[1]{\namedref{Table}{#1}}
\newcommand{\figref}[1]{\namedref{Figure}{#1}}
\newcommand{\footnoteref}[1]{\namedref{Footnote}{#1}}

\makeatletter
\def\mr@ignsp#1 {\ifx\:#1\@empty\else #1\expandafter\mr@ignsp\fi}%
\newcommand{\multiref}[1]{\begingroup
\xdef\mr@no@sparg{\expandafter\mr@ignsp#1 \: }%
\def\mr@comma{}%
\@for\mr@refs:=\mr@no@sparg\do{\mr@comma\def\mr@comma{,}\ref{\mr@refs}}%
\endgroup}
\makeatother
\renewcommand{\eqref}[1]{(\multiref{#1})}

\newcommand{\suprm}[1]{^{\text{#1}}}
\newcommand{\subrm}[1]{_{\text{#1}}}
\newcommand{\alg}[1]{\mathfrak{#1}}
\newcommand{\grp}[1]{\mathrm{#1}}
\newcommand{\mathematica}{\textsc{Mathematica}\@\xspace}

\newcommand{\filename}[1]{\texttt{#1}}
\newcommand{\ancfile}[1]{\href{https://arxiv.org/src/2512.19780/anc/#1}{\filename{#1}}}
\colorlet{c1}{Black}
\colorlet{c2}{Black}
\colorlet{c3}{Black}
\colorlet{c4}{Black}
\colorlet{c5}{Black}

\RequirePackage[extdef]{delimset}
\makeatletter
\providecommand{\brkleft}[1][r]{\begingroup\def\dlm@use{\delim(.}%
\if r#1 \def\dlm@use{\delim(.}\fi%
\if s#1 \def\dlm@use{\delim[.}\fi%
\if c#1 \def\dlm@use{\delim\{.}\fi%
\if a#1 \def\dlm@use{\delim<.}\fi%
\expandafter\endgroup\dlm@use}
\providecommand{\brkright}[1][r]{\begingroup\def\dlm@use{\delim.)}%
\if r#1 \def\dlm@use{\delim.)}\fi%
\if s#1 \def\dlm@use{\delim.]}\fi%
\if c#1 \def\dlm@use{\delim.\}}\fi%
\if a#1 \def\dlm@use{\delim.>}\fi%
\expandafter\endgroup\dlm@use}
\makeatother

\DeclareMathOperator{\tr}{tr}

\DeclareMathOperator{\Li}{Li}

\DeclareMathOperator*{\Res}{Res}

\newcommand{\op}[1]{\mathcal{#1}}
\newcommand{\order}[1]{\mathcal{O}(#1)}

\newcommand{\beq}{\begin{equation}}
\newcommand{\eeq}{\end{equation}}
\newcommand{\bba}{\begin{align}}
\newcommand{\eea}{\end{align}}

\newcommand{\ii}{\mathrm{i}}

\newcommand{\superN}{\mathcal{N}}
\newcommand{\Nc}{N\subrm{c}}

\newcommand{\dd}[2][]{\mathinner{\mathrm{d}\ifx#1\empty\else{^#1}\fi#2}}

\newcommand{\Lint}{L\subrm{int}}

\newcommand{\polygon}{M}
\newcommand{\intpoly}{\mathbb{M}}
\newcommand{\intface}{\mathbb{F}}
\newcommand{\gammaCusp}{\Gamma\subrm{cusp}}
\newcommand{\gammaOct}{\Gamma\subrm{oct}}

\newcommand{\twentyprime}{\mathbf{20'}}

\title{Higher-Point Correlators in \texorpdfstring{$\mathcal{N}=4$}{N=4}
SYM:\texorpdfstring{\\}{ }Ten-Dimensional Null Polygons}

\author{%
Till Bargheer\texorpdfstring{$^1$}{},
Albert Bekov\texorpdfstring{$^1$}{},
Carlos Bercini\texorpdfstring{$^{1,2}$}{},
Frank Coronado\texorpdfstring{$^3$}{}}

\myabstract{We consider higher-point generalizations
of the ``octagon'' large-charge four-point function
in planar $\mathcal{N}=4$ super Yang--Mills theory.
These $n$-point polygon correlators are defined as
ten-dimensional null limits of generating functions of $n$-point
correlators of protected scalar operators with arbitrary R-charge, and
are conjecturally dual to massive scattering amplitudes of W-bosons on
the Coulomb branch of the theory.
We present a twistor-based direct computation of the polygon loop
integrands, and determine the two-loop integrands
for any number of external operators. The result is expressed as a
linear combination of conformal integrals, with coefficients that are
rational functions of the spacetime distances and operator
polarizations. We obtain an exact match with a one-loop integrability
computation, and study the near-massless limit of the pentagon function.}

\subject{Mathematical Physics}


\begin{document}

\pdfbookmark[1]{Title Page}{title}

\thispagestyle{empty}
\setcounter{page}{0}

\renewcommand{\thefootnote}{\fnsymbol{footnote}}
\setcounter{footnote}{0}

\hfill
\texttt{DESY-25-196}

\mbox{}
\vfill

\begin{center}

{\Large\textbf{\mathversion{bold}\thetitle}\par}

\vspace{1cm}

\textsc{\theauthor}

\bigskip

\begingroup
\footnotesize\itshape

${}^1$ Deutsches Elektronen-Synchrotron DESY,
Notkestr.~85, 22607 Hamburg, Germany

${}^2$ Department of Mathematics, King's College London, The Strand, London WC2R 2LS, UK

${}^3$ Institut f\"ur Theoretische Physik, ETH Zurich, CH-8093 Z\"urich, Switzerland

\endgroup

\bigskip

\begingroup
\small\ttfamily
\email{till.bargheer@desy.de},
\email{albert.bekov@desy.de},\\
\email{carlos.bercini@desy.de},
\email{fcidrogo@gmail.com}
\endgroup
\par

\vspace{1cm}

\textbf{Abstract}
\vspace{5mm}

\begin{minipage}{12cm}
\theabstract
\end{minipage}

\end{center}

\vfill
\vfill
\newpage

\renewcommand{\thefootnote}{\arabic{footnote}}

\hrule
\pdfbookmark[1]{\contentsname}{contents}
\setcounter{tocdepth}{3}
\microtypesetup{protrusion=false}
\tableofcontents
\microtypesetup{protrusion=true}
\vspace{3ex}
\hrule

\section{Introduction}

Correlation functions of local operators are the most natural
observables to consider in a conformal field theory. This is no
different in planar $\mathcal{N}=4$ super Yang--Mills theory, where the most
studied correlation functions are those among protected single-trace operators.
Supersymmetry prevents their two- and three-point functions to receive
quantum corrections, their four- and higher-point functions are
extremely non-trivial, and contain a wealth of information about the theory.

The four-point correlator of the lightest protected scalar operators
(which are part of the stress-tensor multiplet) is known to
various orders in both the
weak-coupling~\cite{Eden:2011we,Drummond:2013nda} and
strong-coupling expansions~\cite{Arutyunov:2000py,Goncalves:2014ffa,Alday:2023mvu}.
Also higher-point correlators of these operators have been computed
at weak~\cite{Drukker:2008pi,Bargheer:2022sfd} and strong
coupling~\cite{Goncalves:2019znr,Goncalves:2025jcg}. These
observables provide access to non-protected structure
constants through OPE limits~\cite{Eden:2012fe, Eden:2012rr,
Bercini:2021jti, Bercini:2024pya, Bargheer:2025uai}, and
connect to Wilson loops~\cite{Alday:2010zy} and gluon scattering
amplitudes~\cite{Alday:2007hr, Berkovits:2008ic, Beisert:2008iq,
Bern:2008ap, Drummond:2008aq, Mason:2010yk, Caron-Huot:2010ryg,
Eden:2010zz} via light-like polygon limits and T-duality.

In this paper, we focus on the opposite regime of large-charge operators.
Their correlators are particularly amenable to integrability in the
form of hexagonalization~\cite{Basso:2015zoa, Fleury:2016ykk,
Eden:2016xvg, Fleury:2017eph}.
In fact, their four-point
function in the large-charge limit (in a suitable charge polarization)
has been computed to all orders in perturbation
theory~\cite{Coronado:2018ypq, Coronado:2018cxj, Kostov:2019stn, Kostov:2019auq},
and even at finite
coupling~\cite{Belitsky:2019fan, Belitsky:2020qrm}.
The five-point counterpart is known to two loops in weak-coupling
perturbation theory~\cite{Fleury:2020ykw, Bercini:2024pya}.
More structure emerges once we
unify both the light and the large-charge correlators into a
single common object, the so-called generating
function~\cite{Caron-Huot:2021usw, Caron-Huot:2023wdh}, which encodes not only
these correlators, but in fact all possible correlation functions of
half-BPS scalar operators with arbitrary charges. The generating
function is
\begin{equation}
G_n = \, \sum_{\mathclap{k_1,\dots,k_n = 2}}^{\infty} \;
\langle \mathcal{O}_{k_1}(x_1,y_1) \dots \mathcal{O}_{k_n}(x_n,y_n)\rangle
\,, \qquad
\mathcal{O}_k(x,y)=\frac{1}{k}\tr\left(y\cdot\phi(x)\right)^k
\,,
\label{eq:GenIntro}
\end{equation}
where~$\phi$ are the six scalars of the theory, and~$y_i$ are
six-dimensional null polarization vectors, $y_i\cdot y_i=0$.
Four-point correlators of arbitrary charges have been computed at
three~\cite{Chicherin:2015edu} and five
loops~\cite{Chicherin:2018avq}, and collected in the generating
function $G_4$ in~\cite{Caron-Huot:2021usw}.
At five and six points, the generating functions $G_5$ and $G_6$ are known
to two and one loops, respectively~\cite{Bargheer:2025uai}.
These generating functions display two remarkable properties:
\begin{itemize}
\item
After factoring out the universal four-point superinvariant, the
reduced four-point generating function displays a hidden
ten-dimensional conformal symmetry~\cite{Caron-Huot:2021usw}. The same
symmetry has been
observed at strong values of the coupling, in the tree-level supergravity regime of
the bulk dual~\cite{Caron-Huot:2018kta,Fernandes:2025eqe}.%
\footnote{A similar $8$d symmetry was recently observed for
correlators of $1/2$ BPS operators in $4$d $\superN=4$
SQCDs~\cite{Du:2024xbd}.}
At higher points, there is
so far only limited evidence (but also no contraindication) of this
symmetry~\cite{Bargheer:2025uai}.
Since the symmetry holds at
the level of the integrands, it is yet unclear how the weak-coupling
and strong-coupling instances are related.
The symmetry is broken at integrated level at weak coupling,
and it is broken by stringy corrections to supergravity. Hence it is not a
symmetry that interpolates between weak and strong coupling, and it
might not have the same origins at weak and strong coupling.
\item
After taking a ten-dimensional null polygon limit, the
generating functions are conjecturally dual to massive scattering
amplitudes on the Coulomb branch of $\superN=4$ super Yang--Mills
theory~\cite{Caron-Huot:2021usw,Bork:2022vat,Belitsky:2025bgb}.
\end{itemize}
To date, there is no first-principles derivation of the
ten-dimensional symmetry. However, in the planar limit, the generating function integrands
organize themselves in terms of ten-dimensional distances
\begin{equation}
D_{ij} = -\frac{y_{ij}^2}{X_{ij}^2}
\,,\qquad
X_{ij}^2 = x_{ij}^2+y_{ij}^2
\end{equation}
that re-sum geometric series $D_{ij}=d_{ij}+d_{ij}^2+d_{ij}^3+\dots$ of
four-dimensional propagators $d_{ij}=-y_{ij}^2/x_{ij}^2$.
In terms of these ten-dimensional distances, focusing on a correlator
that has a large charge flow between points $x_i$ and $x_j$ ($k_i,k_j
\to \infty$, with a large power of $y_{ij}^2$) is
equivalent to taking the ten-dimensional null limit $X_{ij}^2 \to 0$
of the generating function.

Thus by taking the
combined limit $X_{i,i+1}^2 \to 0$ with $i=1,\dots,n$, one focuses on
correlators where all operators have large charge, $k_i\to\infty$, and
moreover the limit
enforces a large charge flow between each pair of neighboring
operators $i$, $i+1$.
This is the ten-dimensional null polygon limit introduced in~\cite{Caron-Huot:2021usw}.
The most important aspect of the resulting $10$d null polygon correlators is
their factorization property. This can be easily understood in
perturbation theory. In the free theory, the large charge flow means
that every pair of neighboring operators is connected by a large
number of propagators, which, due to the planar limit, all lie in
parallel without crossing (in color space), forming a bundle (``ribbon'' or
``bridge''). These bridges cut the color sphere into two disk-like
regions, an ``inside'' and an ``outside''. For any virtual particle to
couple the inside with
the outside (in the planar limit), that particle must cross a large
bundle of propagators, and thus such processes are suppressed by a
factor $g^{2k}$, where $g$ is the coupling, and $k$ is the width of
the bundle.

At the same time, virtual processes that are internal to a single propagator
bundle cancel each other out due to
supersymmetry,
since all operators are BPS and thus have protected
two-point functions.
Therefore, sending the number of propagators between neighboring operators to
infinity (by taking the ten-dimensional null limit) decouples the
inside and the outside to all orders in perturbation theory, making
the large-charge correlator factorize into the square of a simpler object,
the \emph{polygon correlator}~$\intpoly_n$:
\begin{equation}
\lim\limits_{X_{i,i+1}^2 \to 0}
\left( \mspace{2mu} \prod_{i=1}^n \frac{X^2_{i,i+1}}{x_{i,i+1}^2}\right)
\Nc^{n-2} \, G_n
= \mathbb{M}_n \times \mathbb{M}_n
\,.
\label{eq:polyDefIntro}
\end{equation}
The simplest non-trivial polygon is the ``octagon'' $\intpoly_4$
studied in~\cite{Coronado:2018cxj,Coronado:2018ypq,Caron-Huot:2021usw}.
But the limit~\eqref{eq:polyDefIntro} defines $n$-sided polygon
correlators $\intpoly_n$ for any~$n$, and these will be the primary
objects of study in this work.

More precisely, we focus on the $\ell$-loop \emph{integrands}
$M_{n,\ell}$ of the polygon correlators $\intpoly_n$, defined via the
Lagrangian insertion procedure (see~\eqref{eq:PolygonIntegrated}
below). We explicitly compute these loop
integrands at two loops in perturbation
theory, by two different methods. Firstly by taking the
ten-dimensional null polygon limit~\eqref{eq:polyDefIntro} of the
generating functions computed in~\cite{Bargheer:2025uai}. Secondly, we
notice that the twistor Feynman rules devised to compute the
integrands of the generating functions~$G_n$~\cite{Caron-Huot:2023wdh} can be adapted to directly compute the polygon
integrands~$\polygon_{n,\ell}$ themselves, by restricting to graphs
with disk topology.
The latter method allows us to compute the two-loop polygon integrands
$\polygon_{n,2}$ for up to $n=10$ points. They are expressed in a
rational basis of integrands of conformal integrals. From the data up
to $n=10$, we could derive a general formula for the coefficients of
this expansion that holds for any~$n$.
We verified its correctness by comparing against the twistor computation
at $n=11$ points.
Our result is available in the ancillary \mathematica file \ancfile{polygons.m}.

An important property of the polygon correlators are their
factorization limits. This follows in
the same way as the factorization of the full generating functions
into the product of two polygons: Taking a ten-dimensional null limit
on any diagonal of a polygon correlator focuses on a large charge flow
on this diagonal, and thus factorizes the polygon into the product
of two smaller polygons.
The computation
of any polygon is thus reduced to assembling all factorization channels
from products of smaller polygons, and computing a
remaining function that is free of ten-dimensional poles
(see~\eqref{eq:MdecompF} below).
We find closed-form expressions for all these pole-free functions at
one and two loops (see \eqref{eq:oneLoopFaces} and
\eqref{eq:F2IntegralDecomp2} below), from which
we reconstruct the two-loop integrand $\polygon_{n,2}$ for any
$n$-sided polygon.
We also compare this decomposition with a similar
expansion from integrability at one loop~\cite{Bargheer:2018jvq}, and find
complete agreement.

Finally, we consider the integration of the polygon integrands
$\polygon_{n,\ell}$ to the interacting polygon correlators
$\intpoly_n$. Unfortunately, the required conformal integrals are
mostly unknown, hence our analysis
is limited to five-point polygons $\intpoly_5$ at
two-loops and in nearly massless kinematics. After
evaluating the contributing integrals, we find perfect
agreement with the four-dimensional polygonal null limit $x_{i,i+1}^2
\to 0$ computed in \cite{Bercini:2024pya,Belitsky:2025bgb} and the
leading logarithm limit (also known as ``stampedes'' limit) considered in
\cite{Olivucci:2021pss,Olivucci:2022aza}.

This work is organized as follows. In
\secref{sec:ten-dimensional-null}, we introduce the polygon correlators
as limits of the full generating functions in ten-dimensional null
kinematics, and highlight their many interesting properties. In
\secref{sec:PolyFromG}, we compute five-point and six-point
polygon integrands from the ten-dimensional null limit of the
respective generating functions. In \secref{sec:twistor-rules-10d}, we
explain how to compute the polygon integrands directly from the twistor
Feynman rules of~\cite{Caron-Huot:2023wdh}. In
\secref{sec:AllPoly}, we introduce the faces decomposition, and find
all polygon integrands at two loops. In \secref{sec:IntegratedPoly}, we
consider the integrated polygon correlators, and compare them with
previously studied near-massless limits. In
\secref{sec:oneloop-hexagonalization}, we compare the faces decomposition
with a similar result from integrability, finding an exact match, and
highlighting the differences between the two expansions.
We conclude by discussing interesting future directions in \secref{sec:discussion}.

\section{Ten-Dimensional Null Limit}
\label{sec:ten-dimensional-null}

In this section, we present the higher-point generalizations of the
\emph{octagon} function~\cite{Coronado:2018cxj,Coronado:2018ypq} (here
renamed as \emph{square}). These higher-point polygons are defined as
the residues of the $n$-point generating functions on a polygonal set of
ten-dimensional poles~\cite{Caron-Huot:2021usw}.

\paragraph{Generating Function.}

In $\mathcal{N}=4$ super Yang--Mills theory, correlation functions of
half-BPS scalar operators of any charges can be naturally re-packaged into
correlation functions of ``master''
operators~\cite{Caron-Huot:2021usw,Caron-Huot:2023wdh}
\begin{equation}
\mathcal{O}(x,y)= \sum_{k=2}^\infty\, \mathcal{O}_k(x,y)
\,,\qquad
\mathcal{O}_k(x,y)=\frac{1}{k}\tr\left[y\cdot\phi(x)\right]^k + \text{multi-traces}
\,.
\label{eq:Oscalar}
\end{equation}
where $\phi$ are the six scalars of the theory, and the polarization vectors $y_i$
satisfy the BPS condition
\begin{equation}
y\cdot y \equiv y^{AB} y_{AB} = 0 \quad \text{with}\quad A,B = 1,2,3,4
\,.
\end{equation}
The operators $\op{O}_k$ have fixed R-charge $k$: The operator with
$k=2$ is the scalar component of the stress-tensor multiplet, which is
dual to the $\grp{AdS}_5$ graviton, and the higher-$k$ operators are
dual to the higher Kaluza--Klein modes.

Just like the correlation functions of fixed-charge operators
$\op{O}_k$, the correlator of master operators~\eqref{eq:Oscalar} can be
written via the Lagrangian insertion procedure as
\begin{equation}
\avg*{\prod_{i=1}^{n}\mathcal{O}(x_i,y_i)}_{\!\text{SYM}}=
\sum_{\ell=0}^\infty\frac{(-g^{2})^{\ell}}{\ell!}
\int\brk[s]3{\,\prod_{k=1}^\ell\frac{\dd[4]{x_{n+k}}}{\pi^2}}
G_{n,\ell}\,,
\label{eq:integratedcorrelator}
\end{equation}
where $G_{n,\ell}$ is the correlator of $n$ operators $\op{O}$ and
$\ell$ chiral interaction Lagrangian operators~$\Lint$:
\begin{equation}
\label{eq:Gnl-integrand}
G_{n,\ell} \equiv \left\langle \prod_{i=1}^{n}\mathcal{O}(x_i,y_i)  \prod_{i=1}^{\ell}\Lint(x_i)  \right\rangle\subrm{\!SDYM}
\,.
\end{equation}
where the subscript ``SDYM'' signifies that this correlation function
is evaluated with the self-dual part of the $\superN=4$ SYM action
(which is equivalent to evaluating the correlator at Born level in
the full theory).

The correlator~\eqref{eq:Gnl-integrand} by definition of the
operators~\eqref{eq:Oscalar} serves as a generating function for the
integrands of fixed-charge correlators,
\begin{equation}
G_{n,\ell}=
\sum_{\mathclap{k_1,\dots,k_n=2}}^{\infty}
\,\avg{k_1\dots k_n}_\ell
\,,
\label{eq:GnlFromFixedCharges}
\end{equation}
where the correlators $\avg{k_1\dots k_n}$ of fixed-charge operators
$\op{O}_k$ can be extracted from $G_{n,\ell}$ via
\begin{align}
\brk[a]*{k_1k_2\cdots k_n}_\ell
&\equiv
\brk[a]*{\prod_{i=1}^{n}\mathcal{O}_{k_i}(x_i,y_i) \prod_{i=1}^{\ell}\Lint(x_i)}\subrm{\!SDYM}
\nn\\ &=
\eval*{
\frac{1}{k_1!}
\frac{\partial^{k_1}}{\partial t_1^{k_1}}
\cdots
\frac{1}{k_n!}
\frac{\partial^{k_n}}{\partial t_n^{k_n}}
\,G_{n,\ell}(x_i, t_i\,y_{i})
}_{t_i\to0}
\,.
\label{eq:Rweightcomponent}
\end{align}
%

\paragraph{Ten-Dimensional Poles.}

One notable feature of the generating
function~\eqref{eq:Gnl-integrand} is that it is expressed in terms of
ten-dimensional propagators
\begin{equation}
D_{ij} = -\frac{y_{ij}^2}{X_{ij}^2}
\,,\qquad
X_{ij}^2 = x_{ij}^2+y_{ij}^2
\,,
\label{eq:Dij}
\end{equation}
where
\begin{equation}
x_{ij}^2\equiv (x_{i}-x_{j})^2
\,,\quad
y_{ij}^2\equiv (y_{i}-y_{j})^2 = -2\, y_{i}\cdot y_{j}
\end{equation}
are the four-dimensional space-time and the six dimensional R-charge distances between operators $i$ and $j$.
These propagators $D_{ij}$ connect the external operators~\eqref{eq:Oscalar}.%
\footnote{The correlator~\eqref{eq:Gnl-integrand} is a component of a
parent supercorrelator of $n+\ell$ superoperators $\mathbb{O}$, in
which all propagators take the ten-dimensional form~\eqref{eq:Dij}, see \cite{Caron-Huot:2023wdh}. The component is recovered by a superprojection and the limit
$y_i\to0$, $i=n+1,\dots,\ell$, which demotes all propagators that connect to
Lagrangian operators $\Lint$ to ordinary four-dimensional form $d_{ij}$.}
Expanding around $y_{ij}^2=0$, they become
a geometric sum of arbitrary powers of ordinary four-dimensional propagators $d_{ij}$:
\begin{equation}
D_{ij}=d_{ij}+d_{ij}^2+d_{ij}^3+\dots
\,,\qquad
d_{ij}=-\frac{y_{ij}^2}{x_{ij}^2}
\,.
\label{eq:DijSeries}
\end{equation}
The component correlators~\eqref{eq:Rweightcomponent} are
recovered by performing this expansion and keeping only the appropriate powers
of the various~$y_i$ polarizations.

\begin{figure}[t]
\centering
\includegraphics[align=c,width=0.9\textwidth]{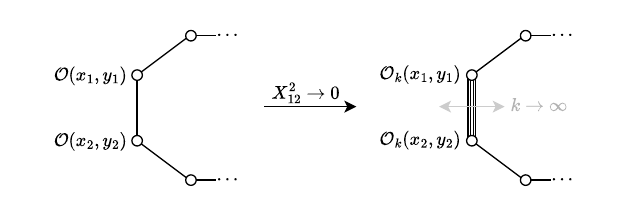}
\caption{The ten-dimensional null limit of the generating function
projects onto an infinite number of four-dimensional propagators
between the two operators.}
\label{fig:10dExample}
\end{figure}

Correlators of operators $\op{O}_k$ with small R-charge $k$ receive
contributions from the first terms in the series~\eqref{eq:DijSeries}.
Conversely, in the planar limit, correlators of operators with
parametrically large R-charge, come from the infinite tail of the
geometric series. This can be understood by noting that the Feynman
rules only admit propagators $D_{ij}$ that are homotopically distinct
from each other on the color sphere~\cite{Caron-Huot:2023wdh}. For a
finite number of operators, only finitely many such propagators can be
drawn on the color sphere. To accommodate for large-charge operators,
at least some of the propagators $D_{ij}$ must thus be expanded to
large orders. As already noted in~\cite{Caron-Huot:2021usw}, one can
focus on such large-charge correlators directly at the level of the
generating function~\eqref{eq:GnlFromFixedCharges} by taking the
ten-dimensional null limit: $X_{ij}^2 \to 0$ (or equivalently
$d_{ij}\to1$). By considering this limit, we project the generating
function to correlators where an infinite number of parallel
four-dimensional propagators $d_{ij}$ connect the operators at points
$x_i$ and $x_j$, as depicted in \figref{fig:10dExample}.

\paragraph{Polygons.}

\begin{figure}
\centering
\includegraphics[align=c,width=1\textwidth]{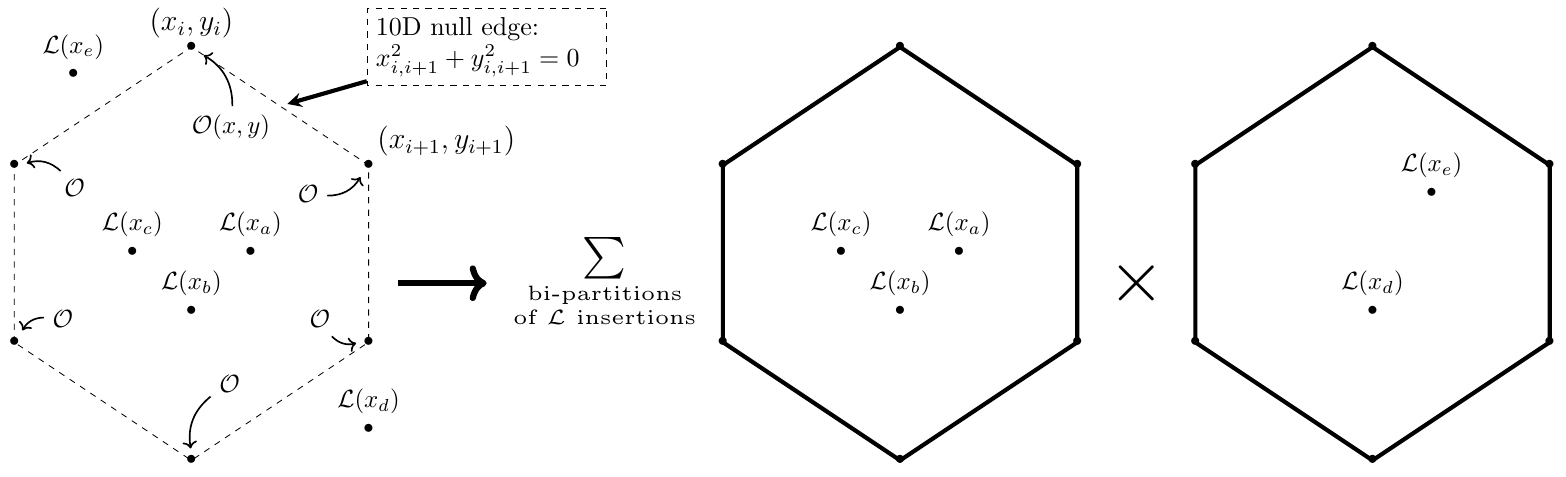}
\caption{The ten-dimensional null polygon limit factorizes the
correlator into two polygons.}
\label{fig:10dnullfactorization}
\end{figure}

A particularly interesting limit is the \emph{ten-dimensional null
polygon limit}, in which we simultaneously take $X_{i,i+1}^2 \to 0$,
$i=1,\dots,n$, such that the $n$ operators~\eqref{eq:Oscalar} sit at
the cusps of an $n$-sided polygon, where the edges are ``thick''
propagators of large (essentially infinite) R-charge that connect
neighboring operators. This thick propagator perimeter drastically
simplifies the correlator, since the ``inside'' and the ``outside'' of the perimeter decouple to all orders in perturbation theory.
The correlator thus factorizes into the product of two
simpler disk-like objects~\cite{Caron-Huot:2021usw}, the
\emph{polygons} $\polygon_{n,\ell}$:
\begin{equation}
\lim\limits_{\substack{d_{i,i+1}\to 1\\i=1,\dots,n}}
\,\prod_{i=1}^{n}(1-d_{i,i+1})\times \Nc^{n-2}\,G_{n,\ell}
=\sum_{\mathbold{k}}
\polygon_{n,\mathbold{k}}\times \polygon_{n,\mathbold{\bar{k}}}
\label{eq:PolygonEquation}
\end{equation}
where the factor
$\Nc^{n-2}$ is chosen such that $\polygon_{n,\mathbold{k}}\sim\Nc^0$, and the sum
on the right-hand side runs over all possible ways of distributing the $\ell$ Lagrangians onto the two polygons: $\mathbold{k}\mathbin{\dot\cup}\mathbold{\bar{k}}=\set{n+1,\dots,n+\ell}$. Thus, the $n$-point polygons
$\polygon_{n,\mathbold{k}}$ are correlators with disk topology, with
$n$ large-charge
operators inserted at the boundary of the disk, and the set
$\mathbold{k}$ of Lagrangian operators
inserted in the interior, see \figref{fig:10dnullfactorization}.
In the case where all Lagrangians are inserted in a single polygon, $\mathbold{k} =\set{n+1,\dots,n+\ell}$, we use the simplified notation $M_{n,\mathbold{k}}\equiv M_{n,\ell}$. These polygons $M_{n,\ell}$ define the integrands of the $n$-point interacting polygon correlator
\begin{equation}
\intpoly_n=
\sum_{\ell=0}^\infty\frac{(-g^{2})^{\ell}}{\ell!}
\int\brk[s]3{\,\prod_{k=1}^\ell\frac{\dd[4]{x_{n+k}}}{\pi^2}}
\polygon_{n,\ell}
\,.
\label{eq:PolygonIntegrated}
\end{equation}
Using~\eqref{eq:integratedcorrelator}, the
factorization~\eqref{eq:PolygonEquation} can be written at
integrated level as:
\begin{equation}
\lim\limits_{\substack{d_{i,i+1}\to 0\\i=1,\dots,n}}
\,\prod_{i=1}^{n}(1-d_{i,i+1}) \times \Nc^{n-2}\,
\avg*{\prod_{i=1}^{n}\mathcal{O}(x_i,y_i)}_{\!\text{SYM}}
= \intpoly_n \times \intpoly_n
\,.
\label{eq:PolygonEquationIntegrated}
\end{equation}
The four-point and five-point polygons $\intpoly_4$ and $\intpoly_5$ have
been conventionally called \emph{octagon} and \emph{decagon} in the
literature. The origin of these names lies in their
integrability-based description in terms of hexagon form
factors~\cite{Basso:2015zoa,Fleury:2016ykk}, in which the cusps
acquire a non-zero ``length''. In this work, we will call these
objects \emph{square} and \emph{pentagon}, respectively, and consider their
higher-point counterparts.

\medskip
\noindent
The polygon correlators $\polygon_{n,\ell}$ and $\intpoly_n$ satisfy
several interesting properties:
%
\begin{description}

\item[Factorization.] Upon taking additional ten-dimensional null
limits on their diagonals, the polygons $\polygon_{n,\ell}$
\emph{factorize} into lower-point polygons:
\begin{equation}
\lim_{d_{ij}\to0} (1-d_{ij}) \, \polygon_{n,\ell}
=\sum_{\mathbold{k}}
\polygon_{\mathbold{m},\mathbold{k}}\times \polygon_{\mathbold{\bar{m}},\mathbold{\bar{k}}}
\,,
\label{eq:PolygonFactorization}
\end{equation}
where $\mathbold{m}=\set{i,\dots,j\,(\mathrm{mod}\;n)}$ and
$\mathbold{\bar{m}}=\set{j,\dots,i\,(\mathrm{mod}\;n)}$ are the external operators
``left'' and ``right'' of the line $(i,j)$, and the sum runs over
all bipartitions
$\mathbold{k}\mathbin{\dot\cup}\mathbold{\bar{k}}=\set{n+1,\dots,n+\ell}$
of the Lagrangian insertion points onto the two factors.
Here, we employ the slight abuse of notation
$M_{\mathbold{m},\mathbold{k}}=M_{\abs{\mathbold{m}},\mathbold{k}}$
with point labels $\mathbold{m}$.
For example, the following limit splits the two-loop pentagon
$M_{5,2}$ into a square and a triangle:
\begin{equation}
\lim_{d_{14}\rightarrow0} (1-d_{14}) \, M_{5,2}
= M_{\mathbold{m},\emptyset}M_{\mathbold{\bar{m}},\set{6,7}}
+ M_{\mathbold{m},\set{6}}M_{\mathbold{\bar{m}},\set{7}}
+ M_{\mathbold{m},\set{7}}M_{\mathbold{\bar{m}},\set{6}}
+ M_{\mathbold{m},\set{6,7}}M_{\mathbold{\bar{m}},\emptyset}
\end{equation}
with $\mathbold{m}=\set{1,2,3,4}$ and $\mathbold{\bar{m}}=\set{1,4,5}$.
See \figref{fig:FactorizePolygon} for an illustration.
\begin{figure}
\centering
\includegraphics[width=\linewidth]{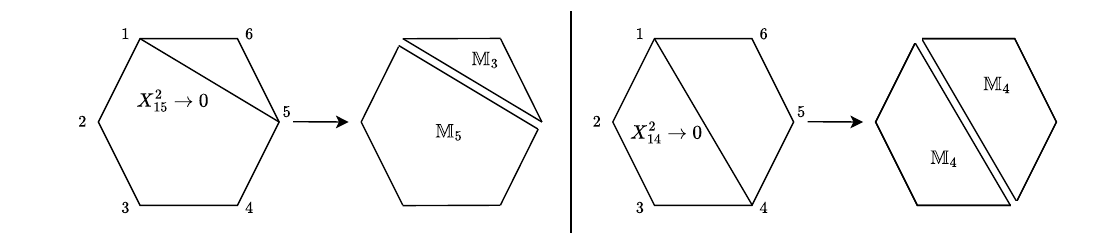}
\caption{Factorization of the hexagon into lower polygons after taking
ten-dimensional null limits on the diagonals.}
\label{fig:FactorizePolygon}
\end{figure}
The factorization~\eqref{eq:PolygonFactorization} follows by the same
reasoning as the factorization
of the generating function $G_{n,\ell}$ into two polygon
factors~\eqref{eq:PolygonEquation}: Taking the $X_{ij}^2\to0$ limit
projects to the tail of the geometric series~\eqref{eq:DijSeries},
thus it limits to terms that include a parametrically large number of
parallel propagators $d_{ij}$ that separates the polygon into two
subregions that do not interact with each other.
The factorization straightforwardly lifts to the integrated polygons:
\begin{equation}
\lim_{d_{ij}=0} (1-d_{ij}) \, \intpoly_n
= \intpoly_{\mathbold{m}} \times \intpoly_{\mathbold{\bar{m}}}
\,,
\label{eq:PolygonFactorizationIntegrated}
\end{equation}
with $\mathbold{m}$ and $\mathbold{\bar{m}}$ as in~\eqref{eq:PolygonFactorization}.

\item[No Double Poles.] The generating function $G_{n,\ell}$ in
general contains double and higher poles in the ten-dimensional
distances $X_{ij}^2$. These originate in Feynman graphs where external
operators $i$ and $j$ are connected by multiple homotopically distinct
propagators $D_{ij}$. In order to be homotopically distinct, any two
such propagators connecting the same two operators must enclose at
least one third operator.%
\footnote{Two propagators $D_{ij}$ are also homotopically distinct if
they only enclose Lagrangian operators, but no further external
operators. Such terms however evaluate to zero, which is in accordance
with the fact that two-point functions of scalar BPS operators are
protected from quantum corrections.}
The polygon disk-topology excludes such
diagrams, and thus the polygon correlators $\polygon_{n,\ell}$ only
contain at most simple poles in the ten-dimensional distances $X_{ij}^2$.
\item[Direct Computability.]
As we will see below in \secref{sec:twistor-rules-10d}, the polygons
$\polygon_{n,\ell}$ can be computed directly from the twistor Feynman rules
of~\cite{Caron-Huot:2023wdh}, avoiding the computation (and subsequent ten-dimensional polygon limit) of the much
more complicated generating functions $G_{n,\ell}$ altogether.

\item[Massive Amplitudes.] There is a
well-established duality in $\mathcal{N}=4$ super Yang--Mills that
relates the four-dimensional null polygon limit $x_{i,i+1}^2 \to 0$
of the $n$-point correlation function of the lightest half-BPS
operators, $k_i=2$ in~\eqref{eq:GnlFromFixedCharges}, with the
$n$-point \emph{massless} MHV gluon scattering
amplitude~\cite{Eden:2010zz, Alday:2010zy, Mason:2010yk, Eden:2010ce,
Caron-Huot:2010ryg, Adamo:2011dq, Eden:2011ku, Eden:2011yp}.
A less firmly established
duality, but with corroborating evidence in the case of
four~\cite{Caron-Huot:2021usw} and five
points~\cite{Belitsky:2025bgb,Bork:2022vat}, is that something similar
also happens in the large R-charge limit.
The statement is that the $n$-point polygons $\intpoly_n$ are dual to
\textit{massive} $n$-point W-boson scattering amplitudes on the
Coulomb branch of $\mathcal{N}=4$ super Yang--Mills theory~\cite{Caron-Huot:2021usw}.
Upon additionally taking the \emph{massless} null polygon limit $x_{i,i+1}^2 \to 0$,
these further reduce to the massless MHV gluon amplitudes.
See \figref{fig:amplitudeDiagram} for a diagram of the various limits.

More precisely, the polygons $M_{n,\ell}$ reduce to the $\ell$-loop
integrands $\mathcal{A}_{n,\ell}$ of the \emph{massless} MHV scattering
amplitudes when one sets all internal diagonals $d_{ij}$ to zero and
passes to the four-dimensional null limit:
\begin{equation}
\polygon_{n,\ell}
\to
\mathcal{A}_{n,\ell}/\mathcal{A}_{n,0}
\qquad \text{for} \qquad
x_{i,i+1}^2\to0, \; i=1,\dots,n
\quad \text{and} \quad
d_{ij}\to0 \; \forall \; i,j
\,,
\label{eq:masslessAmpLimit}
\end{equation}
where $\mathcal{A}_{n,\ell}$ is the $\ell$-loop massless MHV amplitude
integrand, and $\mathcal{A}_{n,0}$ its tree-level expression. The
relation~\eqref{eq:masslessAmpLimit} follows by construction when both
the amplitude integrands $\mathcal{A}_{n,\ell}$ and the polygons
$M_{n,\ell}$ are computed from Lagrangian insertions, see
\eg~\cite{Adamo:2011dq,Eden:2011ku}.
The limit can simply be achieved by setting all $y_{ij}^2$ to zero,
noticing that $x_{i,i+1}^2=-y_{i,i+1}^2\to0$ on the perimeter of the
polygon, due to the ten-dimensional null limit
$0=X_{i,i+1}^2=x_{i,i+1}^2+y_{i,i+1}^2$.
\begin{figure}
\centering
\includegraphics[width=.75\linewidth]{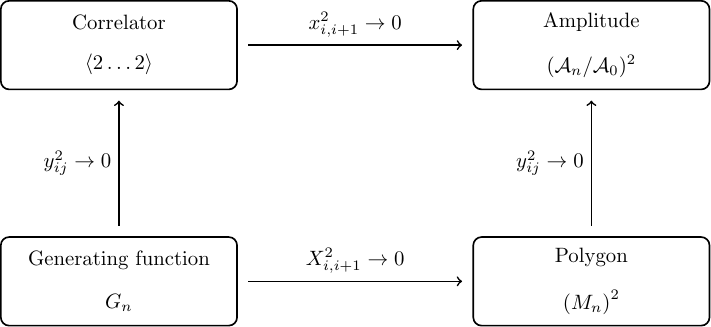}
\caption{Illustrating of the relation between
correlator–amplitude duality (top) and generating functions/polygons
(bottom)~\cite{Caron-Huot:2021usw}. Arrows denote schematic limits:
the left arrow projects to the leading term as $y_{ij}^2\to0$, yielding
the $n$-point $\twentyprime$ integrand, while top and bottom limits
extract the leading divergent contribution in the 4d/10d null limit.}
\label{fig:amplitudeDiagram}
\end{figure}
%

\item[Integrability.] The polygons $\mathbb{M}_n$ are the most natural
objects to be computed from integrability via hexagonalization, \ie
decomposition into hexagon form factors~\cite{Basso:2015zoa,Fleury:2016ykk,Eden:2016xvg}.
Due to their disk topology, they are free of mirror excitations that
wrap operator insertions that require careful treatment~\cite{Basso:2015eqa}.
The polygon configurations in fact make
hexagonalization conceptionally very similar to the
pentagon OPE for massless amplitudes (null polygon Wilson
loops)~\cite{Basso:2013vsa,Basso:2013aha,Basso:2014koa,Basso:2014nra},
and in the four-dimensional null limit, the two should become equivalent.
In the case of the square (\aka octagon),
the simplifications of the ten-dimensional null limit enabled its
bootstrap to all loops~\cite{Coronado:2018cxj,Kostov:2019stn,Kostov:2019auq}, and a finite-coupling
resummation in terms of a Fredholm
determinant~\cite{Belitsky:2019fan,Belitsky:2020qrm}.
The pentagon (\aka decagon) has been computed to two loops~\cite{Fleury:2020ykw}.
Also more general
correlators become most computable in limits where they can be
decomposed into polygons~\cite{Bargheer:2019kxb},
even at subleading non-planar orders of the planar
limit expansion~\cite{Bargheer:2017nne,Bargheer:2018jvq}.

\end{description}
%

\paragraph{Relation with Fixed-Charge Correlators.}

The full generating function $G_{n,\ell}$ contains the complete
information on all fixed-charge correlators
via~\eqref{eq:Rweightcomponent}. Since the ten-dimensional null
polygon limit captures only a specific part of the generating
function, it also only contains partial information on any given
fixed-charge correlator. Without involving the full generating
function, any single generic fixed-charge correlator cannot easily be
projected to the part captured by polygons. An exception to this are
certain correlators at four, five, and six points, where one can
engineer specific operators of large (but fixed) charge, such that the
correlator equals the square, pentagon, or hexagon, up to a
loop order that is bounded by the operator charges. The operators are
R-symmetry descendants of the fixed-charge operators
$\op{O}_k$~\eqref{eq:Oscalar}, designed so that they form a
square/pentagon/hexagon, where only neighboring operators can contract
with free propagators. This is how the square was
originally defined~\cite{Coronado:2018ypq} (see
also~\cite{Bargheer:2019kxb}, in particular Figure~11 there).

\begin{figure}
    \centering
    \includegraphics[width=\linewidth]{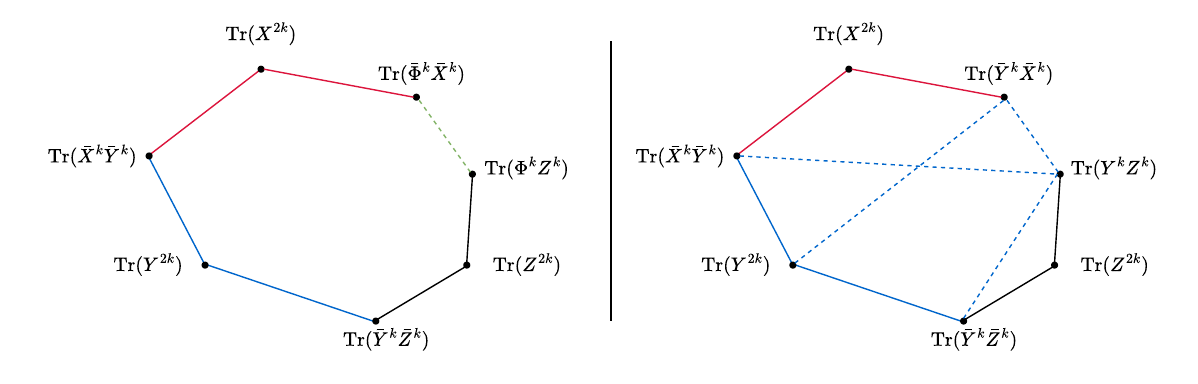}
    \caption{Seven point correlation functions of large charge operators. On the left what one would like to engineer with a field $\Phi$ so that the correlator is given by a thick perimeter and factorizes into two heptagons. There is no choice of fields $\Phi= \{X,\bar{X},Z,\bar{Z},Y,\bar{Y}\}$ that results into this single configuration. As example, on the right we have the case of $\Phi = Y$ where one has to sum over all the possible ways of distributing the $Y\bar{Y}$ propagators (blue), only one of these configurations will be factored out in the product of two heptagons.}
    \label{higherPointsFig}
\end{figure}

Since $\mathcal{N}=4$ super Yang-Mills has only six scalars that
transform in the $\grp{SO}(6)$ R-symmetry, this procedure only works
up to six points. Beyond six points, there is not enough space in the
$\grp{SO}(6)$ to ensure large propagator bundles between neighboring
operators. Therefore, any higher-point function with sufficiently
large charges will have contributions that factorize into polygons,
but will also have other contributions that do not factorize, as exemplified in \figref{higherPointsFig}.

Even though there is no one-to-one map between polygons and
large-charge correlators beyond six points, these polygons are interesting objects on
their own --- as corroborated by their properties listed above; in
particular their identification as massive amplitudes and their
particularly nice integrability representation in terms of hexagon
form factors. Computing these
objects in perturbation theory is the first step towards better
understanding them and it is what we turn to next.

\section{Polygons from Generating Functions}
\label{sec:PolyFromG}

Before addressing their direct construction from twistor Feynman
rules, we will compute the first few one-loop and two-loop polygons
by taking the ten-dimensional null limits~\eqref{eq:PolygonEquation}
of the generating functions $G_{n,\ell}$ determined
in~\cite{Caron-Huot:2023wdh,Bargheer:2025uai}. To this end, we first
define the null polygon residue
\begin{equation}
\label{eq:10dnull}
R_{n,\ell} \equiv
\Res_{\underset{i=1,\dots,n}{d_{i,i+1}=1}} \Nc^{n-2}\,G_{n,\ell}
= \lim_{d_{i,i+1}\to 1}\,\prod_{i=1}^n (1-d_{i,i+1})
\times \Nc^{n-2}\,G_{n,\ell}
\end{equation}
Looking at~\eqref{eq:PolygonEquation}, one can see that the polygons
$M_{n,\ell}$ can be determined from $R_{n,\ell}$ loop order by loop
order. Explicitly:
\begin{equation}
\label{eq:computePolygons}
M_{n, 0} = (R_{n, 0})^{1/2}
\,,\quad
M_{n, 1} = \frac{R_{n,1}}{2\,M_{n,0}}
\,,\quad
M_{n, 2} = \frac{R_{n,2}-2\,(M_{n,1})^2}{2\,M_{n,0}}
\,,\quad
\ldots
\,.
\end{equation}
The simplest polygon is the triangle. Since three-point functions of
half-BPS scalar operators are protected from quantum corrections, the
generating function is simply given by its leading-order term
$G_{3,0}=D_{12}D_{13}D_{23}$, and therefore the triangle correlator is
trivial:
\begin{equation}
\intpoly_3=\polygon_{3,0}=1
\,.
\end{equation}
%

\subsection{The Square (\texorpdfstring{\aka}{a.k.a.} Octagon)}

Both the four-point generating function $G_{n,\ell}$ and its
ten-dimensional null polygon limit were computed
in~\cite{Caron-Huot:2021usw}. Collecting the results, the square
correlator reads
\begin{align}
M_{4,0} \,&=\, 
\frac{x_{13}^2x_{24}^2 -y_{13}^2 y_{24}^2}{X_{13}^2 X_{24}^2}
\,,\nn\\
M_{4,1} \,&=\, M_{4,0}\, \frac{X_{13}^2 X_{24}^2}{X_{15}^2 X_{25}^2
X_{35}^2 X_{45}^2}\,\bigg{|}_{y_{5}\to 0}
\,,\nn\\
M_{4,2} \,&=\, M_{4,0}\, \frac{X_{13}^2 X_{24}^4\,X_{15}^2 X_{36}^2 \,+\,X_{13}^4 X_{24}^2\,X_{45}^2 X_{26}^2 + (X_5\leftrightarrow X_6)}{\left(X_{15}^2 X_{25}^2 X_{35}^2 X_{45}^2\right) X_{56}^2\left(X_{16}^2 X_{26}^2 X_{36}^2 X_{46}^2\right)}\,\bigg{|}_{y_{5},y_{6}\to 0}
\label{eq:M4lintro}
\end{align}
Before taking the $y_5,y_6\to0$ limits, these functions include the
dependence on all higher Kaluza--Klein modes of the Lagrangian operator.
Written in this form, the ratios $M_{4,\ell}/M_{4,0}$ display a 10D (dual)
conformal symmetry, which only gets broken by taking the $y_i\to0$
limit to project to the chiral interaction Lagrangian $\Lint(x)$. It
corresponds to the loop-integrand in higher-dimensional SYM, and by
dimensional reduction gives the integrand of a massive amplitude on
the Coulomb branch~\cite{Caron-Huot:2021usw}.

Performing the $y_5,y_6\to0$ projection, the square correlator can
alternatively be expanded in a basis of integrands of conformal integrals:%
\footnote{\label{fn:integralLabels}The labels $\mathcal{I}_5$, $\mathcal{I}_6$,
and $\mathcal{I}_7$ for the conformal integrals displayed
in~\eqref{eq:M41}--\eqref{eq:pentagonIntegrals} are inherited
from~\cite{Bargheer:2025uai}.}
\begin{align}
M_{4,0} \,&=\, \frac{1-d_{13}d_{24}}{(1-d_{13})(1-d_{24})} \,, \\
M_{4,1} \,&=\, (1-d_{13}d_{24})x_{13}^2x_{24}^2 B^{[1,2,3,4;\,5]}
\,, \label{eq:M41} \\
M_{4,2} \,&=\, (1-d_{13}d_{24})x_{13}^2x_{24}^2 \left[\mathcal{I}_{5}^{[1,3|2,4;\,5,6]}\,X_{13}^2+\mathcal{I}_{5}^{[2,4|1,3;\,5,6]}\,X_{24}^2\right]
\,,
\label{eq:square2loop}
\end{align}
where (see also \figref{fig:onshellprops})
\begin{align}
B^{[1,2,3,4;\,a]}
&= \frac{1}{x_{1a}^2x_{2a}^2x_{3a}^2x_{4a}^2}
\label{eq:oneloopbox}
\,,\\
\mathcal{I}_{5}^{[1,2|3,4;\,a,b]}
&=\frac{1}{(x_{1a}^2 x_{2a}^2 x_{3a}^2) x_{ab}^2(x_{1b}^2 x_{2b}^2 x_{4b}^2)}
+ (a\leftrightarrow b)
\,.
\label{eq:twoloopladder}
\end{align}
Note that the square correlators $\polygon_{4,\ell}$, $\ell>0$ take
the form of a universal prefactor $(1-d_{13}d_{24})x_{13}^2x_{24}^2$,
multiplied by a combination of conformal integrands with
ten-dimensional coefficients. This form follows from the
ten-dimensional nature of the parent expressions~\eqref{eq:M4lintro},
and it continues at higher loop orders.
The prefactor $(1-d_{13}d_{24})x_{13}^2x_{24}^2$ in fact originates
from the null polygon limit of the universal four-point
supersymmetry invariant $R_{1234}$ that appears as an overall factor in the
four-point generating function~\cite{Eden:2011we}:
$G_{4,\ell}=R_{1234}\mathcal{H}_{4,\ell}$, with $\mathcal{H}_{4,\ell}$
the reduced correlator \cite{Caron-Huot:2021usw}:
\begin{equation}
\lim_{d_{i,i+1}\to1} R_{1234}
=(1-d_{13}d_{24})^2 x_{13}^2 x_{24}^2
\,.
\label{eq:RsquareLimit}
\end{equation}
%

\subsection{The Pentagon}

\begin{figure}[t]
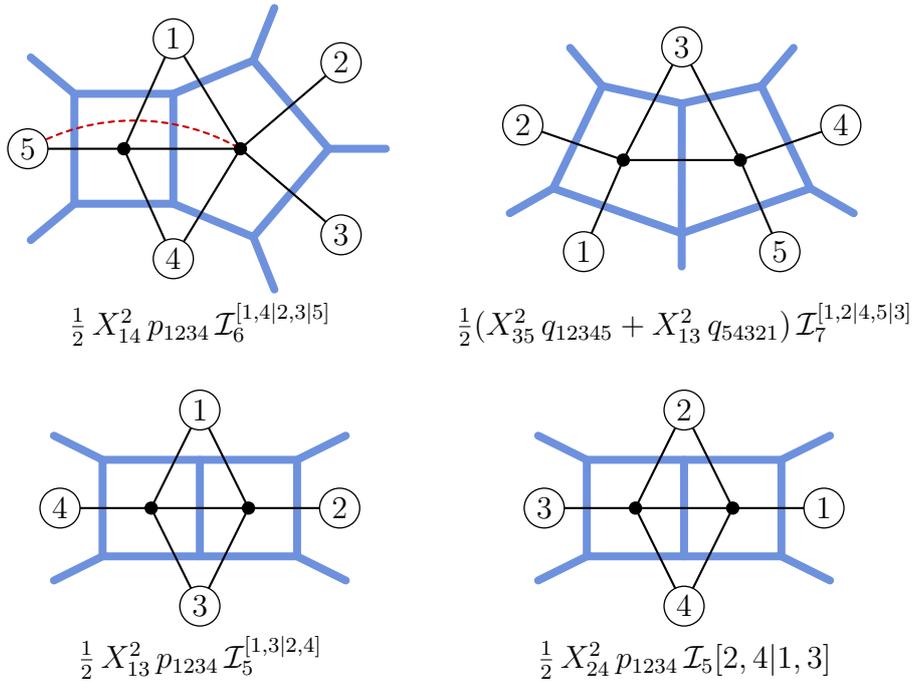

\centering
\begin{tabular}{c@{\qquad}c}
\includegraphics[align=c]{FigInt6WithDual} &
\includegraphics[align=c]{FigInt7WithDual} \\ \rule{0pt}{3ex}%
$\frac{1}{2}\,X_{14}^2\,p_{1234}\,\mathcal{I}_6^{[1,4|2,3|5]}$ &
$\frac{1}{2}(X_{35}^2\,q_{12345}+X_{13}^2\,q_{54321})\,\mathcal{I}_7^{[1,2|4,5|3]}$ \\[3ex]
\includegraphics[align=c]{FigInt5WithDual1} &
\includegraphics[align=c]{FigInt5WithDual2} \\ \rule{0pt}{3ex}%
$\frac{1}{2}\,X_{13}^2\,p_{1234}\,\mathcal{I}_5^{[1,3|2,4]}$ &
$\frac{1}{2}\,X_{24}^2\,p_{1234}\,\mathcal{I}_5{[2,4|1,3]}$
\end{tabular}
\\
\caption{Conformal integrals that appear in the two-loop pentagon
$M_{5,2}^\mathrm{even}$, including their coefficient functions. On the first
row, we present genuine five-point integrals, the second row shows the
two instances of the four-point double-box. The definitions of the integrals are in
\protect\eqref{eq:twoloopladder},
\protect\eqref{eq:pentagonIntegrals}, and the factors $p_{1234}$ and
$q_{12345}$ are defined in
\protect\eqref{eq:p}, \protect\eqref{eq:q}.
Taking an extra residue, only the four-point integrals
survive, and the pentagon $M_{5,2}^\mathrm{even}$ descends to the square
$M_{4,2}$~\protect\eqref{eq:PentagonToSquare}.}
\label{fig:onshellprops}
\end{figure}

Using the five-point generating functions provided
in~\cite{Bargheer:2025uai} (see (6.16) and (6.19) there), we compute
the one-loop and two-loop pentagon by applying~\eqref{eq:10dnull}
and~\eqref{eq:computePolygons}:\footnoterepeat{fn:integralLabels}
\begin{align}
\polygon_{5,0} &=
\frac{1}{5} + \frac{d_{13}}{(1-d_{13})(1-d_{14})}+(\grp{C}_5\text{ perms})
\label{eq:M50}
\\
\polygon_{5,1} &= \frac{p_{1234}}{2}\,B^{[1,2,3,4;\,6]}-2i\,\mathcal{I}_\mathrm{odd}^{[1,2,3,4;\,6]}+(\grp{C}_5 \text{ perms})\,,
\label{eq:M51}
\\
\polygon_{5,2}^\mathrm{even} &=  \frac{p_{1234}}{2}\left(\mathcal{I}_5^{[1,3|2,4;\,6,7]}\, X_{13}^2+\mathcal{I}_5^{[2,4|1,3;\,6,7]}\, X_{24}^2 +\mathcal{I}_{6}^{[1,4|2,3|5;\,6,7]}\,X_{14}^2\right)  \nonumber\\
&\quad +\frac{X_{35}^2\,q_{12345} +X_{13}^2\,q_{54321}}{2}\,\mathcal{I}_{7}^{[1,2|4,5|3;\,6,7]} + (\grp{C}_5 \text{ perms})\,,
\label{eq:M52}
\end{align}
The two-loop pentagon is restricted to its parity-even part.
The parity-odd part of the one-loop pentagon is written as a sum of
four-point parity-odd integrands
\begin{equation}
\mathcal{I}_\mathrm{odd}^{[1,2,3,4;\,0]} = \frac{\epsilon_{\mu\nu\rho\sigma}x^\mu_{10}x^\nu_{20}x^\rho_{30}x^\sigma_{40}}{x_{10}^2x_{20}^2x_{30}^2x_{40}^2}\,,
\label{eq:OddIntegral}
\end{equation}
and we introduce the following conformal integrals for the two-loop pentagon
(see \figref{fig:onshellprops})
\begin{align}
\mathcal{I}_{6}^{[1,2|3,4|5;\,a,b]}
&= \frac{x_{5a}^2}{(x_{1a}^2 x_{2a}^2 x_{3a}^2 x_{4a}^2 )x_{ab}^2 (x_{1b}^2 x_{2b}^2 x_{5b}^2 ) }+(a\leftrightarrow b)
\,,\nonumber\\
\mathcal{I}_{7}^{[1,2|3,4|5;\,a,b]}
&= \frac{1}{(x_{3a}^2 x_{4a}^2 x_{5a}^2 ) x_{ab}^2 (x_{1b}^2 x_{2b}^2 x_{5b}^2   ) } +(a\leftrightarrow b)
\,.
\label{eq:pentagonIntegrals}
\end{align}
The result~\eqref{eq:M52} for the two-loop pentagon $M_{5,2}$ is
consistent with the known expressions for the
``decagon'' function~\cite{Fleury:2020ykw,Bercini:2024pya}.

Similar to the case of the square, one can see that the pentagon
expressions at one and two loops are
written in terms of four-dimensional structures $p$ and $q$ that
multiply linear combinations of conformal integrals with
ten-dimensional coefficients.
The four-dimensional structures are
\begin{align}
    p_{1234} &= x_{12}^2 x_{34}^2 -x_{14}^2 x_{23}^2 +\frac{(1+d_{14}-2d_{13}d_{24})}{1-d_{14}}x_{13}^2 x_{24}^2\,,
    \label{eq:p} \\
    q_{12345} &= x_{12}^2 x_{34}^2 - x_{14}^2 x_{23}^2- \frac{d_{35}(1-d_{13}d_{24})}{1-d_{35}} x_{13}^2 x_{24}^2
    \,.
    \label{eq:q}
\end{align}
The interpretation of these structures is that they are polygon limits
of components of superconformal invariants that occur in the
five-point generating functions, similar to~\eqref{eq:RsquareLimit} at
four points. This picture is further supported by the fact that $p$
can be traced back to the invariants $\mathcal{R}_{1234}$ and
$\mathcal{R}_{1234,5}$ that appear in the five-point generating
function as follows:%
\footnote{At one loop, the
combination~\eqref{eq:Rcombination} is apparent in (6.16)
of~\cite{Bargheer:2025uai}. At two loops, it appears more
subtly in (6.21) when dressing the integral $\mathcal{I}_5$ with $f_5$
given in (6.24).}
\begin{equation}
\Res_{d_{i,i+1}=1} \frac{\mathcal{R}_{1234,5}-C^{(5)}_{1234;5}\,\mathcal{R}_{1234}}{\prod_{i<j=1}^5 w_{ij}} \, = \, M_{5,0}\,\times\, p_{1234}
\,,
\label{eq:Rcombination}
\end{equation}
where $\mathcal{R}_{1234}$, $\mathcal{R}_{1234,5}$, and
$C^{(5)}_{1234;5}$ are given in (5.4), (6.14), and (6.12)
of~\cite{Bargheer:2025uai}.

While the four-point generating functions $G_{4,\ell}$ are purely
parity-even, the higher-point generating functions develop parity-odd
parts, which also induce
parity-odd parts for the respective polygons. Since the parity-odd parts
integrate to zero, one might be tempted to ignore them.
However, when one takes products of polygons to re-construct the null
polygon correlators as in~\eqref{eq:PolygonEquation}, products of
parity-odd parts are essential, as they contribute to the parity-even
part of the correlator. In other words, the
factorization~\eqref{eq:PolygonEquation} is only consistent if one
includes the parity-odd parts of both the generating functions
$G_{n,\ell}$ and the polygons~$M_{n,\ell}$. The parity-odd
parts are similarly essential for the factorization
formula~\eqref{eq:PolygonFactorization}.

With the expressions above, we can explicitly verify the factorization
formula~\eqref{eq:PolygonFactorization}: Taking the ten-dimensional
null limit of the diagonal $X_{14}^2\to0$ (or equivalently $d_{14}\to1$
in terms of the four-dimensional propagator), the pentagon factorizes
into a square and a (trivial) triangle
\begin{equation}
\Res_{d_{14}=1}\,\polygon_{5,\ell}
= \polygon_{4,\ell} \times \polygon_{3,0}
= \polygon_{4,\ell} \times 1
\,, \qquad
\ell=0,1,2
\,,
\label{eq:PentagonToSquare}
\end{equation}
at leading order, one-loop, and two-loop order.
Consistently, the parity-odd part of the one-loop pentagon vanishes in this limit.
Notice that the poles in $p$ and $q$ in the two-loop pentagon are canceled by the factors
$X_{ij}^2$ dressing the penta-box integral $\mathcal{I}_{6}$ and the five-point
double-box $\mathcal{I}_{7}$, whereas the poles of $p$ are still present
for the four-point box $\mathcal{I}_{5}$, thus one can see that
$\mathcal{I}_6$, $\mathcal{I}_7$, and $q$ do not contribute to the limit~\eqref{eq:PentagonToSquare}.

\medskip
\noindent
Before moving on to the hexagon, in the following we provide some more
details on the derivation of the pentagon from the generating function.

\paragraph{One Loop.}

At five points and one loop, a more general correlator than $G_{5,1}$
has been computed~\cite{Caron-Huot:2023wdh,Bargheer:2025uai}: The generating
function $\tilde{G}_{5,1}$ that also includes all higher-R-charge components of
the chiral Lagrangian operator (see~(6.11)
in~\cite{Bargheer:2025uai} for the expression). The correlator
$G_{5,1}$~\eqref{eq:Gnl-integrand} is obtained from $\tilde{G}_{5,1}$
by taking the $y_6\to0$ limit. Applying the 10d null polygonal limit
\eqref{eq:10dnull}, the generating function factorizes according
to~\eqref{eq:PolygonEquation} and~\eqref{eq:10dnull}, $\tilde R_{5,1} = 2 M_{5,0}\tilde
M_{5,1}$, where $\tilde{M}_{5,1}$ is a generalized one-loop pentagon
that includes all higher-R-charge components of the chiral Lagrangian:
\begin{align}
\tilde{M}_{5,1} &=
\frac{
X_{56}^2\,p_{1234}
+ d_{56}x_{56}^2\brk1{(1-d_{13}d_{24})x_{13}^2 x_{24}^2-(1-d_{14})x_{14}^2 x_{23}^2}
+ (\grp{C}_5 \text{ perms})
}{
2\,X_{16}^2 X_{26}^2 X_{36}^2 X_{46}^2 X_{56}^2
}
\nn \\ & \quad
-\frac{
4 \ii \eval[s]1{X^{\mathrm{anti}}_{123456} -
Y^{\mathrm{anti}}_{123456}}_{\text{10d null}}
}{
X_{16}^2 X_{26}^2 X_{36}^2 X_{46}^2 X_{56}^2
}\,,
\end{align}
with $p_{1234}$ given in~\eqref{eq:p}.
$X^\mathrm{anti}$ and $Y^\mathrm{anti}$ are antisymmetric scalars
that can be defined using the six-dimensional projective vectors $X^I$
and $Y^I$, which transform in the fundamental representation of the
conformal group $\grp{SO}(2,4)$ and R-symmetry group $\grp{SO}(6)$,
respectively:
\begin{align}
X^\mathrm{anti}_{123456} &= \varepsilon_{IJKLMP}X^I_1X^J_2X^K_3X^L_4X^M_5X^P_6\,,
\nonumber \\
Y^\mathrm{anti}_{123456} &= \varepsilon_{IJKLMP}Y^I_1Y^J_2Y^K_3Y^L_4Y^M_5Y^P_6 \,.
\end{align}
The subscript ``10d null'' in the above equation indicates that the
$X^I$ and $Y^I$ vectors are not independent, but satisfy the 10d null
constraint on the boundary of the pentagon: $X_i^I
X^I_{i+1}-Y^I_iY^I_{i+1}=0$.

By setting $y_6=0$, we project out all higher-R-charge partners of the
Lagrangian, and thus, obtain the loop-integrand of the pentagon at
one-loop order~\eqref{eq:M51}.
The scalar $Y\suprm{anti}_{12345}$ projects to zero, and the
antisymmetric $X$-scalar becomes the parity-odd integral~\eqref{eq:OddIntegral},
\begin{equation}
X^\mathrm{anti}_{123456} = \frac{1}{2}x_{16}^2 x_{26}^2 x_{36}^2 x_{46}^2x_{56}^2\,I_\mathrm{odd}^{[1,2,3,4;\,6]} + (\grp{C}_5 \text{ perms})\,.
\end{equation}
%

\paragraph{Two Loops.}

Obtaining the two-loop pentagon using the last equation
in~\eqref{eq:computePolygons} requires a careful treatment of the
parity-even and -odd parts in the one-loop pentagon. Starting from the
parity-even part of five-point two-loop generating function, we
compute the parity-even two-loop pentagon via
\begin{equation} \label{eq:M52even}
    M_{5,2}^\mathrm{even} = \frac{R_{5,2}^\mathrm{even}-2\,(M_{5,1}^\mathrm{even})^2+2\,(M_{5,1}^\mathrm{odd})^2}{2\,M_{5,0}}\,.
\end{equation}
The parity-odd part of the one-loop pentagon squared yields an even
contribution to the residue $R_{5,2}^\mathrm{even}$ that needs to be
subtracted in order to obtain the two-loop pentagon. Since
$M_{5,1}^\mathrm{odd}$ vanishes upon integration of $x_6$, its
contribution to~\eqref{eq:M52even} can also be omitted after lifting
the expression to the integrated level. In this case, however, an
integral identity has to be applied, which is exactly given by
$(M_{5,1}^\mathrm{odd})^2=0$ (after integration). In fact, this
integral identity is a generalization of (6.23) in \cite{Eden:2010zz},
where it was found in the context of the correlator/amplitude duality
in the 4d null limit, where $x_{i,i+1}^2=0$.

Plugging the generating function $G_{5,2}$, provided in (6.19)
in~\cite{Bargheer:2025uai}, into~\eqref{eq:M52even}, we obtain the
parity-even two-loop pentagon~\eqref{eq:M52}.
Graphically, we depict this expression in terms of the conformal
integrals~\eqref{eq:twoloopladder} and~\eqref{eq:pentagonIntegrals}
in~\figref{fig:onshellprops}.
%

\subsection{The Hexagon}

Taking the parity-even part of one-loop six-point generating function
(cf.~(7.1) in \cite{Bargheer:2025uai}), we consider its 10d null
hexagon limit and apply~\eqref{eq:computePolygons} to find the
parity-even part of the one-loop hexagon%
\footnote{For the full expressions of all one-loop polygons, that
include both the parity-even as well as the parity-odd parts, we refer
to~\secref{sec:oneLoopPolygons}.}
\begin{align}
M_{6,1}^\mathrm{even} &=  \frac{(1-d_{15}d_{24})}{4\,(1-d_{15})(1-d_{24})}\left(x_{12}^2 x_{45}^2 - x_{15}^2 x_{24}^2 +\frac{(1+d_{15}d_{24}-2d_{14}d_{25})}{1-d_{15}d_{24}}x_{14}^2 x_{25}^2\right)B^{[1,2,4,5;\,7]}
\nonumber\\
&\quad - \frac{(d_{15}-d_{35})}{2\,(1-d_{15})(1-d_{35})}\left(x_{12}^2 x_{35}^2 - x_{15}^2 x_{23}^2 - \frac{(d_{15}+d_{35}-2d_{13}d_{25})}{d_{15}-d_{35}}x_{13}^2 x_{25}^2\right)B^{[1,2,3,5;\,7]}
\nonumber \\
&\quad + \frac{(1-d_{15}d_{46})}{2\,(1-d_{15})(1-d_{46})}\,p_{1234}\, B^{[1,2,3,4;\,7]}
+ (\grp{C}_6 \text{ perms})\,.
\label{eq:M61}
\end{align}
We present the hexagon in terms of one-loop box
integrals~\eqref{eq:oneloopbox}, and notice the appearance of
$p_{1234}$ that we already encountered previously in the one- and two-loop
pentagon.

We verify the factorization of the hexagon into lower-point polygons
by taking the following residues (see \figref{fig:FactorizePolygon}
for illustrations of these two limits):
\begin{align}
\Res_{d_{15}=1}\,M_{6,1}^\mathrm{even}
&= M_{5,1}^\mathrm{even} \times 1
\,,
\nn \\
\Res_{d_{36}=1}\,M_{6,1}^\mathrm{even}
&= M_{\set{1236},0} \times M_{\set{3456},1}
+ M_{\set{1236},1} \times M_{\set{3456},0}
\,,
\end{align}
where we recover the one-loop pentagon~\eqref{eq:M51} (multiplying the
trivial triangle) and products of the one-loop square~\eqref{eq:M41}
(with different point labels).

\medskip
\noindent
The explicit computations carried out in this section verify the two
types of factorizations we considered: The factorization of the
generating function into a product of
polygons~\eqref{eq:PolygonEquation},
\eqref{eq:PolygonEquationIntegrated}, and the factorization of
polygons into products of smaller
polygons~\eqref{eq:PolygonFactorization},
\eqref{eq:PolygonFactorizationIntegrated}.
Although straightforward, first evaluating the full generating
function to then take its ten-dimensional null polygon limit to
factorize it into polygons appears like a long detour. The generating
function contains much more information than the polygons, and its
computation is accordingly demanding. The polygons have disk topology
and contain much less information, hence there should be an easier way
to compute them. Indeed it turns out there is a way to compute them
directly, which is what we turn to next.

\section{Twistor Rules for 10d Null Polygons}
\label{sec:twistor-rules-10d}

So far, we computed polygons $M_{n,\ell}$ by taking the square root of
a 10d null polygonal limit $X_{i,i+1}^2\rightarrow0$, $i=1,\dots,n$, of the
corresponding generating function $G_{n,\ell}$. Deriving the
generating function using the twistor formulation~\cite{Chicherin:2014uca,Caron-Huot:2023wdh}, however, is
computationally quite demanding as both the number of graphs as well
as the size of the ansatz grows quickly for
higher points and/or higher loops.

The full generating function $G_{n,\ell}$ contains information of all
underlying fixed-weight correlators, whereas the polygons only
correspond to a specifically polarized large-charge limit. Hence, computing
the polygons directly -- without having to take a detour deriving the
generating function -- should be significantly more efficient. In fact, it is
possible to compute the polygons directly in the twistor formalism,
using the same Feynman rules as for the generating function, but
restricting to graphs with disk topology and applying ten-dimensional null
kinematics.

\paragraph{Prescription.}

\begin{table}
\centering
\begin{tabular}{crrrrrrrr}
\toprule
$\ell\backslash n$  & $4$       & $5$        & $6$         & $7$        & $8$          & $9$         & $10$            & $11$ \\
\midrule
$0$                 & $2$       & $3$        & $11$        & $29$       & $122$        & $479$       & $2\,113$        & $9\,369$ \\
$1$                 & $9$       & $36$       & $176$       & $830$      & $4\,125$     & $20\,632$   & $104\,924$ & $538\,746$ \\
$2$                 & $135$     & $740$      & $4\,203$    & $23\,273$  &  $130\,113$  & $726\,250$  & $4\,069\,167$   & $22\,838\,831$ \\
$3$                 & $2\,683$  & $17\,210$  & $106\,904$  &            &              &             &                 & \\
$4$ & $61\,395$ & $425\,819$ & & & & & & \\
\bottomrule
\end{tabular}
\caption{Counting of single-particle twistor graphs with disk topology
for various number of external points $n$ and internal insertions $\ell$. These are
ribbon (or fat) graphs with $n$ external vertices forming the
perimeter of an $n$-sided polygon, and $\ell$ internal vertices
located inside the disk. The perimeter consists of simple edges, and
only graphs that are inequivalent under $\grp{C}_n\times\grp{S}_\ell$
permutations of the vertices are counted as distinct.
The first line ($\ell=0$) counts the number of dissections of the
$n$-gon by diagonals, see \href{https://oeis.org/A003455}{OEIS
Sequence A003455}.}
\label{tab:graphcounting}
\end{table}

To compute the full generating function $G_{n,\ell}$ in twistor space,
one has to sum over all planar diagrams, \ie graphs with sphere
topology. Once we take the ten-dimensional null polygon limit, the
generating function gets projected to a configuration of large-charge
operators that factorizes into two polygons --- in color space, the
sphere factorizes to two disks, and accordingly the polygons have disk
topology. It turns out that this can be directly implemented at the
level of the twistor Feynman rules:
The polygons $M_{n,\ell}$ can be computed directly by only considering
a much smaller set of graphs $\Gamma_{n,\ell}^\text{disk}$, which
consists of all planar graphs with disk topology whose perimeter is
formed by $n$ external operators~\eqref{eq:Oscalar}, with two adjacent operators being
connected by a single fat propagator. On the inside of the disk, $\ell$ Lagrangian
insertions are placed that are connected by edges to each other and the
external operators while preserving planarity. We list the counting of
these graphs in~\tabref{tab:graphcounting}. This should be contrasted
with Table~1 in~\cite{Bargheer:2025uai}. For example at seven points
and two loops, $1\,323\,096$ graphs contribute to the sphere
generating function $G_{7,2}$, whereas the polygon
$\polygon_{7,2}$ can be computed from only $23\,273$ graphs. After applying the twistor Feynman
rules $\Phi$ to all disk-topology graphs, one obtains the polygon $M_{n,\ell}$:%
\footnote{See~\cite{Caron-Huot:2023wdh} for a detailed
explanation on how to apply the twistor Feynman rules to obtain the
loop integrands (also reviewed in Section~3 of~\cite{Bargheer:2025uai}).
These rules are applied in the same manner to the disk-topology
graphs in $\Gamma_{n,\ell}^\mathrm{disk}.$}
\begin{equation}
M_{n,\ell} = \lim_{d_{i,i+1}\rightarrow 1}\,\prod_{i=1}^n (1-d_{i,i+1})\times\sum_{\gamma\in\Gamma_{n,\ell}^\text{disk}}\Phi(\gamma)\,.
\label{eq:twistorPolygon}
\end{equation}
The limit manifests the 10d null polygonal kinematics that defines the
polygons in the first place. It can be readily implemented when
evaluating the twistor expression numerically by choosing 10d
null kinematics with $X_{i,i+1}^2=0$. A review of the twistor Feynman rules
of~\cite{Chicherin:2014uca,Caron-Huot:2023wdh} can be found
in~\cite{Bargheer:2025uai}. For an exemplary review of their
application in~\eqref{eq:twistorPolygon}, see
\appref{sec:twist-feynm-rules}.

\paragraph{Derivation.}

\begin{figure}[t]
\centering
\includegraphics[page=1,scale=1,align=c]{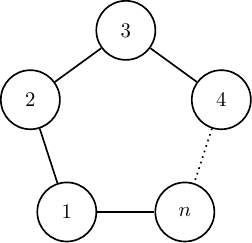}
\hspace{5mm}
\includegraphics[page=3,scale=1,align=c]{sumofgraphs.pdf}
\hspace{5mm}
\includegraphics[page=2,scale=1,align=c]{sumofgraphs.pdf}
\caption{Depicted are exemplary graphs that contribute to an
$n$-point generating function. In the first set of graphs on the left,
adjacent external operators are connected by a simple fat propagator.
Not depicted are the Lagrangians or propagators connecting them to
each other or to external operators, or propagators connecting
non-adjacent operators. In the second set of graphs, at least one pair
of non-adjacent operators is not connected -- here, the operators 1
and 2. And finally on the right, at least one propagator on the
perimeter is split into two edges, which is only possible if they
enclose at least one Lagrangian insertion.}
\label{fig:sumofgraphs}
\end{figure}

The reasoning for considering only graphs with disk topology is as
follows: On the level of graphs, the generating function $G_{n,\ell}$
can be understood as a sum of planar ribbon graphs whose $(n+\ell)$
vertices are at least of valency two. Furthermore, multiple edges
between a single pair of vertices are allowed if these edges are homotopically distinct
(preserving planarity).
After permuting over all distinct labels of the vertices, we partition
the set of graphs into three groups (cf.~\figref{fig:sumofgraphs}):
\begin{enumerate}[label=(\roman*)]
	\item Graphs in which the operators $1,
    \ldots,n$ are connected by simple edges, such that they form a polygon (\ie~an $n$-cycle),
	\item graphs in which at least one pair of adjacent operators $(i,i+1)$ is not connected by a propagator, and
	\item graphs in which the operators $1,\ldots,n$ form an
    $n$-cycle, and at least one pair of adjacent operators $(i,i+1)$
    is connected by two or more propagators.
\end{enumerate}
We argue that only the first set of graphs~(i) reveals a non-zero
contribution after taking the 10d null polygonal limit of the
generating function. Consider, the second set of
graphs~(ii). These graphs miss at least one edge on the perimeter of
the polygon, and thus, do not obtain a factor
$D_{i,i+1}=x^2_{i,i+1}/X^2_{i,i+1}$ for at least one $i=1,\dots,n$,
after applying the twistor Feynman rules. Consequently, taking the
residue~\eqref{eq:twistorPolygon} at $X_{i,i+1}^2\rightarrow0$ eliminates any contribution
obtained from such graphs.

Summing up the third set of graphs~(iii) will inherently lead to a
zero contribution. Since all external operators form the
boundary of the disk, two adjacent external operators can only be
connected by two or more
propagators if these propagators enclose Lagrangian
insertions. This will effectively lead to an isolated loop integrand of
a two-point function which is zero.

Now that we have restricted the graphs to group~(i), we can reduce the
set further by considering that the polygons are the square roots of
the residues~\eqref{eq:PolygonEquationIntegrated}, \eqref{eq:computePolygons}.
The square root has a natural
interpretation in terms of these graphs. The external operators form a
polygon that cuts the color
surface into two regions with disk topology (an ``inside'' and an ``outside''). Depending on whether
the Lagrangian insertions are located on the inside or outside of the
polygon, or split between the two regions, different contributions of
products of polygon expressions are obtained. At two loops, for
example, the residue can be written schematically as follows:
\begin{equation}
R_{n,2}
\,=\, \includegraphics[page=1,align=c,scale=.9]{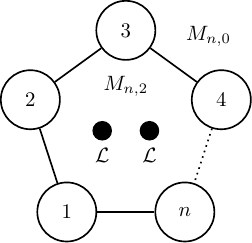}
\,+\, \includegraphics[page=2,align=c,scale=.9]{residue52.pdf}
\,+\, 2\times\includegraphics[page=3,align=c,scale=.9]{residue52.pdf}
\,.
\end{equation}
As can be seen from this equation, all Lagrangian insertions have to
be gathered inside (or outside) of the polygon to obtain the relevant
loop correction, while the propagators on the outside (or inside) will
form the tree-level polygon result. Products of lower loop polygons
are obtained if the Lagrangians are distributed in both regions.
The polygon $M_{n,\ell}$ can thus be isolated by restricting the graphs to disk
topology with the perimeter being single edged propagators connecting
adjacent operators.%
\footnote{While the propagators naturally split between the inside and
outside of the polygon, one might have to worry about the dressing
factors of the vertices $V^i_{j_1\ldots j_k}$ when applying the
Feynman rules. (Here, $i$ denotes the respective vertex and the lower
indices reflect the cyclic ordering of the edges around $i$.) Since
the dressing factor can be written as a product of factors depending
only on neighboring edges
$V^i_{j_1\ldots j_k} = \Delta^i_{j_1j_2}\dots\Delta^i_{j_kj_1}$
(see~(3.33) in~\cite{Caron-Huot:2023wdh}),
they also factorize along the perimeter of the polygon.}

Hence, the polygons are obtained by applying the twistor Feynman rules
to the graphs with disk topology $\Gamma_{n,\ell}^\mathrm{disk}$ and
taking the 10d null polygonal limit as formulated
in~\eqref{eq:twistorPolygon}.

By exactly the same reasoning, the
factorization~\eqref{eq:PolygonFactorization} of polygons into smaller
polygons upon taking further ten-dimensional null limits directly
follows.

\paragraph{Ansatz.}

We have found a way to compute the polygons directly from
twistors, but the general drawback of the twistor formulation still
applies: The result contains the (arbitrary but fixed) reference
twistor. While the final expression is ultimately independent of this
reference twistor, it is virtually impossible to remove the reference twistor
from it to write it in terms of basic invariants
(\ie square distances). As usual, we therefore resort to writing a
suitable manifestly invariant ansatz, and match it against the twistor
result by numerical evaluation.

The ansatz has the general form
\begin{equation}
M_{n,\ell} =
\sum_{a} f_a(x_{ij}^2,y_{ij}^2) \, \mathcal{I}_a(x_{ij}^2)
\,,
\label{eq:ansatz}
\end{equation}
where the $\mathcal{I}_a$ form a basis of rational functions that are
the integrands of conformal integrals planarly embedded inside the
polygon,
and $f_a(x_{ij}^2,y_{ij}^2)$
are polynomial coefficient functions. An obstacle in the computation
is that there appears to be no way to bound the coefficients $f_a$ from
first principles, so we have to partially resort to educated guesses
to formulate ansätze that are both sufficiently general and not too
big to constrain.

\section{All 10d Null Polygons to Two Loops}
\label{sec:AllPoly}

Using the twistor Feynman rules on graphs with disk topology enables
us to compute the polygons $M_{n,\ell}$ directly, without having to
find the full $n$-point $\ell$-loop generating function first.
At one-loop order, the explicit expression of the square~\eqref{eq:M41}, the
pentagon~\eqref{eq:M51}, and the hexagon~\eqref{eq:M61}
are sufficient to state a formula for all one-loop polygons
that includes both the parity-even and parity-odd parts. We have verified this
proposal numerically up to nine points by matching against the
respective twistor expressions.

At two loops, we focus on the parity-even part of the polygons, by
numerically matching appropriate ansätze against the twistor
computation, as described above. In
this way, we successively compute two-loop polygons starting from six
points. We find systematics in their expressions that further constrain the
ansatz for respective higher-point polygons. Reaching the
ten-sided polygon, we could fix all patterns in the expression, and
thus conjecture
a formula for the parity-even part of all polygons $M_{n,2}$ at two
loops. We checked that proposal explicitly at eleven points, and find
agreement with the twistor result.

The final result for all polygons $\polygon_{n,\ell}$ for any $n$ and
up to $\ell=2$ loops is attached in the ancillary \mathematica file
\ancfile{polygons.m}.

Before presenting the formulae for all one-loop and two-loop polygons
in~\eqref{eq:MdecompF}, \eqref{eq:oneLoopFaces},
and~\eqref{eq:F2IntegralDecomp2} below,
let us describe how the factorization property of the polygons can be used
to organize them systematically.

\subsection{Factorization Channels}

The polygon integrands are rational functions of ten-dimensional
distances $X_{ij}^2$ as well as four-dimensional distances $x_{ij}^2$
(four-dimensional propagators $d_{i,j}$ appear only polynomially). In
particular, they display ten-dimensional poles. Like
all rational functions, they are fixed by the residues at their
poles, and their behavior at infinity. The poles are precisely the
factorization channels, on which the polygons factorize into products
of lower polygons~\eqref{eq:PolygonFactorizationIntegrated}.
After taking these residues into account, what remains are terms that
are free of ten-dimensional propagators $D_{ij}=x_{ij}^2/X_{ij}^2$ on
the diagonals of the polygon (the ``empty'' polygon).

The factorization property~\eqref{eq:PolygonFactorizationIntegrated}
as well as the above picture inspire the following natural
decomposition of the polygons $\intpoly_n$ into more
basic building blocks $\mathbb{F}_f$ that we call \emph{faces}:
\begin{equation}
\mathbb{M}_{n}=
\mathbb{F}_{1,\dots,n}
+\sum_{\substack{\Gamma\in\,\text{tree}\\\text{graphs}}}
\;
{\prod_{\substack{\text{edges}\\(ij)\in\Gamma}}\!\!\!D_{ij}}
\times
{\prod_{\substack{\text{faces}\\[0.3ex]f\in\Gamma}}\!\mathbb{F}_{f}}
\,.
\label{eq:MdecompF}
\end{equation}
Here, the sum runs over all tree graphs that contribute to
$M_{n,0}$, excluding the empty polygon graph (whose contribution is
captured by the first term).%
\footnote{These graphs are exactly the planar graphs one can draw on the disk
with~$n$ punctures on the boundary (and no internal vertices), \ie all
graphs that tessellate the $n$-gon by diagonals.
See the first line of \tabref{tab:graphcounting}, as well
as~\cite{Krasko:2015aa} for their counting modulo dihedral permutations.}
The formula~\eqref{eq:MdecompF}
\begin{itemize}
\item
defines the faces $\mathbb{F}_{1,\dots,n}$
recursively in terms of the polygons $\intpoly_n$ as well as the
lower-point and lower-loop faces,
\item makes the polygon factorization
property~\eqref{eq:PolygonFactorizationIntegrated} manifest,%
\footnote{\label{fn:facesAmbiguity}%
Starting with the factorization property of the polygons
$\mathbb{M}_n$, there is some freedom in the
definition of the faces $\mathbb{F}$, in the sense that one could write
variations of the formula~\eqref{eq:MdecompF}, that would differ by
re-arrangements of the
geometric series in the fat propagators, e.g.~$D_{ij}/d_{ij} = 1 +
D_{ij}$. We find that our choice~\eqref{eq:MdecompF} yields the most
natural definition.}
%
\item and it reduces the polygon functions $\intpoly_n$ to yet simpler
functions, the faces $\intface_{1,\dots,n}$.
\end{itemize}
In order to apply~\eqref{eq:MdecompF} perturbatively, we expand both
the polygons~$\mathbb{M}_n$ (cf.~\eqref{eq:PolygonIntegrated}) and the faces $\mathbb{F}_{1,\dots,n}$ in the coupling,
\begin{equation}
\mathbb{F}_{1,\dots,n} = 1 +
\sum_{\ell=1}^\infty\frac{(-g^{2})^{\ell}}{\ell!}
\int\brk[s]3{\,\prod_{k}\frac{\dd[4]{x_{k}}}{\pi^2}}
F_{1,\dots,n}^{(\ell)}
\,,
\label{eq:FacesExpansion}
\end{equation}
where the product over $k$ is taken over all $\ell$ loop variables.
We wrote the decomposition~\eqref{eq:MdecompF} in terms of the
integrated objects $\intpoly_n$ and $\intface_{1,\dots,n}$, but we
emphasize that they hold at the level of the \emph{integrands}, \ie
the integrations can be removed from the equation. At integrand level,
one must symmetrize each term over all distributions of integration points
(Lagrangian insertions) on the factors $F_f$, exactly as it happens
in~\eqref{eq:PolygonEquation} and~\eqref{eq:PolygonFactorization}.

Notably, the leading order $F_{1,\dots,n}^{(0)}$ of the faces are
trivial by construction, and the triangles receive no loop
corrections:
\begin{equation}
	\mathbb{F}_{i_1,i_2, i_3} = 1
    \,.
\end{equation}
Let us exemplify the decomposition~\eqref{eq:MdecompF} by explicitly writing the
expressions for the square, pentagon and hexagon.
The former two are expressed as
\begin{align}
	\mathbb{M}_4 &= \mathbb{F}_{1,2,3,4} + D_{13} \times \mathbb{F}_{1,2,3}\,\mathbb{F}_{1,3,4}+ D_{24} \times \mathbb{F}_{1,2,4}\,\mathbb{F}_{2,3,4} \,, \\
	\mathbb{M}_5 &= \mathbb{F}_{1,2,3,4,5} +
	\brk[s]*{
        \frac{1}{2}D_{14}\times \mathbb{F}_{1,2,3,4}\,\mathbb{F}_{1,4,5}
        + \frac{1}{2}D_{14}\,D_{13}\times \mathbb{F}_{1,2,3}\, \mathbb{F}_{1,3,4}\, \mathbb{F}_{1,4,5}
        + (\grp{D}_5\text{ perms})
    }\,,
	\end{align}
while the hexagon is split into tree-level faces as follows:
\begin{align}
	\mathbb{M}_{6}&=
	\mathbb{F}_{1,2,3,4,5,6}
	\qquad\qquad\qquad &
	\includegraphics[align=c]{FigHexaEmpty}
	\nonumber\\ &\qquad
	+ \Big{[}
	\frac{1}{2}D_{13}\times \mathbb{F}_{1,2,3}\, \mathbb{F}_{3,4,5,6,1}
	\qquad\qquad\qquad &
	\includegraphics[align=c]{FigHexaD13}
	\nonumber \\ &\qquad\qquad
	+ \frac{1}{4} D_{14}\times \mathbb{F}_{1,2,3,4}\, \mathbb{F}_{4,5,6,1}
	\qquad\qquad\qquad &
	\includegraphics[align=c]{FigHexaD14}
	\nonumber \\ &\qquad\qquad
	+ D_{13} D_{14}\times \mathbb{F}_{1,2,3} \, \mathbb{F}_{3,4,1} \, \mathbb{F}_{4,5,6,1}
	\qquad\qquad\qquad &
	\includegraphics[align=c]{FigHexaD13D14}
	\nonumber \\ &\qquad\qquad
	+ \frac{1}{2}D_{13} D_{15}\times \mathbb{F}_{1,2,3} \, \mathbb{F}_{5,6,1} \, \mathbb{F}_{3,4,5,1}
	\qquad\qquad\qquad &
	\includegraphics[align=c]{FigHexaD13D15}
	\nonumber \\ &\qquad\qquad
	+ \frac{1}{4} D_{15} D_{24} \times \mathbb{F}_{2,3,4} \, \mathbb{F}_{5,6,1} \, \mathbb{F}_{1,2,4,5}
	\qquad\qquad\qquad &
	\includegraphics[align=c]{FigHexaD15D24}
	\nonumber\\ &\qquad\qquad
	+ \frac{1}{2}D_{13}D_{14}D_{15}\times
      \mathbb{F}_{1,2,3}\, \mathbb{F}_{1,3,4}\, \mathbb{F}_{1,4,5}\, \mathbb{F}_{1,5,6}
    + \dots
	\qquad\qquad\qquad &
	\includegraphics[align=c]{FigHexaD13D14D15}
	\nonumber\\ &\qquad\qquad
	+ (\grp{D}_6\text{ perms})
    \Big{]}\,,
    \label{eq:hexFaces}
\end{align}
where the dots stand for two more terms that are products of three
propagators $D_{ij}$ and four triangle faces $\mathbb{F}_{i_1,i_2,i_3}$.
Expanding these expressions at one-loop order
using~\eqref{eq:FacesExpansion}, and passing to the integrands yields
\begin{align}
    M_{4, 1} \,=&\, F^{(1)}_{1,2,3,4} \,, \label{eq:squareoneloopexpansion}\\
	M_{5,1} \,=&\, F^{(1)}_{1,2,3,4,5} + \Big{[}
	\frac{1}{2} D_{14}\times \, F^{(1)}_{1,2,3,4} + (\grp{D}_5\text{ perms})\Big{]}\,, \label{eq:penoneloopexpansion}\\
	M_{6,1} \,=&\,
	F^{(1)}_{1,2,3,4,5,6} +
    \Big{[}
	\frac{1}{2}D_{13}\times \, F^{(1)}_{3,4,5,6,1}
    + \frac{1}{4} D_{14}\times (F^{(1)}_{1,2,3,4}
    +  F^{(1)}_{4,5,6,1} )
    + D_{13} D_{14}\times F^{(1)}_{4,5,6,1}
    \nonumber \\
	&\qquad\qquad\qquad
    + \frac{1}{2}D_{13} D_{15}\times F^{(1)}_{3,4,5,1}
    + \frac{1}{4} D_{15} D_{24} \times F^{(1)}_{1,2,4,5}
    + (\grp{D}_6\text{ perms})
    \Big{]}
    \label{eq:hexoneloopexpansion}
    \,.
\end{align}
At two-loop order, these expressions almost stay the same, up to the
substitution $F^{(1)}\to F^{(2)}$. Only the two-loop hexagon receives
an extra term that is a product of two one-loop squares
(from the third line of~\eqref{eq:hexFaces}):
\begin{equation}
	M_{6,2}=
    \eqref{eq:hexoneloopexpansion}\big{|}_{F^{(1)}\to F^{(2)}}
    +\brk[s]*{
    \frac{1}{2}D_{14}\times F^{(1)}_{1,2,3,4} F^{(1)}_{1,4,5,6}
    + (\grp{D}_6\text{ perms})
    }
    \,.
    \label{eq:hextwoloopexpansion}
\end{equation}
Importantly, terms of this type can contain
products of parity-odd terms that contribute to the
parity-even part of $M_{n,2}$. The four-point
faces $F_{1,2,3,4}^{(\ell)}$ are free of parity-odd parts, but
$F^{(1)}_{i_1,\dots,i_{n\ge5}}$ does contain parity-odd terms, and
thus starting from $M_{8,2}$, products of such terms
do contribute to the decomposition~\eqref{eq:MdecompF}.
This highlights that it is important
to include both parity-even and parity-odd parts in the definition of
the polygons $M_n$ and the faces $F_{1,\dots,n}$ in order to make the
decomposition~\eqref{eq:MdecompF} consistent. The same happens for
massless amplitudes, where parity-odd terms have to be included in
order to consistently establish the
super-correlator/super-amplitude duality
\cite{Eden:2011yp,Eden:2011ku}.

\subsection{One-Loop Result} \label{sec:oneLoopPolygons}

The expressions
for the square~\eqref{eq:M41}, the pentagon~\eqref{eq:M51}, and the hexagon~\eqref{eq:M61} suffice to conjecture the general
structure of both the even and odd part of all one-loop polygons. The
corresponding one-loop faces that dress the tree-level graphs for any $n\ge4$ are
\begin{align}
	F^{(1)}_{i_1,i_2\,,\cdots, i_n} & =
    \mspace{30mu}
	\sum_{\mathclap{\underset{\text{non-adjacent}}{[i_{m},i_{m+1}]\,,[i_{l},i_{l+1}]}}}
    \mspace{20mu}
	\frac{P_{i_m i_{l+1},i_{m+1}
    i_l}-4\ii\,\epsilon_{\mu\nu\rho\sigma}x^\mu_{i_{m}0}x^\nu_{i_{m+1}0}x^\rho_{i_l0}x^\sigma_{i_{l+1}0}}{2\,x_{i_{m}0}^2
    x_{i_{m+1}0}^2 x_{i_{l}0}^2 x_{i_{l+1}0}^2 }
    \nonumber \\[-1ex] & 
    \mspace{25mu} - \mspace{5mu}
    \mspace{1.5mu}
    \sum_{\mathclap{1\le j<k<l<m\le n}}
    \mspace{37mu}
    \frac{y_{i_ji_l}^2 y_{i_ki_m}^2}{x_{i_j0}^2 x_{i_k0}^2 x_{i_l0}^2 x_{i_m0}^2}
    \,,
    \label{eq:oneLoopFaces}
\end{align}
where the first sum runs over all pairs of non-adjacent edges at the
perimeter of the polygon.
The imaginary term proportional to $\epsilon_{\mu\nu\rho\sigma}$ is
parity-odd, all other terms are parity-even, $x_0$ is the coordinate
of the Lagrangian insertion, and we defined the combination
\begin{equation}
	P_{ij,kl}=-x_{ij}^2x_{kl}^2+x_{ik}^2x_{jl}^2+x_{il}^2x_{jk}^2\,.
    \label{eq:Pijkl}
\end{equation}
We have verified this proposal by numerically comparing it to the
respective twistor expressions~\eqref{eq:twistorPolygon} up to nine points.

\subsection{Two-Loop Result}
\label{sec:two-loop-result}

The formula for the two-loop faces
$F_{1,\dots,n}^{(2)}$ is more intricate, and we only computed its
parity-even part. Hence also the two-loop polygons $\polygon_n$
are restricted to their parity-even parts. We will not indicate the
superscript ``even'', with the understanding that all two-loop
expressions refer to their parity-even parts only.

\paragraph{Result.}

We decompose the two-loop $n$-point faces in exactly the same way as was done
for the integrand of the two-loop $n$-point massless gluon amplitude~\cite{Vergu:2009tu}:%
\footnote{The two-loop faces~\eqref{eq:F2IntegralDecomp2} as well as the
resulting two-loop polygons~\eqref{eq:MdecompF} are constructed in the
ancillary \mathematica file \ancfile{polygons.m}. The coefficients
$f_{k,\mathbold{e},a,b}$ are stored in the file
\ancfile{twoLoopCoefficients.m}.}
\begin{equation}
F^{(2)}_{1,\dots,n}
=
\sum_{k,\mathbold{e}}
f_{k,\mathbold{e},a,b}(x_{ij}^2, y_{ij}^2)
\, \mathcal{J}_k^{\mathbold{e},a,b}
+(a \leftrightarrow b)
\,.
\label{eq:F2IntegralDecomp2}
\end{equation}
Here, the sum over $k$ runs over three different propagator topologies
$\mathcal{J}_k^{\mathbold{e},a,b}$, the sum over $\mathbold{e}$ runs
over their planar embeddings in the face $F_{1,\dots,n}$, and the
coefficients $f_{k,\mathbold{e},a,b}$ are
polynomials in the basic scalar invariants $x_{ij}^2$ and
$y_{ij}^2$.

The following three
propagator topologies occur in the decomposition:
\begin{alignat}{2}
\mspace{50mu}
\includegraphics[align=c,hsmash=c]{FigPentaPentaJ1}
&&\mspace{120mu}
\mathcal{J}_1^{1,\dots,8,{a},{b}}&=
\frac{1}{
x_{1{a}}^2 x_{2{a}}^2 x_{3{a}}^2 x_{4{a}}^2
\, x_{{a}{b}}^2 \,
x_{5{b}}^2 x_{6{b}}^2 x_{7{b}}^2 x_{8{b}}^2
}
\,,\\
\mspace{50mu}
\includegraphics[align=c,hsmash=c]{FigPentaBoxJ2}
&&\mspace{120mu}
\mathcal{J}_2^{1,\dots,7,a,b}&=
\frac{1}{
x_{1{a}}^2 x_{2{a}}^2 x_{3{a}}^2
\, x_{{a}{b}}^2 \,
x_{4{b}}^2 x_{5{b}}^2 x_{6{b}}^2 x_{7{b}}^2
}
\,,\\
\mspace{50mu}
\includegraphics[align=c,hsmash=c]{FigDoubleBoxJ3}
&&\mspace{120mu}
\mathcal{J}_3^{1,\dots,6,a,b}&=
\frac{1}{
x_{1{a}}^2 x_{2{a}}^2 x_{3{a}}^2
\, x_{{a}{b}}^2 \,
x_{4{b}}^2 x_{5{b}}^2 x_{6{b}}^2
}
\,.
\end{alignat}
Note that the structures $\mathcal{J}_1$, $\mathcal{J}_2$,
$\mathcal{J}_3$ do not include numerator factors, which means that they do
not absorb all dependence on the Lagrangian points $a$, $b$.%
\footnote{See \appref{sec:DCIFaces} for an equivalent decomposition in
terms of integrands of conformal integrals that include the numerator factors.}
Therefore, the coefficients $f_{k,\mathbold{e},a,b}$ must not only contain
factors $x_{ij}^2$ between external points $i,j\leq n$, but also
among external points $i\leq n$ and Lagrangian points
$j\in\set{a,b}$. However, they may not contain factors that cancel any
of the denominator factors in the definitions of the integrands
$\mathcal{J}_1$, $\mathcal{J}_2$, $\mathcal{J}_3$. The only exception to
this is the coefficient $f_{1,\mathbold{e},a,b}$ of the double-penta
integrand, which may include terms with a factor $x_{ab}^2$ that
turns $\mathcal{J}_1$ into the product of two one-loop box integrands.

The sum over $\mathbold{e}$ runs over all planar embeddings of the
integrands $\mathcal{J}_1$, $\mathcal{J}_2$, $\mathcal{J}_3$ into the
polygonal face $F_{1,\dots,n}$. For example, the double-penta
integrand $\mathcal{J}_{3}^{\mathbold{e}}$ is summed over embeddings
\begin{equation}
\mathbold{e}=(i_1,\dots,i_8)
\,,\qquad
i_1<i_2<i_3<i_4\leq i_5<i_6<i_7<i_8\leq i_1
\;(\text{mod}\,n)
\,.
\end{equation}
Making use of the dihedral $\grp{D}_n$ symmetry, we can fix $i_1=1$.
Due to the planar embedding, we can equally sum over the distances
$\ell_k$ of neighboring points that attach to the integrand:
\begin{equation} \label{eq:distanceLabels}
\mathbold{\tilde{e}}=[\ell_1,\dots,\ell_8]
\,,\qquad
\ell_k=i_k-i_{k-1}
\;\;(i_0\equiv i_8)
\,,\qquad
\ell_k\geq
\begin{cases}
1 & k=2,3,4,6,7,8, \\
0 & k=1,5 \,.
\end{cases}
\end{equation}
We use round/square brackets to distinguish the two notations.
The coefficient functions $f_{k,\mathbold{e},a,b}$ are polynomials in
$x_{ij}^2$, $i,j\in\set{1,\dots,n,a,b}$, and $y_{ij}^2$, $i,j\in\set{1,\dots,n}$.
For $y_{ij}^2=0$, they reduce to the coefficients of the massless
gluon amplitude integrand~\eqref{eq:masslessAmpLimit}, which was
determined in~\cite{Vergu:2009tu}.%
\footnote{Setting $y_{i,j}^2=0$ for
all $i,j$ in particular includes passing to the four-dimensional null limit
$x_{i,i+1}^2 = 0$ at the perimeter of
the polygon, due to the ten-dimensional null condition
$0 = X_{i,i+1}^2 = x_{i,i+1}^2 + y_{i,i+1}^2$.}
The coefficient $f_{1,\mathbold{e},a,b}$ of the double-penta integrand
$\mathcal{J}_1$ can be written in a relatively compact way:%
\footnote{For compactness, we use both the $i_k$ and the $\ell_k$
labels in this formula.}
\begin{multline} \label{eq:f1}
4\,f_{1,i_1,\dots,i_8,a,b}=
{x^2_{ab}}
\big(\delta_{\ell_2,1}\delta_{\ell_4,1}P_{i_1i_4,i_2i_3}-2y_{i_1i_3}^2y_{i_2i_4}^2\big)
\big(\delta_{\ell_6,1}\delta_{\ell_8,1}P_{i_5i_8,i_6i_7}-2y_{i_5i_7}^2y_{i_6i_8}^2\big)
\\
-\delta_{\ell_2,1}\delta_{\ell_4,1}\delta_{\ell_6,1}\delta_{\ell_8,1}
\begin{bmatrix}
i_1 & i_2 & i_3 & i_4 & a \\
i_5 & i_6 & i_7 & i_8 & b
\end{bmatrix}
+\delta_{\ell_1,0}\delta_{\ell_3,1}\delta_{\ell_5,0}\delta_{\ell_7,1}
\,y_{i_1i_4}^2\, h_{\text{DP}}
\,,
\end{multline}
with the factor $h_{\text{DP}}$ given by
\begin{multline}
h_\text{DP}=
\begin{bmatrix}
i_1 & i_2 & i_3 & a \\
i_4 & i_6 & i_7 & b
\end{bmatrix}
+
\begin{bmatrix}
i_2 & i_3 & i_4 & a \\
i_6 & i_7 & i_1 & b
\end{bmatrix}
-x_{i_1i_4}^2
\begin{bmatrix}
i_2 & i_3 & a \\
i_6 & i_7 & b
\end{bmatrix}
\\
-x_{ab}^2\left(
x_{i_2i_3}^2P_{i_1i_7,i_4i_6}
+x_{i_6i_7}^2P_{i_1i_2,i_3i_4}
-x_{i_1i_4}^2x_{i_2i_3}^2x_{i_6i_7}^2
\right)
\,,
\label{eq:factorDP}
\end{multline}
and where the square brackets denote determinants:
\begin{equation}
\begin{bmatrix}
i_1 & i_2 & \dots \\
j_1 & j_2 & \dots \\
\end{bmatrix}
=
\det \,\{x_{ij}^2\}_{\substack{i=i_1,i_2,\dots\\j=j_1,j_2,\dots}}\,.
\end{equation}

The numerator functions $f_{k,\mathbold{e}}$ encode all two-loop faces
$F_{1,\dots,n}^{(2)}$, and thereby all two-loop polygons $M_{n,2}$.
Through the dependence on the embedding $\mathbold{e}$, one could
imagine that the number of functions $f_{k,\mathbold{e}}$ proliferates
as $n$ increases. However, we see in the example
of~$f_1$~\eqref{eq:f1} that the dependence on the embedding is
relatively moderate, in the sense that the expression stabilizes once
all $\ell_j$ are sufficiently large. We observe that the same
saturation property also holds for the other coefficients: In general,
the coefficients cease to depend on the distances $\ell_j$ once
$\ell_j\geq2$:%
\footnote{In a strict sense, the saturation means the following:\\
Given two planar embeddings $\mathbold{e}=(i_1,\ldots,i_j, i_{j+1},
\ldots,i_{n_k})$ and $\mathbold{\hat{e}} = (i_1,\ldots,i_j, i_{j+1}+1,
\ldots,i_{n_k}+1)$ that only differ by increasing the distance between
a single pair of indices by one, then
\begin{equation*}
f_{k,\mathbold{\hat{e}},a,b} \big|_{\mathbold{\hat{e}} \rightarrow \mathbold{e}} = f_{k,\mathbold{e},a,b} \quad\text{if}\quad i_{j+1}-i_j\ge2
\,.
\end{equation*}}
\begin{equation}
    f_{k,[\dots,\,\ell_j,\,\dots]} = f_{k,[\dots,\,2,\,\dots]}
    \quad \text{for} \quad
    \ell_j\geq2
    \,.
    \label{eq:fSaturation}
\end{equation}
The only exception to this pattern are the distances
$\ell_1=i_1-i_6$ and $\ell_4=i_4-i_3$ of the coefficients $f_{3,
\mathbold{\tilde e}}$, where this saturation only occurs at
$\ell_{1,4}\geq3$. This is also the case for the integrand of the
massless gluon amplitude~\cite{Vergu:2009tu}.
Furthermore, the coefficient functions display the following label symmetries that
are inherited from the symmetries of the respective propagator
topologies $\mathcal{J}_1$, $\mathcal{J}_2$ and
$\mathcal{J}_3$:
\begin{equation}
\begin{aligned}[t]
           & f_{1,[\ell_1,\ell_2,\ell_3,\ell_4,\ell_5,\ell_6,\ell_7,\ell_8],a,b} \\
=\mathord{}& f_{1,[\ell_1,\ell_8,\ell_7,\ell_6,\ell_5,\ell_4,\ell_3,\ell_2],b,a} \\
=\mathord{}& f_{1,[\ell_5,\ell_6,\ell_7,\ell_8,\ell_1,\ell_2,\ell_3,\ell_4],b,a} \\
=\mathord{}& f_{1,[\ell_5,\ell_4,\ell_3,\ell_2,\ell_1,\ell_8,\ell_7,\ell_6],a,b} \,,
\end{aligned}
\qquad
\begin{aligned}[t]
           & f_{2,[\ell_1,\ell_2,\ell_3,\ell_4,\ell_5,\ell_6,\ell_7],a,b} \\
=\mathord{}& f_{2,[\ell_4,\ell_3,\ell_2,\ell_1,\ell_7,\ell_6,\ell_5],a,b} \,,
\end{aligned}
\qquad
\begin{aligned}[t]
           & f_{3,[\ell_1,\ell_2,\ell_3,\ell_4,\ell_5,\ell_6],a,b} \\
=\mathord{}& f_{3,[\ell_1,\ell_6,\ell_5,\ell_4,\ell_3,\ell_2],b,a} \\
=\mathord{}& f_{3,[\ell_4,\ell_5,\ell_6,\ell_1,\ell_2,\ell_3],b,a} \\
=\mathord{}& f_{3,[\ell_4,\ell_3,\ell_2,\ell_1,\ell_6,\ell_5],a,b} \,.
\end{aligned}
\label{eq:fCoeffSym}
\end{equation}
Assuming the saturation property as well as the label
symmetries, we can list all potentially different coefficient
functions $f_{k,\mathbold{e}}$. In this way, we find that there are a
priori $180$ different coefficients $f_{1,\mathbold{\tilde e}}$, $156$ different
coefficients $f_{2,\mathbold{\tilde e}}$, and $88$ different coefficients $f_{3,\mathbold{\tilde
e}}$ corresponding to the three propagator topologies.

As seen above, the coefficients $f_{1,\mathbold{e}}$ of the
double-penta topology combine elegantly
into~\eqref{eq:f1}. In contrast, the other two families
$f_{2,\mathbold{e}}$ and $f_{3,\mathbold{e}}$ exhibit a considerably
more intricate structure, with a
subtler dependence on the embeddings. Their explicit forms can be
found in the ancillary file \ancfile{twoLoopCoefficients.m}.

\paragraph{Two-Loop Ansatz Construction.}

Having laid out the structure of the result, we now step back to
outline how it was obtained. In order to match the twistor
expression of the polygons to a manifestly Lorentz-invariant
representation, it is necessary to formulate a compact ansatz.
This ansatz is provided by the functions $f_{k,\mathbold{e}}$, but
with a priori undetermined numerical coefficients. Combining these functions with
the denominators in~\eqref{eq:F2IntegralDecomp2}, and using the
factorization decomposition~\eqref{eq:MdecompF} yields an ansatz for
the respective two-loop polygon.

Due to conformal symmetry, the coefficients $f_{k,\boldsymbol{a}}$ are
polynomials in $x_{ij}^2$ and $d_{ij}$, constructed such that each
term becomes conformally invariant in $x_{ij}^2$ when combined with
the corresponding denominator $\mathcal{J}_k$.%
\footnote{The terms become conformally invariant after integration, so
the integrands are of conformal weight zero in the external points and
of weight 4 in the integration variables $a$ and $b$.}
This property is directly inherited from the generating function
$G_{n,\ell}$, which is itself conformal in the variables $x_{ij}^2$
when expressed as a function of $d_{ij}$ and~$x_{ij}^2$. Consequently,
this symmetry constrains the possible
monomials in $x_{ij}^2$ that can appear in the polynomials
$f_{k,\boldsymbol{a}}$. Heuristically, we also observe that the
appearance of the factors $d_{ij}$ is similarly restricted: Each
factor $d_{ij}$ must be accompanied by a corresponding factor
$x_{ij}^2$, ensuring that $f_{k,\boldsymbol{a}}$ remains a polynomial
in the combined variables $(x_{ij}^2,y_{ij}^2=x_{ij}^2 d_{ij})$.
In other words, the functions $f_{k,\mathbold{e}}$ are polynomials in
$x_{ij}^2$ and $y_{ij}^2$ such that the conformal and
$R$-charge weights of the products
$f_{k,\mathbold{e}}\mathcal{J}_k^{\mathbold{e}}$
match those of the polygon correlator:
\begin{equation}
f_{k,\mathbold{e},a,b}(\lambda_i\lambda_j x_{ij}^2, \lambda_i\lambda_j y_{ij}^2)
\,\mathcal{J}_k^{\mathbold{e},a,b}(\lambda_i\lambda_jx_{ij}^2)
= \frac{f_{k,\mathbold{e},a,b}(x_{ij}^2,y_{ij}^2)\,\mathcal{J}_k^{\mathbold{e},a,b}(x_{ij}^2)}{\lambda_a^4\lambda_b^4}\,.
\end{equation}
Moreover, the functions $f_{k,\mathbold{e}}$ do not contain
distances $x_{ij}^2$ that cancel factors in the denominator of $\mathcal{J}_k^{\mathbold{e}}$, with the
exception of $f_{1,\mathbold{e}}$, that may contain factors
$x_{ab}^2$, as was already mentioned above. Finally, the functions
$f_{k,\mathbold{e}}$ must obey the label symmetries~\eqref{eq:fCoeffSym}, which further
restricts the size of the ansätze. In this way, we find the number of
unknown coefficients to be $1242$, $240$, and $50$ for the functions
$f_{1,\mathbold{e}}$, $f_{2,\mathbold{e}}$, and $f_{3,\mathbold{e}}$,
respectively (for fixed embeddings~$\mathbold{e}$).
However, when inserting the ansätze into a specific face
function $F$ in an $n$-sided polygon $M_{n,2}$, the number of
coefficients typically reduces, for three reasons: First,
factors $y_{ij}^2$ are identified with $x_{ij}^2$ for $j=i+1$. Second,
part of the ansatz collapses if points in the embedding $\mathbold{e}$
are equal to another. Third, there might be identifications in the
ansatz due to the dihedral symmetry~$\grp{D}_n$ of the respective
polygon.

The functions $f_{k,\mathbold{e}}$ can now be probed in order of their
appearance, beginning from the square $M_{4,2}$, and successively increasing the
size of the polygons by one. At each step, the coefficients in the
polynomials $f_{k,\mathbold{e}}$ are fixed by comparison to the
twistor expression, and subsequently fed into
the succeeding ansatz. In this process, we can make use of the saturation
property~\eqref{eq:fSaturation}.
To probe all degrees of freedom of the coefficient functions $f_k$, we
would in principle have to push this process all the way to the
two-loop hexadecagon $M_{16,2}$, such that all distances $\ell_j$
become saturated according to~\eqref{eq:fSaturation}. However,
the pattern of the various embeddings often becomes apparent
much earlier. This is most notable for the double-penta
functions $f_{1,\mathbold{e}}$, which neatly combine into~\eqref{eq:f1}.
In practice, we find the complete forms of all functions
$f_{k,\mathbold{e}}$ by matching the respective ansätze to polygons
$M_{n,2}$ up to the decagon with $n=10$. This yields a complete formula for all two-loop
polygons. We have checked the resulting formula for the hendecagon
$M_{11,2}$ by matching against the twistor computation,
and found complete agreement.
This verifies the correctness of the result for general~$n$ beyond
reasonable doubt.

\paragraph{Examples: Square, Pentagon, and Hexagon.}

To exemplify the decomposition into different propagator
topologies~\eqref{eq:MdecompF}, we explicitly apply this formula up to
six points. For the square and pentagon, the two-loop faces read:
\begin{align}
    F^{(2)}_{1,2,3,4} &= \frac{1}{4}f_{3,[0,1,1,0,1,1]}\mathcal{J}_3^{\mathbold{e}} + (\grp{D}_4\text{ perms})\,,
    \label{eq:F42}
    \\
    F^{(2)}_{1,2,3,4,5} &= f_{3,[0,1,1,0,1,2]}\mathcal{J}_3^{\mathbold{e}}+\frac{1}{2}f_{3,[0,1,1,1,1,1]}\mathcal{J}_3^{\mathbold{e}}+\frac{1}{2}f_{2,[0,1,1,0,1,1,1]}\mathcal{J}_2^{\mathbold{e}} + (\grp{D}_5\text{ perms})\,,
    \label{eq:F52}
\end{align}
with the coefficient functions given by
\begin{align}
		f_{3,[0,1,1,0,1,1]}
		&= (x_{13}^2 - y_{13}^2)\,\big(x_{13}^2 x_{24}^2 - y_{13}^2 y_{24}^2\big)\,, \nonumber \\
		f_{3,[0,1,1,0,1,2]}
		&= \frac{1}{2}(x_{13}^2 - y_{13}^2)\,\big(P_{15,23} - 2 y_{13}^2 y_{24}^2\big)\,, \nonumber\\
		f_{3,[0,1,1,1,1,1]}
		&= -\frac{1}{2}P_{12,45}\,x_{13}^2
		+ \frac{1}{2}P_{12,35}\,(x_{14}^2 - y_{14}^2)
		- \frac{1}{2}y_{13}^2\big(P_{15,24} - 2 y_{14}^2 y_{25}^2\big)\,, \nonumber \\
		f_{2,[0,1,1,0,1,1,1]}
		&= \frac{1}{2}x_{2b}^2\Big(P_{16,35}\,y_{13}^2
		+ x_{13}^2\big(P_{13,56} - 2 y_{14}^2 y_{35}^2\big)\Big)\,,
\end{align}
where $P_{ij,kl}$ is defined in~\eqref{eq:Pijkl}.
In the above equations, the coefficients $f_{k,\mathbold{\tilde{e}}}$ are labeled by the distances
$\ell_j$ between neighboring points in the embedding, as in~\eqref{eq:distanceLabels}. It is implied
that the corresponding point labels start with point one, that is
$\mathbold{e}=(1,1+\ell_2,\dots)$, and that the respective factors
$\mathcal{J}_k^{\mathbold{e}}$ are defined with this embedding $\mathbold{e}$.

According
to~\eqref{eq:squareoneloopexpansion}, the two-loop square $M_{4,2}$ equals
the two-loop four-point face, and~\eqref{eq:F42} consistently matches
the expression~\eqref{eq:square2loop} for $M_{4,2}$ obtained from
the generating function. In order to obtain
the two-loop pentagon $M_{5,2}$, the five-point face~\eqref{eq:F52}
must be combined with the
four-point face as in~\eqref{eq:penoneloopexpansion}. Once
more, the expression~\eqref{eq:M52} derived from the generating
function is correctly reproduced.

The six-point face at two-loop order already consists of a much larger
number of propagator topologies:
\begin{align}
    F^{(2)}_{1,2,3,4,5,6} &=
     f_{3,[0,1,1,0,1,3]}\mathcal{J}_3^{\mathbold{e}}
	+\frac{1}{2}f_{3,[0,1,1,0,2,2]}\mathcal{J}_3^{\mathbold{e}}
	+f_{3,[0,1,1,1,1,2]}\mathcal{J}_3^{\mathbold{e}}
	+f_{3,[0,1,1,1,2,1]}\mathcal{J}_3^{\mathbold{e}}
   	\nonumber\\ & \quad
	+\frac{1}{2}f_{3,[0,1,1,2,1,1]}\mathcal{J}_3^{\mathbold{e}}
	+\frac{1}{2}f_{3,[0,1,2,0,1,2]}\mathcal{J}_3^{\mathbold{e}}
	+\frac{1}{2}f_{3,[0,1,2,0,2,1]}\mathcal{J}_3^{\mathbold{e}}
	+\frac{1}{4}f_{3,[1,1,1,1,1,1]}\mathcal{J}_3^{\mathbold{e}}
    \nonumber\\ & \quad
	+f_{2,[0,1,1,0,1,1,2]}\mathcal{J}_2^{\mathbold{e}}
	+\frac{1}{2}f_{2,[0,1,1,0,1,2,1]}\mathcal{J}_2^{\mathbold{e}}
	+f_{2,[0,1,1,1,1,1,1]}\mathcal{J}_2^{\mathbold{e}}
	+f_{2,[0,1,2,0,1,1,1]}\mathcal{J}_2^{\mathbold{e}}
    \nonumber\\ & \quad
    +\frac{1}{4}f_{1,[0,1,1,1,0,1,1,1]}\mathcal{J}_1^{\mathbold{e}} + (\grp{D}_6\text{ perms})\,.
\end{align}
In particular, the double-penta topology $\mathcal{J}_1$ appears for
the first time.
Again, the coefficients are labeled by to the distances between
neighboring points, and are given by
\begin{align}
	f_{3,[0,1,1,0,1,3]}	&= \frac{1}{2}(x_{13}^2 - y_{13}^2)\,\big(P_{15,23} - 2 y_{13}^2 y_{24}^2\big)\,, \nonumber\\
	f_{3,[0,1,1,0,2,2]}
	&= -y_{13}^2\,y_{25}^2\,(x_{13}^2 + y_{13}^2)\,, \nonumber \\
	f_{3,[0,1,1,1,1,2]}
	&= -\frac{1}{2}x_{15}^2(x_{13}^2 x_{24}^2-y_{13}^2 y_{24}^2)
	+\frac{1}{4} P_{12,35}(x_{14}^2 - y_{14}^2)
	+\frac{1}{2}y_{13}^2\big(2 y_{14}^2 y_{25}^2-P_{15,24}\big)\,,\nonumber \\
	f_{3,[0,1,1,1,2,1]}
	&= -\frac{1}{4} P_{15,23} x_{14}^2
	+ \frac{1}{2}x_{12}^2 x_{13}^2 x_{46}^2
	- \frac{1}{2}x_{16}^2 y_{13}^2 y_{24}^2
	-\frac{1}{4} y_{14}^2\big(P_{12,35} - 4 y_{13}^2 y_{26}^2\big)\,, \nonumber\\
	f_{3,[0,1,1,2,1,1]}
	&= -\frac{1}{2}P_{15,24} y_{13}^2
	- \frac{1}{2}y_{15}^2\big(P_{12,35} - 2 y_{13}^2 y_{26}^2\big)\,, \nonumber \\
	f_{3,[0,1,2,0,1,2]}
	&= \frac{1}{2}(x_{14}^2 - y_{14}^2)\Big(\frac{1}{2}P_{15,23} + x_{12}^2 x_{45}^2 - 2 y_{14}^2 y_{25}^2\Big)\,, \nonumber\\
	f_{3,[0,1,2,0,2,1]}
	&= \frac{1}{4}(x_{14}^2 - y_{14}^2)\big(P_{15,23} - 2 x_{12}^2 x_{46}^2 - 4 y_{14}^2 y_{26}^2\big)\,,\nonumber \\
	f_{3,[1,1,1,1,1,1]}
	&= \frac{1}{4}\bigg(
	\begin{bmatrix}
		2 & 3 & 4 \\
		5 & 6 & 1
	\end{bmatrix}+
	\begin{bmatrix}
		6 & 1 & 2 \\
		3 & 4 & 5
	\end{bmatrix}
	+ P_{25,46} x_{13}^2 + P_{13,25} x_{46}^2 + P_{25,36} y_{14}^2+ P_{14,25} y_{36}^2  \nonumber\\
	&\quad -2 x_{56}^2 y_{13}^2 y_{24}^2 - 2 x_{45}^2 y_{13}^2 y_{26}^2
	 - 2 x_{23}^2 y_{15}^2 y_{46}^2 -2 x_{12}^2 y_{35}^2 y_{46}^2 + 4 y_{13}^2 y_{25}^2 y_{46}^2
	\bigg)\,, \nonumber\\
	f_{2,[0,1,1,0,1,1,2]}
	&=\frac{1}{2} x_{2b}^2 \big(P_{16,35} y_{13}^2 - 2 x_{13}^2 y_{14}^2 y_{35}^2\big)\,,\nonumber \\[6pt]
	f_{2,[0,1,1,0,1,2,1]}
	&= \frac{1}{2}x_{13}^2 x_{2b}^2\big(P_{13,56} - 2 y_{14}^2 y_{36}^2\big)\,, \nonumber\\
	f_{2,[0,1,1,1,1,1,1]}
	&= \frac{1}{4}\bigg(
	-\begin{bmatrix}
		1 & 2 & 3 & 4 \\
		5 & 6 & 1 & b
	\end{bmatrix}-
	\begin{bmatrix}
		1 & 2 & 3 \\
		5 & 6 & b
	\end{bmatrix} y_{14}^2
	\nonumber\\
	&\quad + (x_{13}^2 x_{2b}^2 - x_{12}^2 x_{3b}^2)\big(P_{14,56} + x_{56}^2 y_{14}^2 - 2 y_{15}^2 y_{46}^2\big)
	\bigg)\,, \nonumber \\
	f_{2,[0,1,2,0,1,1,1]}
	&= \frac{1}{4} x_{2b}^2\big(P_{16,35} y_{14}^2 + x_{14}^2 (P_{13,56} - 2 y_{15}^2 y_{46}^2)\big)\,, \nonumber \\
    f_{1,[0,1,1,1,0,1,1,1]}
    &= \frac{1}{4}{x^2_{ab}}
    \big(P_{14,23}-2y_{13}^2y_{24}^2\big)
    \big(P_{14,56}-2y_{15}^2y_{46}^2\big)
    -\frac{1}{4}
    \begin{bmatrix}
    1 & 2 & 3 & 4 & a \\
    4 & 5 & 6 & 1 & b
    \end{bmatrix}
    +
    \frac{1}{4}y_{14}^2\, h_{\text{DP}}
    \,,
\end{align}
where again it is understood that $\mathbold{e}=(1,1+\ell_2,\dots)$
starts with the point label~$1$. The last (double-penta) coefficient
$f_{1,\dots}$ can be obtained from~\eqref{eq:f1}, with the factor
$h_{\text{DP}}$ given in~\eqref{eq:factorDP}.
In accordance
with the saturation property~\eqref{eq:fSaturation}, one can see that
$f_{3,[0,1,1,0,1,3]}=f_{3,[0,1,1,0,1,2]}$.
The two-loop hexagon $M_{6,2}$ can then be
obtained by using~\eqref{eq:hextwoloopexpansion}, and plugging in the
expressions for the two-loop four-point and five-point faces as well
as the product of one-loop squares.

\section{10d Null Polygons at Integrated Level}
\label{sec:IntegratedPoly}

We obtain the coupling-dependent polygonal correlator
$\intpoly_n$~\eqref{eq:PolygonIntegrated} by integrating over the
positions $x_{n+1},\dots,x_{n+\ell}$ of the Lagrangian
insertions that appear in the integrands $M_{n,\ell}$ discussed so far.
Explicitly, up to two loops we have
\begin{equation}
\mathbb{M}_{n}
=\, M_{n,0}
- \frac{1}{2}g^{2}\int \frac{d^4x_{n+1}}{\pi^2}\,M_{n,1}
+ \frac{1}{2}g^{4}\int\hspace{-0.5em}\int
  \frac{d^4x_{n+1}}{\pi^2} \frac{d^4x_{n+2}}{\pi^2}
  \,M_{n,2}
+ \order{g^6}
\,.
\label{eq:GnIntegrated}
\end{equation}
In the following, we analyze the coupling-dependent polygon correlators,
focusing on the particular case of the pentagon at one and two
loops.

\subsection{Pentagon at Integrated Level}
\label{sec:integrated}

Very little is known about the integrated pentagon in perturbation
theory. At two loops, the (completely off-shell) five-point conformal integrals
$\mathcal{I}_6$ and $\mathcal{I}_7$~\eqref{eq:pentagonIntegrals} are not known for
general space-time configurations. But they have been computed in some
physically relevant kinematic limits~\cite{Bercini:2024pya}, which we address below.
Collecting~\eqref{eq:M51} and~\eqref{eq:M52}, the two-loop pentagon can be
written as
\begin{align}
    \mathbb{M}_5 & = \frac{1}{5}M_{5,0}
    -\frac{g^2}{2}p_{1234}
    \mathbb{I}_1^{[1234]}+\frac{g^4}{2}p_{1234}x_{14}^2(1-d_{14})\mathbb{I}_6^{[14|23|5]}
    \nonumber \\ & \quad
    +\frac{g^4}{2}p_{1234}\left(x_{13}^2(1-d_{13})\mathbb{I}_5^{[13|24]}
    + x_{24}^2(1-d_{24})\mathbb{I}_5^{[24|13]} \right)
    \nonumber \\ & \quad
    +\frac{g^4}{2}\left(
    x_{35}^2(1-d_{35})q_{12345}+x_{13}^2(1-d_{13})q_{54321}\right)\mathbb{I}_7^{[12|45|3]}
    + (\grp{C}_5\;\text{perms}) + \mathcal{O}(g^6)
    \,,
    \label{eq:G5Integrated}
\end{align}
where $+(\grp{C}_5\;\text{perms})$ stands for summing over cyclic
permutations, the factor $1/5$ compensates for over-counting the
tree-level contribution, and the factors $p$ and $q$ are defined in~\eqref{eq:p},~\eqref{eq:q}. The conformal integrals appearing above
follow from the respective integrands~\eqref{eq:oneloopbox},
\eqref{eq:twoloopladder}, and~\eqref{eq:pentagonIntegrals}, and are
given by
\begin{align}
    \mathbb{I}_1^{[1234]} &
    = \int \frac{d^4 x_6}{\pi^2} \frac{1}{x_{16}^2 x_{26}^2 x_{36}^2 x_{46}^2}
    = \frac{F_1\left(z_1,\bar{z}_1\right)}{x_{13}^2x_{24}^2} \label{int1} \,,\\
    \mathbb{I}_5^{[12|34]} &
    = \int \hspace{-0.5em} \int \frac{d^4 x_6}{\pi^2} \frac{d^4 x_7}{\pi^2}
      \frac{1}{(x_{16}^2 x_{26}^2 x_{36}^2) x_{67}^2 (x_{17}^2 x_{27}^2 x_{47}^2)}
    = \frac{F_2\left(\frac{1}{z_1},\frac{1}{\bar{z}_1}\right)}{x_{12}^2 x_{34}^2}
    \label{int2} \,,\\
    \mathbb{I}_6^{[12|34|5]} &
    = \int \hspace{-0.5em} \int \frac{d^4 x_6}{\pi^2} \frac{d^4 x_7}{\pi^2}
      \frac{x_{56}^2}{(x_{16}^2 x_{26}^2 x_{36}^2 x_{46}^2) x_{67}^2 (x_{17}^2 x_{27}^2 x_{57}^2)}
    \label{int3}  \,,\\
    \mathbb{I}_7^{[12|34|5]} &
    = \int \hspace{-0.5em} \int \frac{d^4 x_6}{\pi^2} \frac{d^4 x_7}{\pi^2}
      \frac{1}{(x_{36}^2 x_{46}^2 x_{56}^2) x_{67}^2 (x_{17}^2 x_{27}^2 x_{57}^2)}
    \label{int4}
    \,.
\end{align}

\begin{figure}[t]
    \centering
    \includegraphics[width=\linewidth]{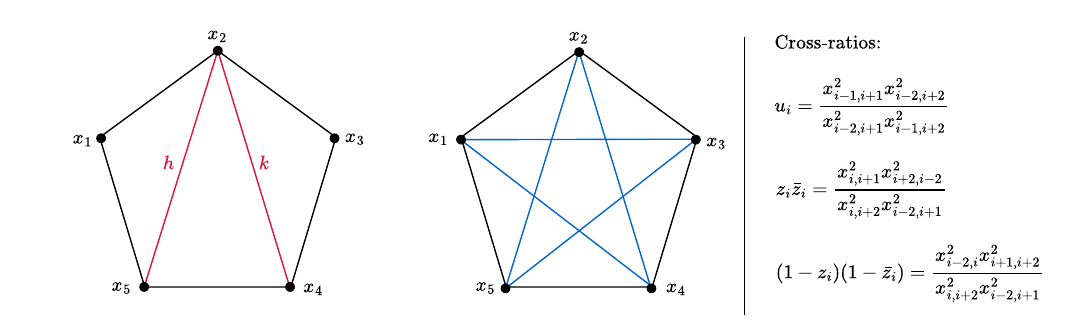}
    \caption{The two type of pentagon configuration we consider: on the first the point $x_2$ has
    $h$ and $k$ bridges to the points $x_5$ and $x_4$, respectively. On the second, we will argue that even when all highlighted bridges are non-zero the pentagon is still planar and non-trivial. On the right, the cross-ratios we use to parametrize this five
    point correlation function.}
    \label{fig:PentagonBridges}
\end{figure}

Due to conformal symmetry, five-point correlators and the integrals
\eqref{int1}--\eqref{int4} are functions of five independent conformal
cross-ratios $u_i$, which we define in \figref{fig:PentagonBridges}. There, we also
introduce an over-complete ten-element set of
cross-ratios $z_i,\bar{z}_i$, $i=1,\dots,5$, which will be useful to write the
four-point sub-correlators of the polygons (and four-point integrals) which
depend on just a pair of cross-ratios. For example, each of the
first two integrals \eqref{int1} and \eqref{int2} depend only on a
single pair of cross ratios.
These are the well-known one-loop and two-loop ladder
integrals, which are given by singled-valued
polylogarithms~\cite{Usyukina:1993ch}
\begin{equation}
    F_{p}(z,\bar{z}) = \sum_{j=0}^{p}\frac{(-1)^{j}(2p-j)!}{p!(p-j)!j!}\,\log(z\bar{z})^{j}\times\frac{\Li_{2p-j}(z)-\Li_{2p-j}(\bar{z})}{z-\bar{z}}\,.
    \label{BoxIntegrals}
\end{equation}
For the two five-point conformal integrals \eqref{int3} and
\eqref{int4}, no closed-form expressions are known
for generic configurations of the five insertion points. However,
they have been computed in a particular kinematic regime that will
be important for us~\cite{Bercini:2020msp}: The four-dimensional null
polygon limit, also known as double
light-cone limit. Evaluating the pentagon in
this limit is what we turn to now.

\subsection{4d Null Polygon Limit}

In the four-dimensional null polygon limit, the insertion points of
the operators approach the cusps of a four-dimensional null polygon, such
that all neighboring points become light-like separated ($x_{i,i+1}^2 \to 0$).
In this limit, all five-point cross-ratios $u_i$ defined in
\figref{fig:PentagonBridges} go to zero, $u_i \to 0$.
The conformal integrals develop logarithmic
divergences, their leading behavior was computed
in~\cite{Bercini:2024pya} and is given by
\begin{align}
    x_{13}^2x_{24}^2\mathbb{I}_1^{[1234]} &
    =\ell_5(\ell_1+\ell_4)+2\zeta_2
    \label{eq:NullStart}
    \,, \\
    x_{13}^4 x_{24}^2\mathbb{I}_5^{[13|24]} &
    =\frac{1}{4}\ell_5^2(\ell_1^2+\ell_4)^2 +\frac{1}{2}(\ell_5^2
    +4\ell_5(\ell_1+\ell_4) +
    (\ell_1+\ell_4)^2)\zeta_2+\frac{21}{2}\zeta_4
    \,, \\
    x_{13}^2x_{14}^2x_{24}^2 \mathbb{I}_6^{[14|23|5]} & =
    \frac{1}{4}\ell_5\left(\ell_1^2\ell_2+\ell_3\ell_4^2
    +2\ell_1\ell_2\ell_3 + 2\ell_2\ell_3\ell_4\right) -
    \frac{1}{2}\big(\ell_1^2+\ell_4^2+2\ell_1\ell_3+2\ell_2\ell_4
    \\ \nonumber & \quad
    -2\ell_2\ell_5-2\ell_3\ell_5 -4\ell_2\ell_3 - 4\ell_1\ell_5 -
    4\ell_4\ell_5\big)\zeta_2 + (\ell_2+\ell_3-2\ell_5)\zeta_3+5\zeta_4
    \,, \\
    x_{13}^2x_{24}^2x_{35}^2\mathbb{I}_7^{[12|34|5]} & =
    \frac{1}{4}\ell_1\ell_5(\ell_1+2\ell_4)(\ell_5+2\ell_2) +
    (\ell_2\ell_4+2\ell_1\ell_2+2\ell_1\ell_5+2\ell_4\ell_5)\zeta_2
    \nonumber \\ & \quad
    -(\ell_2+\ell_4)\zeta_3 + \frac{31}{4}\zeta_4
    \label{eq:NullFinal}
    \,.
\end{align}
where $\ell_i = \log(u_i)$. Subleading terms will contain
powers of $u_i$ and are not included.

In the four-dimensional null polygon limit, we can therefore express the two-loop
pentagon~\eqref{eq:G5Integrated} as a function of the (divergent)
logarithms $\ell_i$ and the finite propagators $d_{ij}$, which
encode the number of internal bridges in the pentagon.

The polygons $\intpoly_n$ are limits of generating functions in the
charges of the external operators. As such, the polygons themselves
are also generating functions.%
\footnote{The polygon limit projects the full generating function to
terms that have large powers of four-dimensional propagators
$d_{i,i+1}$ on the perimeter. At five and six points, one can engineer
large-charge operators ($\mathcal{O}_k$ and their
descendants~\eqref{eq:Oscalar}) such that the resulting correlator is
exclusively given by the square of the respective polygon. In general, this is
not the case: Correlators of operators $\mathcal{O}_k$, even with $k$
large, will include terms that are squares of polygons, but it will
also contain other terms. See the discussion at the end of
\secref{sec:ten-dimensional-null}.}
In order to study the four-dimensional null limit, we further project
the polygons down to terms that contain specific powers of propagators
$d_{ij}$ on the diagonals $(ij)$ of the polygon. Specifically, we
focus on the terms displayed in \figref{fig:PentagonBridges}, which
are defined as
\begin{equation}
    \mathbb{M}_{5}^{h,k} = \eval[s]*{
      \text{coefficient of } (d_{25}^{h}d_{24}^k)
      \text{ in } \mathbb{M}_{5}
    }_{d_{i,j}\to 0}
    \label{eqFormalExtraction}\,.
\end{equation}
The object~$\intpoly^{h,k}_5$ is the part of the pentagon~$\intpoly_5$
that has~$h$ propagators (or units of R-charge flux, also called ``bridges'') between
operators~$2$ and~$5$, and~$k$ propagators between operators~$2$
and~$4$. The coefficients~$\intpoly^{h,k}_5$ were studied
earlier~\cite{Olivucci:2022aza}, and can easily be extracted
from~\eqref{eq:G5Integrated} and be evaluated in the four-dimensional
null limit using~\eqref{int1}--\eqref{int4}. For example, the
pentagon with two bridges of length one is given by
\begin{multline}
\mathbb{M}_5^{1,1}
=1
-\frac{g^4}{4}\Big(\ell_1^2(\ell_2+\ell_5)^2+\ell_3^2(\ell_2+\ell_4)^2
+ 2\zeta_2\big(\ell_1^2+2\ell_2^2+\ell_3^2+\ell_4^2+\ell_5^2
\\
+4\ell_1\ell_2 +4\ell_2\ell_3 +2\ell_2\ell_4 +4\ell_3\ell_4
+4\ell_1\ell_5 +2\ell_2\ell_5\big) + 84 \zeta_4\Big)
+\order{g^6}
\,.
\end{multline}
Notice that due to the factorization
property~\eqref{eq:PolygonFactorizationIntegrated} of the polygons,
the loop corrections to this object only start at $\order{g^4}$. More
generally, the pentagon coefficients $\mathbb{M}_5^{h,k}$ will only
receive loop corrections at order
$g^{2(\min\set{h,k}+1)}$. From this point of view, the pentagon with no internal bridges
$\mathbb{M}_5^{0,0}$ is the most non-trivial object, and it has been
studied before~\cite{Fleury:2020ykw,Bork:2022vat,Bercini:2024pya,Belitsky:2025bgb}.
In the four-dimension null polygon limit, we find that it exponentiates to
\begin{equation}
\log{\left(\mathbb{M}_5^{0,0}\right)}
= - 5g^2\zeta_2 + \frac{135}{4}g^4\zeta_4
- \sum_{n=1}^{5}\left((g^2-2g^4\zeta_2)\ell_n\ell_{n+1}-g^4\zeta_2\ell_n(\ell_n+\ell_{n+2})\right)
+\mathcal{O}(g^6)
\,,
\end{equation}
which agrees with~\cite{Bercini:2024pya}, and with the all-loop
conjecture~\cite{Belitsky:2025bgb} for the null pentagon. In our
notation, the latter reads
\begin{equation}
    \log{\left(\mathbb{M}_5^{0,0}\right)}
    = \mathcal{C}_0(g)
    - \sum_{n=1}^{5}\left(\frac{\gammaCusp(g)}{4}\ell_n\ell_{n+1}
    +\frac{\gammaOct(g)-\gammaCusp(g)}{8}\ell_n(\ell_n+\ell_{n+2})\right)
    \,.
    \label{eq:M00NullConjecture}
\end{equation}
For comparison, the square $\intpoly_4^{0,0}$ in the four-dimensional
null limit expands to
\begin{equation}
    \log{\left(\mathbb{M}_4^{0,0}\right)}
    = \mathcal{D}_0(g)
    - \sum_{n=1}^4 \left(
        \frac{\Gamma_{\text{oct}}(g)+4g^2}{32}\ell_n\ell_{n+1}
        + \frac{\Gamma_{\text{oct}}-4g^2}{64}\ell_n(\ell_n + \ell_{n+2})
    \right)
    \,,
\end{equation}
where $\ell_i=\log(u_i)$, and $u_1=u_3=u$, $u_2=u_4=v$.
Here, $\mathcal{C}_0(g)$ and $\mathcal{D}_0(g)$ are finite functions
of the coupling.
The perturbative expansions of the cusp and octagon anomalous
dimensions are $\gammaCusp(g) = 4g^2 - 8\zeta_2g^4+\order{g^6}$
and $\gammaOct(g) = 4g^2 - 16\zeta_2g^4+\order{g^6}$.
Note that our two-loop data can barely distinguish between
$\gammaCusp$ and $\gammaOct$, thus our result only lends marginal
support to the conjecture~\eqref{eq:M00NullConjecture}.
It would be
nice to evaluate the pentagon at higher orders in perturbation theory
to test this conjecture, and understand if its null limit is
controlled by $\gammaCusp(g)$ and $\gammaOct(g)$,
or other tilted cusps $\Gamma_{\alpha}(g)$, as is the case in the
origin limits of gluon scattering
amplitudes~\cite{Basso:2020xts,Basso:2022ruw}.

\subsection{Leading Logarithms}

On top of the null polygonal limit, there is another, more simplifying limit that
one can consider: The leading logarithm or ``stampedes''
double-scaling limit~\cite{Belitsky:2019fan,Olivucci:2021pss,Olivucci:2022aza}. This
limit is attained by taking both $g \to 0$ and $u_i \to 0$,
$i=1,\dots,5$, such that the so-called cusp times
\begin{equation}
    t_n^2 \equiv g^2\log(u_{n-2})\log(u_{n+2}) = g^2\ell_{n-2}\ell_{n+2}
    \label{cuspTimes}
\end{equation}
are held fixed. This limit projects to the
most divergent terms in the four-dimensional null polygon limit, which are
the terms with the highest powers of logarithms
in~(\ref{eq:NullStart}-\ref{eq:NullFinal}). For example,
\begin{equation}
g^2x_{13}^2x_{24}^2\mathbb{I}_1^{[1234]}
\to (t_2^2 + t_3^2)
\qquad  \text{and} \qquad
g^4x_{13}^2x_{24}^2x_{35}^2\mathbb{I}_7^{[12|34|5]}
\to \frac{(t_3^2+2t_2^2)(t_3^2+2t_4^2)}{4}
\,.
\end{equation}

Akin to the ordinary four-dimensional null polygon limit, we can also compute the stampedes
limit of the pentagons $\mathbb{M}_5^{h,k}$ with arbitrary bridge
lengths. For example, for the two cases $\intpoly^{0,0}_5$ and
$\intpoly^{1,1}_5$ considered before, the stampedes limit
is given by
\begin{align}
    \mathbb{M}_5^{0,0} &\to 1
    -(t_1^2+t_2^2+t_3^2+t_4^2+t_5^2)+\frac{1}{2}(t_1^2+t_2^2+t_3^2+t_4^2+t_5^2)^2
    +\mathcal{O}(t_i^6)
    \label{eq:M00Stampedes2Loops}
    \\
    \mathbb{M}_5^{1,1} &\to 1 +(t_3^2+t_4^2)^2+(t_1^2+t_5^2)^2 +\mathcal{O}(t_i^6)
\end{align}
Remarkably, these two stampedes-limit pentagons are not
independent~\cite{Olivucci:2021pss}. In fact, all coefficients
$\intpoly^{h,k}_5$ in the stampedes limit
are related via a set of coupled lattice Toda differential equations,
see equation~(10) in~\cite{Olivucci:2022aza}. The seed of these
equations is the empty pentagon $\mathbb{M}_5^{0,0}$, which as we saw already
exponentiates in the four-dimensional null polygonal limit, and in the
stampedes limit simply becomes
\begin{equation}
    \mathbb{M}_5^{0,0} = e^{-(t_1^2 + t_2^2 + t_3^2 + t_4^2 + t_5^2)}\,.
\end{equation}
The two-loop expression~\eqref{eq:M00Stampedes2Loops} of
$\mathbb{M}_5^{0,0}$ computed above perfectly reproduces the perturbative expansion
of this exponentiation. Moreover, we verified that the pentagon with
length-one bridges $\mathbb{M}_5^{1,1}$ as well as all other pentagons
$\mathbb{M}_5^{h,k}$ with arbitrary bridge lengths $h,k$ computed from
our two-loop result for $\intpoly_5$ also agree perfectly with the
stampedes prediction.

\paragraph{Polygons with Crossing Bridges.}

Finally, let us address the seemingly non-planar configurations like
the ones depicted on the center of \figref{fig:PentagonBridges}. Such
configurations can be extracted in the exact same way as before via
\begin{equation}
    \mathbb{M}_{5}^{h,k,l,m,n} = \text{coefficient of } (d_{25}^{h}d_{24}^k d_{13}^l d_{14}^m d_{35}^n)
      \text{ in } \mathbb{M}_{5}
    \label{eqFormalExtraction2}\,.
\end{equation}
Even when the bridges $d_{25}^h,\dots,d_{35}^n$ overlap each other, these objects are still
planar and thus non-vanishing in perturbation theory, as one can see
from their values in the stampedes limit, \eg
\begin{align}
    \mathbb{M}_{5}^{2,0,1,0,0} &= t_1^2+t_5^2 - \frac{t_1^2(t_2^2+2t_3^2)}{2}-\frac{t_5^2(t_4^2+2t_3^2)}{2} \,, \\
    \mathbb{M}_{5}^{1,1,1,0,0} &= \frac{t_5^2+2t_1^2}{2}\frac{t_5^2+2t_4^2}{2}
    \,.
    \label{M111}
\end{align}
This happens because the bridges in~\eqref{eqFormalExtraction2} can be
identified one-to-one with contractions of the R-charge
polarization vectors $y_i \cdot y_j$ of the operators, but not with
color (and spacetime) propagators.
At tree-level, each factor $d_{ij}$ is truly a propagator, containing
$y_i\cdot y_j$, $1/x_{ij}^2$, and a color propagator factor.
Therefore, terms with overlapping $d_{ij}$ factors are truly
non-planar, \ie they vanish at leading order in the planar limit.
Beyond tree level, loop corrections can re-route the R-charge flow
and change the color structure (as well as the $1/x_{ij}^2$ propagator
structure).  Thus at loop order, the color structure no longer maps
one-to-one to the topology of $d_{ij}$ factors, and terms with
overlapping $d_{ij}$ factors may acquire terms that contribute to the
leading order in the $1/\Nc$ expansion. However, the more the $d_{ij}$
factors overlap, the more loop orders are required for such planar
terms to occur.
Therefore, the objects~\eqref{eqFormalExtraction2} in the planar limit
only start at higher orders
in perturbation theory. For example, $\mathbb{M}_{5}^{1,1,1,0,0}$
starts only at two loops~\eqref{M111}. We can expect that the pentagon with
$h=k=l=m=n=1$, \ie all five diagonals present, is only non-zero
starting from four loops, which is inaccessible
with our present data. And consistently, no such term exists in our
two-loop data for $\intpoly_5$.
To summarize, the pentagon with overlapping
bridges is non-planar in internal R-symmetry space, but may be planar
in color space. This feature is not restricted to the pentagon, in
fact it equally occurs in the square $\intpoly_4$,
where one can similarly define
\begin{equation}
    \mathbb{M}_{4}^{h,k} = \text{coefficient of } (d_{13}^{h}d_{24}^k)
    \text{ in } \mathbb{M}_{4}\,,
    \label{eqFormalExtraction3}
\end{equation}
and terms with both $h$ and $k$ non-zero are ``non-planar'' in
R-symmetry space.
Since the square has disk topology, one cannot think of these two
diagonals as existing ``inside'' and ``outside'' of the large-R-charge
square frame (this may happen for the
full four-point correlator with sphere topology). Therefore, all
$\intpoly^{h,k}$ with $h,k>0$ are zero at tree level, and only start being
non-zero at order $g^{2(h+k-1)}$ in perturbation theory.
For example, using the high-loop data of~\cite{Caron-Huot:2021usw},
we find
\begin{equation}
    \mathbb{M}_4^{1,1} = 0 + t^2 + \frac{t^4}{2}+\frac{t^6}{6} + \frac{5 t^8}{144} + \frac{7 t^{10}}{1440} + \dots
    \,,
\end{equation}
where $t = g^2 \log(z)\log(1-z)$ is the cusp times held fixed.
Expanding to higher orders, up to $\order{t^{40}}$, and re-summing, we find
\begin{equation}
    \mathbb{M}_4^{1,1} = 2t^2 \texttt{I}_0^2(2t)-2t^2 \texttt{I}_1^2(2t) - t\texttt{I}_0(2t)\texttt{I}_1(2t)
    \,,
\end{equation}
where $\texttt{I}_n$ is the modified Bessel function of the first
kind.
Interestingly, we find that $\intpoly_4^{1,1}$ is
related with square with only one bridge (of size two) via
\begin{equation}
    \frac{1}{2 t} \frac{d}{dt} \mathbb{M}_4^{1,1} = \mathbb{M}_4^{2,0}\,.
\end{equation}
It would be nice to further study the stampedes limits of these more
general squares and pentagons with crossing bridges
$\mathbb{M}_{4}^{h,k}$ and $\intpoly_5^{h,k,l,m,n}$. The
relation above suggests that they might also be related by Toda-like
equations, as it happens for polygons with non-crossing bridges
$\mathbb{M}_{4}^{h,0}$ and $\intpoly_5^{h,k,0,0,0}$~\cite{Belitsky:2020qir,Olivucci:2021pss}.

\section{Match with Hexagons}
\label{sec:oneloop-hexagonalization}

The faces decomposition~\eqref{eq:FacesExpansion}, together with the
one-loop faces~\eqref{eq:oneLoopFaces} is
reminiscent of the one-loop hexagonalization
result~\cite{Bargheer:2018jvq}, but not quite equal to it. According
to the hexagonalization proposal~\cite{Fleury:2016ykk,Eden:2016xvg},
the loop corrections to a correlation function of local operators in
the planar limit is obtained by dressing
each tree-level graph by mirror particles. At one-loop
order, the mirror particles are effectively confined to individual
faces of the tree-level (ribbon) graphs. However, the sum over mirror particles
is not only a function of $x_{ij}^2$, but also produces factors
$d_{ij}$ that in particular can occur with both positive and also
\emph{negative} exponents.
The product of $d_{ij}$ factors produced by the underlying tree-level graph thus gets
modified to a non-trivial polynomial. In other words, a given monomial
(product of $d_{ij}$ factors) in the final answer will receive contributions
from different tree-level graphs. Let us see how this works in practice.

The hexagonalization contributions that dress each individual
$n$-point face of each tree graph at one-loop order amount
to~\cite{Fleury:2016ykk,Bargheer:2018jvq}
\begin{align}
H_{i_1,\dots,i_n}
&=1+g^2H_{i_1,\dots,i_n}^{(1)}+\order{g^4}
\,,\nn\\
H_{i_1,\dots,i_n}^{(1)}
&=\sum_{k=1}^{n-2}\sum_{m=k+2}^{n-\delta_{k1}} m_{i_k,i_{k+1},i_m,i_{m+1}}
\,,\qquad
i_{n+1}\equiv{i_1}
\,,
\label{eq:hexOneLoopFaces}
\end{align}
where the sums run over non-overlapping pairs $(i_k,i_{k+1})$,
$(i_m,i_{m+1})$, or equivalently over pairs of non-adjacent edges of the
polygonal face. The function $m$ is given by%
\footnote{This expression has an overall factor $-1/2$ compared
to~\cite{Bargheer:2018jvq}, which is due to differing conventions.}
\begin{equation}
m_{ijkl}=\frac{-1}{2x_{i0}^2x_{j0}^2x_{k0}^2x_{l0}^2}
\brk*{
 x_{il}^2x_{jk}^2\brk*{1-\frac{d_{il}d_{jk}}{d_{ij}d_{kl}}}
-x_{ik}^2x_{jl}^2\brk*{1-\frac{d_{ik}d_{jl}}{d_{ij}d_{kl}}}
}
\,.
\label{eq:mFunction}
\end{equation}
The hexagonalization contributions sum to~\eqref{eq:mFunction}
integrated over $\dd[4]{x_0}$. At one-loop order however, the mapping between the integrated
answer and the integrand~\eqref{eq:mFunction} is unambiguous.%
\footnote{Up to parity-odd terms that integrate to zero.}
The function $m_{ijkl}$ has an $\grp{S_2}\times\grp{S}_2$ symmetry:
\begin{equation}
m_{1234}=m_{2143}=m_{3412}=m_{4321}
\,,
\end{equation}
which implies the expected dihedral symmetry for the faces
$H_{1,\dots,n}$. In addition, the faces satisfy the ``pinching
property''~\cite{Bargheer:2018jvq}
\begin{equation}
H_{\dots,i,j,k,j,l,\dots}
=H_{\dots,i,j,l,\dots}
\,.
\end{equation}
As one can see in~\eqref{eq:mFunction}, the function $m_{ijkl}$ has
factors $d_{ij}$ in the denominator.
However, in the combination~\eqref{eq:hexOneLoopFaces}, only
denominators $d_{ij}$ belonging to the perimeter of the face appear, \ie
$H_{1,\dots,n}$ at most has denominator factors $d_{i,i+1}$, which cancel
against the corresponding numerator factors of the ambient tree graph, such
that the one-loop correlator is polynomial in $d_{ij}\sim y_{ij}^2$, as required.
For example, the one-loop square and pentagon faces become
\begin{align}
\label{eq:hex4gon}
H_{1234}^{(1)}
&=m_{1234}+m_{2341}
\\\nn
&=\frac{1}{2x_{10}^2x_{20}^2x_{30}^2x_{40}^2}
\brk*{
 x_{13}^2x_{24}^2\brk*{1-\frac{d_{13}d_{24}}{d_{14}d_{23}}}
-x_{14}^2x_{23}^2\brk*{1-\frac{d_{14}d_{23}}{d_{12}d_{34}}}
+ (\text{$\grp{C}_4$ perms})
}
\,,\\
\label{eq:hex5gon}
H_{12345}^{(1)}
&=m_{1234}+m_{1245}+m_{2345}+m_{2351}+m_{3451}
\\\nn
&=
 \frac{x_{13}^2x_{24}^2}{x_{10}^2x_{20}^2x_{30}^2x_{40}^2}\brk*{1-\frac{d_{13}d_{24}}{d_{12}d_{34}}}
-\frac{x_{14}^2x_{23}^2}{x_{10}^2x_{20}^2x_{30}^2x_{40}^2}\brk*{1-\frac{d_{14}d_{23}}{d_{12}d_{34}}}
+ (\text{$\grp{C}_5$ perms})
\,.
\end{align}
The one-loop contribution to any correlator is obtained by inserting
the one-loop mirror-particle sums $H^{(1)}_{i_1,\dots,i_n}$ into the
faces of the tree-level graphs of the respective correlator. Therefore, the complete
one-loop generating function $G_{n,1}$ is obtained by inserting the
same functions $H^{(1)}_{i_1,\dots,i_n}$ into the faces of the graphs of the
tree-level generating function $G_{n,0}$. The edges of the tree-level
graphs now produce re-summed (ten-dimensional) propagators $D_{ij}=d_{ij}+d_{ij}^2+d_{ij}^3+\dots$.
Notably, the
structure of the function $m_{ijkl}$ is such that it can produce
numerator factors $1-d_{ij}=D_{ij}/d_{ij}$ when summing over tree
graphs. This may affect the ten-dimensional pole structure of the
ambient graph, since it can modify the product of
$D_{ij}\sim1/X_{ij}^2$ factors produced by the graph. Consequently,
the coefficient of a given $D_{ij}$ product in the final result may
receive contributions from
several different tree graphs ($D_{ij}$ monomials). For example,
\begin{multline}
(1+D_{jk})m_{ijkl}
=\frac{m_{ijkl}}{1-d_{jk}}
\\
=\frac{1}{x_{i0}^2x_{j0}^2x_{k0}^2x_{l0}^2}
\brk[s]*{
    \frac{d_{il}x_{il}^2x_{jk}^2}{d_{ij}d_{kl}}
    +\frac{D_{jk}}{d_{jk}}\brk*{
        \brk*{1-\frac{d_{il}}{d_{ij}d_{kl}}}x_{il}^2x_{jk}^2
        -\brk*{1-\frac{d_{ik}d_{jl}}{d_{ij}d_{kl}}}x_{ik}^2x_{jl}^2
    }
}
\,.
\end{multline}
This property makes it slightly non-trivial to match the
proposal~\eqref{eq:oneLoopFaces} analytically against the prediction
from hexagonalization. But by explicit computation, we do find an exact
match with $M_{n,1}$ for $n=4,5,6,7$ by
inserting the one-loop faces $H_{i_1,\dots,i_m}$ into the graphs of
$M_{n,0}$.%
\footnote{When we do this comparison, we must take the limit $d_{i,i+1}\to1$
for $i=1,\dots,n$ of the one-loop hexagonalization faces
$H_{i_1,\dots,i_m}$, to be consistent with the ten-dimensional polygon
limit.}
Moreover, we also find an exact match with the full generating
functions $G_{4,1}$, $G_{5,1}$, and $G_{6,1}$ after inserting the one-loop faces
into the graphs of the tree-level generating functions $G_{n,0}$.

One might wonder how it can be consistent that inserting the
hexagonalization faces $H_{i_1,\dots,i_n}$ into the graphs of
$\polygon_{n,0}$ gives the same answer as inserting the one-loop faces
$F^{(1)}_{i_1,\dots,i_n}$~\eqref{eq:oneLoopFaces} into the exact same graphs~\eqref{eq:MdecompF}, even
though $H_{i_1,\dots,i_n}$ and $F^{(1)}_{i_1,\dots,i_n}$ are clearly different.
The reason for this are the effects discussed
below~\eqref{eq:hex5gon}, and is also related to the ambiguity pointed out in
\footnoteref{fn:facesAmbiguity}: The two decompositions are
related by shifting terms between different faces and/or different graphs.
Only the sum over all graphs is the same in both decompositions.
The faces $\intface_n$ are still unambiguously
defined by~\eqref{eq:MdecompF} (recursively in $n$), since for the
purpose of defining $\intface_n$, we set
$d_{i,i+1}=0$ in that relation for all $i=1,\dots,n$.

\section{Discussion}
\label{sec:discussion}

Ten-dimensional null polygon correlators are interesting objects for
several reasons. They form a middle ground between planar massless
scattering amplitudes, for which a rich web of interesting structures
and methods have been developed during the past two decades, and
correlation functions of single-trace local operators, which are
significantly more complicated, in particular due to their sphere
topology in color space. The ten-dimensional null polygons~$\intpoly_n$ offer a sweet
spot: They largely tame the complicated structure of general
correlation functions by their disk topology, and at the same time
offer a natural off-shell generalization of massless amplitudes that
is UV and IR finite.

The ten-dimensional null polygons as observables in
$\superN=4$ SYM were established
in~\cite{Caron-Huot:2021usw}.
In this paper, we study the polygon
correlators in more depth, in particular pushing their computation in
the higher-point case with more than four external operators. Our key
results are:
\begin{itemize}
\item A method to directly compute the polygons $\intpoly_n$, which
is derived from the twistor rules for general correlators of arbitrary-charge
BPS scalar operators~\cite{Caron-Huot:2023wdh}.
\item Explicit formulas for the one-loop and two-loop integrands
$\polygon_{n,1}$ and $\polygon_{n,2}$ for any number $n$ of external
operators, based on a tessellation of the integrands into faces that
are free of ten-dimensional poles.
\end{itemize}
In the case of massless scattering amplitudes, the determination of
the two-loop MHV amplitude for any number of
points~\cite{Vergu:2009tu,CaronHuot:2011ky} was an important stepping stone for
many later developments and discoveries. One may hope that many
of the structures found for massless amplitudes will have
generalizations for the polygons~$\intpoly_n$.
Let us highlight some of the potential directions, many of which are
interrelated:
\begin{description}

\item[Massive Amplitudes.]

There is evidence that the polygon correlators are dual to massive
scattering amplitudes on the Coulomb branch of $\superN=4$
SYM, which extends the massless amplitude/correlator duality to an
off-shell setting~\cite{Caron-Huot:2021usw}.

In the $4$d massless limit, the \emph{integrands} $M_{n,\ell}$ reduce to the
$\ell$-loop integrands of $4$d null correlators of $\twentyprime$ operators,
which equal the integrands of massless MHV gluon amplitudes~\eqref{eq:masslessAmpLimit}.
However, this limit does not commute with the integration over the
Lagrangian points. In particular, when taking the massless limit of
the integrated~$\intpoly_n$, its divergences (Sudakov logarithms) are no longer governed by
the cusp anomalous dimension $\gammaCusp$ of massless amplitudes, but
instead by the \emph{octagon anomalous dimension}
$\gammaOct$~\cite{Belitsky:2019fan}. This was first observed for the
square~\cite{Caron-Huot:2021usw}, and later for the
pentagon~\cite{Bercini:2024pya}.%
\footnote{Intriguingly, the octagon anomalous dimension also governs gluon
amplitudes in the ``origin limit''~\cite{Basso:2020xts,Basso:2022ruw}.}

Prompted by the proposed off-shell correlator/amplitude
duality~\cite{Caron-Huot:2021usw}, five-point and six-point amplitudes
of massive W-bosons on the Coulomb branch have been computed~\cite{Bork:2022vat,Belitsky:2025bgb,Belitsky:2025vfc}.
For the five-point amplitude, it was found that its infrared
divergences in the on-shell (massless) limit are governed by the
octagon anomalous dimension~\cite{Bork:2022vat}, which generalizes the
four point result~\cite{Caron-Huot:2021usw}. Moreover, the five-point
amplitude equals the pentagon correlator in the double-scaling
``stampedes'' limit~\cite{Belitsky:2025bgb}. These results further corroborate the
proposed off-shell duality.

However, a full match between off-shell Coulomb-branch amplitudes and
polygon correlators beyond four points is still missing.
The duality is not yet fully formulated,
because we lack a clear organizational principle of the amplitude
polarizations. The latter would be required for a precise dictionary, which
could then be tested (even at integrand level).
Speaking in terms of massless concepts, we are missing the analog of
MHV and N$^k$MHV amplitudes in the massive case.

In its firmly established massless incarnation,
the duality between correlation functions and gluon
amplitudes is in fact a \emph{triality} that also includes
null polygon Wilson loops~\cite{Alday:2007he,Alday:2007hr,Drummond:2007aua,Brandhuber:2007yx,Bern:2008ap,Drummond:2008aq}.
Going off-shell, could there be a Wilson-loop-like object that provides
a bridge between polygon correlators and Coulomb-branch
amplitudes? For first steps in this direction, see~\cite{Belitsky:2021huz}.

\item[Recursion Relations.]

The 10d all-loop
factorization~\eqref{eq:PolygonFactorizationIntegrated} is reminiscent of an
amplitude factorization on an on-shell particle with momentum
$x_{ij}^\mu$ and mass $m^2=-y_{ij}^2$. This lends motivation to look
for generalizations of off-shell Berends--Giele~\cite{Berends:1987me} or on-shell
BCFW~\cite{Britto:2004ap,Britto:2005fq} recursion relations to the
polygon correlators. The factorization
poles~\eqref{eq:PolygonFactorizationIntegrated} are completely
on-shell (from a $10$d perspective). Can one engineer a holomorphic
deformation of the kinematics, such that the pole at infinity can be systematically understood?

\item[Correlahedron.]

It would be interesting to see if there is a geometric description for
the polygon integrands, like the amplituhedron for massless
amplitudes~\cite{Arkani-Hamed:2013jha}, and the correlahedron for
stress-energy correlators~\cite{Eden:2017fow}. The relation of the
polygon correlators to Coulomb-branch amplitudes
indicates that such a description could exist. A geometric description
could also single out a particular
integrand form (like the dlog forms in the massless case), which could
in turn help find good master integrals, or even better a good finite
function space that one could use for a functional bootstrap of the integrated
polygons.
A possible starting point in the search of such a description could be
the recently proposed ``deformed''
amplituhedron~\cite{Arkani-Hamed:2023epq} whose parameters can be
interpreted as vacuum expectation values on the Coulomb branch.
As a first step, one could look for a parametrization of the square correlator
(\aka octagon) that matches the two-loop integrand/amplitude
of~\cite{Arkani-Hamed:2023epq}, and then study the resulting geometry
in this parametrization.

\item[Superpolygons \& 10d Symmetry.]

The bosonic polygon correlators $\mathbb{M}_n$ should have
supersymmetric generalizations where $4(n-4)$ Grassmann variables
$\theta$ are distributed on the perimeter of the polygon. Some
information on this supersymmetrization could be
derived by taking the $n$-point ten-dimensional null limit of the
$m$-point generating function $G_{m,\ell}$, with $m<n$. However, this would
require to also include all higher Kaluza--Klein modes of $(n-m)$
of the Lagrangian insertion points in the generating function, which
has presently only been done for $m=4$, \ie the
square~\cite{Caron-Huot:2021usw}.

The proposed superpolygon correlator should naturally
decompose into basic superconformal invariants. In fact, this
decomposition should derive from a corresponding decomposition
of the full super-generating function (before taking
the ten-dimensional null limit). At four points, there is only one
such superinvariant $\mathcal{R}$~\cite{Eden:2011we}. In the
$\ell$-loop integrands of four $\twentyprime$ operators, $\mathcal{R}$ absorbs all
dependence on the polarizations $y_i$, hence its coefficient function
$\mathcal{H}$ (the ``reduced correlator'') is a function of $x_{ij}^2$
only. And in fact, the full four-point generating functions
$G_{4,\ell}$ (including all KK modes of the Lagrangian) are obtained by promoting all four-dimensional $x_{ij}^2$
in the reduced correlator $\mathcal{H}$ to ten-dimensional
$X_{ij}^2$, which results in a ten-dimensional conformal symmetry~\cite{Caron-Huot:2021usw}.
Projecting $G_{4,\ell}$ to the square correlator $M_{4,\ell}$, this
structure is preserved: The remnant of
$\mathcal{R}$~\eqref{eq:RsquareLimit} is multiplied by a function with
ten-dimensional conformal symmetry~\eqref{eq:M4lintro}.

One can expect that this structure is repeated at higher points, only that
now there will be several superinvariants, each multiplied by a
different $10$d-symmetric ``reduced correlator''. The
decomposition into superinvariants is not known even at five points,
which makes it difficult to look for ten-dimensional symmetry.
However, the pentagon correlator already has a suggestive
form~\eqref{eq:M52}: Two structures $p$ and $q$ multiply combinations
of conformal integrands that could be promoted to functions with $10$d
symmetry (conjecturally including the higher KK modes of the
Lagrangian points).
Further supporting evidence for the existence of such a decomposition
derives from strong coupling, where the five-point correlator in
supergravity could be organized over $10$d denominators, with
coefficients that are linear combinations of
superinvariants~\cite{Fernandes:2025eqe}.

Better understanding the decomposition into superinvariants would also
help make the duality with Coulomb-branch amplitudes more precise.
This was the case for the massless correlator/amplitude duality: The super
Wilson-loop defined in \cite{Caron-Huot:2010ryg, Mason:2010yk} allowed
for a complete duality including all N$^{k}$MHV massless
amplitudes~\cite{Eden:2011yp}.

\item[Integrability.]

Correlation functions of local operators can be computed from
integrability by decomposing them into hexagon form
factors~\cite{Basso:2015zoa,Fleury:2016ykk,Eden:2016xvg}.
Performing such computations in practice is difficult due to the
intricate interactions among all contributing worldsheet excitations,
and in particular due to winding modes that require careful
treatment~\cite{Basso:2015eqa}. Polygon correlators have disk topology
and thus admit no non-trivial cycles, hence winding modes are projected out, which
should simplify their computation.
In this sense, the polygons $\intpoly_n$ are the most natural objects
one can compute from hexagons.
In fact the hexagonalization of the square (\aka octagon)
led to an
all-loop~\cite{Coronado:2018cxj,Kostov:2019stn,Kostov:2019auq} and even finite-coupling
resummation~\cite{Belitsky:2019fan,Belitsky:2020qrm}.

It would be great to study the dynamics of worldsheet excitations on
the polygon correlators in more depth. The present formalism admitted
the computation of the pentagon to two loops~\cite{Fleury:2020ykw},
but is plagued by a complicated matrix structure. Can one find a more
suitable basis of excitations that is specifically tailored for
polygon computations, and that makes the sum over excitations ``more
diagonal''? Perhaps in simplifying kinematic limits?

Of particularly interest is the four-dimensional null limit, in which
one can expect some similarity with the pentagon OPE for massless
amplitudes (null polygon Wilson
loops)~\cite{Basso:2013vsa,Basso:2013aha,Basso:2014koa,Basso:2014nra}.
The two will not be the same, as one can already see from their
distinct anomalous dimensions ($\gammaCusp$ \vs $\gammaOct$).
Moreover, massless amplitudes have no R-charge flowing around their
perimeter, whereas the polygon correlators are surrounded by a large
charge flow.
It would be great to understand the differences and similarities of
the pentagon OPE and the hexagon expansion in the massless limit in
more detail.

Further input can come from strong
coupling: The square correlator could be computed from a clustering
procedure~\cite{Jiang:2016ulr,Bargheer:2019exp}. Can the same be done
for bigger polygons? The square is described by an integral over a
single Y-function~\cite{Bargheer:2019exp}. Is there a Y-system for strongly coupled
polygons, similar to the massless amplitude case~\cite{Alday:2010vh}?

\item[Integration \& Functional Bootstrap.]

In this work, we have focused on the integrands $\polygon_{n,\ell}$ of
the polygon correlators. Of course it would be interesting to
integrate over the Lagrangian points to obtain the fully interacting
$n$-point polygons $\intpoly_n$. For the square, this was possible
because the result is exclusively expressed in terms of well-known
ladder integrals~\cite{Coronado:2018cxj}. At higher points, the
situation is very different. There have been some recent advances in
computing five-point and six-point loop integrals~\cite{Abreu:2020jxa,
Canko:2020ylt, Abreu:2021smk, Abreu:2023rco, Bercini:2024pya,
Rodrigues:2024znq, Henn:2024ngj, Abreu:2024yit, Abreu:2024fei,
Becchetti:2025oyb}, but the completely off-shell higher-point
integrals remain mostly unknown, hence a direct integration looks
infeasible at the moment.

A more promising route to the integrated polygons would be a
functional bootstrap, which has been applied very successfully in the
massless amplitude case (see \eg~\cite{Papathanasiou:2022lan} for a review).
The first requisite for this would be the
knowledge of the leading singularities as well as the symbol alphabet
(assuming that the polygons are expressible in terms of multiple
polylogarithms). At five points, it might be feasible to extract this
information, for example by relating the occurring integrals to
non-conformal off-shell four-point integrals~\cite{He:2022ctv}.
Once the function space is known (\eg leading singularities and
alphabet), one could write a linear ansatz at each loop order, and fix
all remaining freedom by boundary data from hexagonalization.

\item[Yangian Symmetry.]

The polygons are cyclically invariant objects with disk topology,
hence one might expect that they (or their supersymmetric
generalizations) transform in some realization of the
$\alg{psu}(2,2|4)$ Yangian symmetry of $\superN=4$ super Yang--Mills
theory. In the case of massless amplitudes, the space of invariants is
strongly reduced by Yangian symmetry in the form of dual
superconformal invariance, which greatly facilitates their
construction. Could Yangian symmetry play a similar role for the
polygon correlators considered here?

\end{description}
We hope that the large amount of data provided in this work will help
to address some of these interesting questions in the future.

\pdfbookmark[1]{Acknowledgments}{Acknowledgments}
\subsection*{Acknowledgments}

We thank Benjamin Basso, Simon Caron-Huot,
Thiago Fleury, Johannes Henn, Paul Heslop,
Gregory Korchemsky,
Davide Lai,
and Tabea Siebrecht
for interesting discussions.
The work of T.\,B., A.\,B., and C.\,B. was funded by the Deutsche
Forschungsgemeinschaft (DFG, German Research Foundation) Grant No.
460391856.
T.\,B., A.\,B., and C.\,B. acknowledge support from DESY
(Hamburg, Germany), a member of the Helmholtz Association HGF,
and by the Deutsche
Forschungsgemeinschaft (DFG, German Research Foundation) under
Germany's Excellence Strategy -- EXC 2121 ``Quantum Universe'' --
390833306.
A.\,B. is further supported by the Studienstiftung des Deutschen Volkes. The work of C.\,B. was partially supported by the European Research
Council (ERC) under the European Union's Horizon 2020 research and
innovation program - 60 - (grant agreement No.~865075) EXACTC.
The work of F.C. is supported in part by the Simons Foundation
grant 994306 (Simons Collaboration on Confinement and QCD Strings), as well the NCCR
SwissMAP that is also funded by the Swiss National Science Foundation.

\appendix

\section{Two-Loop Faces in Terms of Conformal Integrals}
\label{sec:DCIFaces}

In \secref{sec:two-loop-result}, we decomposed the two-loop faces
$F^{(2)}_{1,\dots,n}$ into a basis of three different propagator
structures $\mathcal{J}_1$, $\mathcal{J}_2$, and $\mathcal{J}_3$ that
are free of numerator factors. In other words, all numerator factors are
absorbed in the coefficients $f_{k,\mathbold{e},a,b}$ of that
decomposition~\eqref{eq:F2IntegralDecomp2}.
Alternatively, and perhaps more naturally if one aims for actually
integrating over the Lagrangian points, is to keep all dependence on
the Lagrangian insertion points together, and expand in a rational
basis of integrands $\mathcal{I}_k$ of planar conformal integrals:
\begin{equation}
F^{(2)}_{1,\dots,n}
=
\sum_{k,\mathbold{e}}
\tilde{f}_{k,\mathbold{e}}(x_{ij}^2, y_{ij}^2)
\, \mathcal{I}_k^{\mathbold{e}}
+(\grp{D}_n\,\text{perms})
\,,
\label{eq:F2IntegralDecomp}
\end{equation}
where the sum over $k$ runs over the $12$ different conformal
integrands $\mathcal{I}_k$ that appear, $\mathbold{e}$ labels
their planar embeddings in the face $F_{1,\dots,n}$, and
$\tilde{f}_{k,\mathbold{e}}$ are rational coefficients that only depend on the spacetime distances
$x_{ij}^2$ and internal (R-charge) distances $y_{ij}^2$ among the
external points $1,\dots,n$.

In order to list the conformal integrands that appear,
we define the following
eight-point double-pentagon ``master'' integrand:
\begin{equation}
\includegraphics[align=c]{FigMasterIntegral}
\qquad
\mathcal{I}_1^{1,\dots,8,{a},{b}}=
\frac{1}{2}\brk*{
\frac{x_{1{b}}^2 x_{5{a}}^2}{
x_{1{a}}^2 x_{2{a}}^2 x_{3{a}}^2 x_{4{a}}^2
\, x_{{a}{b}}^2 \,
x_{5{b}}^2 x_{6{b}}^2 x_{7{b}}^2 x_{8{b}}^2
}
+({a}\leftrightarrow{b})
}
\label{eq:masterIntegral}
\end{equation}
All two-loop integral topologies that appear in the
decomposition~\eqref{eq:F2IntegralDecomp} can be obtained from
permutations and limits of
this master integrand:
\begin{alignat}{2}
\includegraphics[align=c,hsmash=c]{FigPentaPenta7}
&&\mspace{120mu}
\mathcal{I}_2^{1,\dots,7,a,b}&=\mathcal{I}_1^{2,1,3,4,5,6,7,1,a,b}
\,,\\
\includegraphics[align=c,hsmash=c]{FigPentaPenta6}
&&\mspace{120mu}
\mathcal{I}_3^{1,\dots,6,a,b}&=\mathcal{I}_1^{2,1,3,4,5,4,6,1,a,b}
\,,\\
\includegraphics[align=c,hsmash=c]{FigPentaBox7}
&&\mspace{120mu}
\mathcal{I}_4^{1,\dots,7,a,b}&=\mathcal{I}_1^{1,2,3,4,4,5,6,7,a,b}
\,,\\
\includegraphics[align=c,hsmash=c]{FigPentaBox6}
&&\mspace{120mu}
\mathcal{I}_5^{1,\dots,6,a,b}&=\mathcal{I}_1^{2,1,3,4,4,5,6,1,a,b}
\,,\\
\includegraphics[align=c,hsmash=c]{FigPentaBox5}
&&\mspace{120mu}
\mathcal{I}_6^{1,\dots,5,a,b}&=\mathcal{I}_1^{2,1,3,3,3,4,5,1,a,b}
\,,\\
\includegraphics[align=c,hsmash=c]{FigDoubleBox6}
&&\mspace{120mu}
\mathcal{I}_7^{1,\dots,6,a,b}&=\mathcal{I}_1^{1,2,3,4,4,5,6,1,a,b}
\,,\\
\includegraphics[align=c,hsmash=c]{FigDoubleBox5}
&&\mspace{120mu}
\mathcal{I}_8^{1,\dots,5,a,b}&=\mathcal{I}_1^{1,2,3,4,4,5,1,1,a,b}
\,,\\
\includegraphics[align=c,hsmash=c]{FigDoubleBox4}
&&\mspace{120mu}
\mathcal{I}_9^{1,\dots,4,a,b}&=\mathcal{I}_1^{1,2,3,3,3,4,1,1,a,b}
\,.
\end{alignat}
$\mathcal{I}_1$, $\mathcal{I}_2$, and $\mathcal{I}_3$ are the eight-point,
seven-point, and six-point double-penta integrands.
$\mathcal{I}_4$, $\mathcal{I}_5$, and $\mathcal{I}_6$ are the seven-point,
six-point, and five-point penta-box integrands.
And $\mathcal{I}_7$, $\mathcal{I}_8$, and $\mathcal{I}_9$ are the six-point,
five-point, and four-point double-box integrands.
Besides these two-loop integrands, the following products of one-loop
box integrands also contribute to~\eqref{eq:F2IntegralDecomp}:
\begin{alignat}{2}
\mspace{70mu}
\includegraphics[align=c,hsmash=c]{FigBoxBox8}
&&\mspace{120mu}
\mathcal{I}_{10}^{1,\dots,8,a,b}&=
\frac{1}{2}\brk*{
\frac{1}{x_{1{a}}^2 x_{2{a}}^2 x_{3{a}}^2 x_{4{a}}^2}
\frac{1}{x_{5{b}}^2 x_{6{b}}^2 x_{7{b}}^2 x_{8{b}}^2}
+({a}\leftrightarrow{b})
}
\label{eq:I10}
\,,\\
\mspace{70mu}
\includegraphics[align=c,hsmash=c]{FigBoxBox7}
&&\mspace{120mu}
\mathcal{I}_{11}^{1,\dots,7,a,b}&=\mathcal{I}_{10}^{1,2,3,4,5,6,7,1,a,b}
\,,\\
\mspace{70mu}
\includegraphics[align=c,hsmash=c]{FigBoxBox6}
&&\mspace{120mu}
\mathcal{I}_{12}^{1,\dots,6,a,b}&=\mathcal{I}_{10}^{1,2,3,4,4,5,6,1,a,b}
\,.
\end{alignat}
The coefficient functions $\tilde{f}_{k,\mathbold{e}}$ are functions of the
spacetime and internal distances. They depend polynomially on the
internal distances $y_{ij}^2$. Their dependence is further constrained
by the fact that the faces $F_{1,\dots,n}$ are conformally invariant functions of $x_i$ when
expressed in terms of $d_{ij}$ and $x_{ij}^2$. Most importantly,
$F_{1,\dots,n}$ reduce to the integrand of the massless $n$-point
gluon amplitude~\cite{Vergu:2009tu} in the $y_i\to0$ limit~\eqref{eq:masslessAmpLimit}.

\section{Twistor Feynman Rules}
\label{sec:twist-feynm-rules}

In the following, we illustrate how the twistor Feynman rules
$\Phi$ are applied to a given twistor graph in order to compute
the polygon integrand. As an example, we consider one of the~$36$
graphs contributing to the one-loop pentagon integrand $M_{5,1}$
(see~\tabref{tab:graphcounting}), namely
\begin{equation}
\gamma=\includegraphics[align=c,scale=.6]{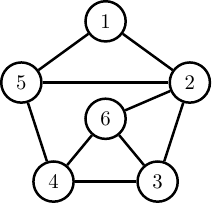}
\,.
\end{equation}
According to~\eqref{eq:twistorPolygon}, the contribution~$M_{5,1}^\gamma$ of this graph is
\begin{equation} \label{eq:examplegraph}
    M_{5,1}^\gamma = \lim_{d_{i,i+1}\rightarrow 1}\prod_{i=1}^{5}(1-d_{i,i+1})\, \Phi(\gamma)
    \,.
\end{equation}
In general, this contribution depends on the space-time points $x_i$, the $R$-charge polarization vectors $y_i$, and the reference twistor $Z_\star$, so that $M_{5,1}^\gamma = M_{5,1}^\gamma(x_i,y_i,Z_\star)$.
The dependence on~$Z_\star$ cancels only after summing over all graphs contributing to the full integrand.

Following the twistor Feynman rules for the supersymmetric operators $\mathbb{O}(x,y,\theta)$, which are the chiral supermultiplets whose bottom component is the ``master'' operator $\mathcal{O}(x,y)$~\cite{Caron-Huot:2023wdh}, each edge connecting two vertices $i$ and $j$ contributes a propagator factor
\begin{equation}
    D_{ij} = \frac{d_{ij}}{1-d_{ij}} \,,
\end{equation}
while each vertex $i$ of valency greater than two is accompanied by a vertex factor
\begin{equation} \label{eq:vertexrules}
    \Delta^i_{j_1\dots j_k}=\Delta^i_{j_1j_2j_3}\Delta^i_{j_1j_3j_4}\dots\Delta^i_{j_1j_{k-1}j_k}
    \,, \qquad
    \Delta^i_{jkl} = 1 + R^i_{jkl}
    \,.
\end{equation}
The vertex $R^i_{jkl}$ was derived in~\cite{Chicherin:2014uca},
see \eg equation~(3.12) in~\cite{Bargheer:2025uai} for its explicit form.
These factors depend on the reference twistor, the space-time points, the polarization vectors, and the Grassmann variables $\theta^a_\alpha$, with $\alpha,a=1,2$, associated with the operators involved. Upon expansion, they are homogeneous polynomials of degree two in the Grassmann variables.
Applying these rules to the graph in~\eqref{eq:examplegraph}, one finds
\begin{equation}
    \label{eq:M51contr1}
    M_{5,1}^\gamma = D_{25}D_{26}D_{36}D_{46}\,\Delta^2_{3156}\Delta^3_{264}\Delta^4_{536}\Delta^5_{142}\Delta^6_{243}
    \,,
\end{equation}
where the vertex factors $\Delta^i_{j_1\dots j_k}$ must be evaluated
on the support of $d_{i,i+1}=1$, that is $x_{ij}^2+y_{ij}^2=0$ for $i=1,\dots,5$.

To evaluate this expression further, we note the following. The
five operators on the perimeter are the bosonic operators
$\mathcal{O}(x,y)$, and therefore do not depend on Grassmann variables.
The insertion operator, on the other hand, is the chiral
Lagrangian $\Lint$, which is obtained from the supersymmetric operator $\mathbb{O}(x,y,\theta)$ by
\begin{equation}
    \Lint(x) = \lim_{y\rightarrow0}\int d^4 \theta\,\mathbb{O}(x,y,\theta)
    \,.
\end{equation}
Thus, in~\eqref{eq:M51contr1} we have to extract the $(\theta_6)^4$ component.
Since only the insertion point carries Grassmann dependence, any $R$-factor not involving point~$6$ vanishes. Using~\eqref{eq:vertexrules}, one therefore obtains
\begin{align}
    \Delta^2_{3156}
    &= \Delta^2_{315}\Delta^2_{356} = 1 \times (1+R^2_{356})
    \,, \nonumber \\
    \Delta^3_{264}
    &= 1 + R^3_{264}
    \,, \nonumber \\
    \Delta^4_{536}
    &= 1 + R^4_{536}
    \,, \nonumber \\
    \Delta^5_{142}
    &= 1
    \,, \nonumber \\
    \Delta^6_{243}
    &= 1 + R^6_{243}
    \,.
\end{align}
Substituting these expressions into~\eqref{eq:M51contr1}, we keep only those terms containing two $R$-factors, since only such terms can produce the full Grassmann structure $(\theta_6)^4$. This gives
\begin{equation}
    M_{5,1}^\gamma =
    D_{25}D_{26}D_{36}D_{46}\,
    \left(
    R^2_{356}R^3_{264}
    +(R^2_{356}+R^3_{264})R^4_{536}
    +(R^2_{356}+R^3_{264}+R^4_{536})R^6_{243}
    \right)
    \,.
\end{equation}

At this stage, one has to be slightly careful when taking the limit $y_6\to0$. Individually, the $R$-factors diverge in this limit, but their dependence on $y_6$ is homogeneous. Under the scaling $y_6\rightarrow t\,y_6$, one finds
\begin{equation}
    R^{i}_{k_1k_2k_3}R^j_{l_1l_2l_3} \sim
    \begin{cases}
        t^{-3} \,, & \text{if } 6 \in \{i,j\} \,, \\
        t^{-2} \,, & \text{otherwise}
    \end{cases}
    \,.
\end{equation}
In practice, this scaling behavior can be determined numerically, or inferred from the heuristic argument given in Footnote~8 of~\cite{Bargheer:2025uai}. On the other hand, the scaling of the propagator factors follows from expanding
\begin{equation}
    D_{ij} = d_{ij} + d_{ij}^2 + \dots \,,
\end{equation}
which implies
\begin{equation}
    D_{25}D_{26}D_{36}D_{46} \sim t^3 + \text{higher orders in } t
    \,.
\end{equation}
Combining both scalings, one can identify the terms that remain finite in the limit $y_6\to0$. The final contribution of this graph to $M_{5,1}$ is therefore
\begin{equation}
    M_{5,1}^\gamma =
    \frac{d_{25}}{1-d_{25}}\,d_{26}d_{36}d_{46}\,
    (R^2_{356}+R^3_{264}+R^4_{536})R^6_{243}
    \,.
\end{equation}
Repeating this procedure for all initial 36 graphs $\gamma\in\Gamma_{5,1}^{\text{disk}}$ and their
inequivalent $D_5$ permutations yields the full one-loop pentagon
integrand as a polynomial in the R-factors. This final form is
evaluated numerically as explained in~\cite{Bargheer:2025uai} in
order to match against a suitable ansatz~\eqref{eq:ansatz} that is exclusively
written in $y_{ij}^2$ and $x_{ij}^2$. Performing this match, we arrive
at the result~\eqref{eq:M51} for the one-loop pentagon.

\bibliographystyle{nb}
\bibliography{references}

\begin{thebibliography}{10}
\addcontentsline{toc}{section}{\refname}
\providecommand{\href}[2]{#2}
\providecommand{\arxivref}[2]{\href{http://arxiv.org/abs/#1}{#2}}
\providecommand{\doiref}[2]{\href{http://dx.doi.org/#1}{#2}}
\providecommand{\nbbstauthor}[1]{#1}
\providecommand{\nbbstjournal}[1]{\textsf{#1}}
\providecommand{\nbbsttitle}[1]{\textit{#1}}
\providecommand{\nbbsturl}[1]{\texttt{#1}}
\providecommand{\nbbsteprint}[1]{\texttt{#1}}
\providecommand{\nbbststyle}{\raggedright\small\parskip0pt}
\nbbststyle

\bibitem{Eden:2011we}
\nbbstauthor{B.~Eden, P.~Heslop, G.~P.~Korchemsky and E.~Sokatchev},
\nbbsttitle{Hidden symmetry of four-point correlation functions and amplitudes
  in {$\mathcal{N}=\mathord{}$4} {SYM}},
\nbbstjournal{\doiref{10.1016/j.nuclphysb.2012.04.007}{Nucl.~Phys.~B~862,~193~(2012)}},
\nbbsteprint{\arxivref{1108.3557}{arxiv:1108.3557}}.

\bibitem{Drummond:2013nda}
\nbbstauthor{J.~Drummond, C.~Duhr, B.~Eden, P.~Heslop, J.~Pennington and
  V.~A.~Smirnov},
\nbbsttitle{Leading singularities and off-shell conformal integrals},
\nbbstjournal{\doiref{10.1007/JHEP08(2013)133}{JHEP~1308,~133~(2013)}},
\nbbsteprint{\arxivref{1303.6909}{arxiv:1303.6909}}.

\bibitem{Arutyunov:2000py}
\nbbstauthor{G.~Arutyunov and S.~Frolov},
\nbbsttitle{Four point functions of lowest weight CPOs in
  {$\mathcal{N}=\mathord{}$4} {SYM}(4) in supergravity approximation},
\nbbstjournal{\doiref{10.1103/PhysRevD.62.064016}{Phys.~Rev.~D~62,~064016~(2000)}},
\nbbsteprint{\arxivref{hep-th/0002170}{hep-th/0002170}}.

\bibitem{Goncalves:2014ffa}
\nbbstauthor{V.~Gon\c{c}alves},
\nbbsttitle{Four point function of {$\mathcal{N}=\mathord{}$4} stress-tensor
  multiplet at strong coupling},
\nbbstjournal{\doiref{10.1007/JHEP04(2015)150}{JHEP~1504,~150~(2015)}},
\nbbsteprint{\arxivref{1411.1675}{arxiv:1411.1675}}.

\bibitem{Alday:2023mvu}
\nbbstauthor{L.~F.~Alday and T.~Hansen},
\nbbsttitle{The {AdS} {Virasoro}-{Shapiro} amplitude},
\nbbstjournal{\doiref{10.1007/JHEP10(2023)023}{JHEP~2310,~023~(2023)}},
\nbbsteprint{\arxivref{2306.12786}{arxiv:2306.12786}}.

\bibitem{Drukker:2008pi}
\nbbstauthor{N.~Drukker and J.~Plefka},
\nbbsttitle{The Structure of n-point functions of chiral primary operators in
  {$\mathcal{N}=\mathord{}$4} super {Yang}--{Mills} at one-loop},
\nbbstjournal{\doiref{10.1088/1126-6708/2009/04/001}{JHEP~0904,~001~(2009)}},
\nbbsteprint{\arxivref{0812.3341}{arxiv:0812.3341}}.

\bibitem{Bargheer:2022sfd}
\nbbstauthor{T.~Bargheer, T.~Fleury and V.~Gon\c{c}alves},
\nbbsttitle{Higher-Point Integrands in {$\mathcal{N}=\mathord{}$4} super
  {Yang}--{Mills} Theory},
\nbbstjournal{\doiref{10.21468/SciPostPhys.15.2.059}{SciPost~Phys.~15,~059~(2023)}},
\nbbsteprint{\arxivref{2212.03773}{arxiv:2212.03773}}.

\bibitem{Goncalves:2019znr}
\nbbstauthor{V.~Gon\c{c}alves, R.~Pereira and X.~Zhou},
\nbbsttitle{$20'$ Five-Point Function from $AdS_5\times S^5$ Supergravity},
\nbbstjournal{\doiref{10.1007/JHEP10(2019)247}{JHEP~1910,~247~(2019)}},
\nbbsteprint{\arxivref{1906.05305}{arxiv:1906.05305}}.

\bibitem{Goncalves:2025jcg}
\nbbstauthor{V.~Gon\c{c}alves, M.~Nocchi and X.~Zhou},
\nbbsttitle{Dissecting supergraviton six-point function with lightcone limits
  and chiral algebra},
\nbbstjournal{\doiref{10.1007/JHEP06(2025)173}{JHEP~2506,~173~(2025)}},
\nbbsteprint{\arxivref{2502.10269}{arxiv:2502.10269}}.

\bibitem{Eden:2012fe}
\nbbstauthor{B.~Eden, P.~Heslop, G.~P.~Korchemsky, V.~A.~Smirnov and
  E.~Sokatchev},
\nbbsttitle{Five-loop {Konishi} in {$\mathcal{N}=\mathord{}$4} {SYM}},
\nbbstjournal{\doiref{10.1016/j.nuclphysb.2012.04.015}{Nucl.~Phys.~B~862,~123~(2012)}},
\nbbsteprint{\arxivref{1202.5733}{arxiv:1202.5733}}.

\bibitem{Eden:2012rr}
\nbbstauthor{B.~Eden},
\nbbsttitle{Three-loop universal structure constants in
  {$\mathcal{N}=\mathord{}$4} susy {Yang}--{Mills} theory},
\nbbsteprint{\arxivref{1207.3112}{arxiv:1207.3112}}.

\bibitem{Bercini:2021jti}
\nbbstauthor{C.~Bercini, V.~Gon\c{c}alves, A.~Homrich and P.~Vieira},
\nbbsttitle{The {Wilson} loop \textemdash{} large spin OPE dictionary},
\nbbstjournal{\doiref{10.1007/JHEP07(2022)079}{JHEP~2207,~079~(2022)}},
\nbbsteprint{\arxivref{2110.04364}{arxiv:2110.04364}}.

\bibitem{Bercini:2024pya}
\nbbstauthor{C.~Bercini, B.~Fernandes and V.~Gon\c{c}alves},
\nbbsttitle{Two-loop five-point integrals: light, heavy and large-spin
  correlators},
\nbbstjournal{\doiref{10.1007/JHEP10(2024)242}{JHEP~2410,~242~(2024)}},
\nbbsteprint{\arxivref{2401.06099}{arxiv:2401.06099}}.

\bibitem{Bargheer:2025uai}
\nbbstauthor{T.~Bargheer, A.~Bekov, C.~Bercini and F.~Coronado},
\nbbsttitle{Higher-Point Correlators in {$\mathcal{N}=\mathord{}$4} {SYM}:
  Generating Functions},
\nbbstjournal{\doiref{10.1007/JHEP02(2026)161}{JHEP~2602,~161~(2026)}},
\nbbsteprint{\arxivref{2509.14332}{arxiv:2509.14332}}.

\bibitem{Alday:2010zy}
\nbbstauthor{L.~F.~Alday, B.~Eden, G.~P.~Korchemsky, J.~Maldacena and
  E.~Sokatchev},
\nbbsttitle{From correlation functions to {Wilson} loops},
\nbbstjournal{\doiref{10.1007/JHEP09(2011)123}{JHEP~1109,~123~(2011)}},
\nbbsteprint{\arxivref{1007.3243}{arxiv:1007.3243}}.

\bibitem{Alday:2007hr}
\nbbstauthor{L.~F.~Alday and J.~M.~Maldacena},
\nbbsttitle{Gluon scattering amplitudes at strong coupling},
\nbbstjournal{\doiref{10.1088/1126-6708/2007/06/064}{JHEP~0706,~064~(2007)}},
\nbbsteprint{\arxivref{0705.0303}{arxiv:0705.0303}}.

\bibitem{Berkovits:2008ic}
\nbbstauthor{N.~Berkovits and J.~Maldacena},
\nbbsttitle{Fermionic {T}-Duality, Dual Superconformal Symmetry, and the
  Amplitude/{Wilson} Loop Connection},
\nbbstjournal{\doiref{10.1088/1126-6708/2008/09/062}{JHEP~0809,~062~(2008)}},
\nbbsteprint{\arxivref{0807.3196}{arxiv:0807.3196}}.

\bibitem{Beisert:2008iq}
\nbbstauthor{N.~Beisert, R.~Ricci, A.~A.~Tseytlin and M.~Wolf},
\nbbsttitle{Dual Superconformal Symmetry from {AdS$_5$ $\times$ S$^5$}
  Superstring Integrability},
\nbbstjournal{\doiref{10.1103/PhysRevD.78.126004}{Phys.~Rev.~D78,~126004~(2008)}},
\nbbsteprint{\arxivref{0807.3228}{arxiv:0807.3228}}.

\bibitem{Bern:2008ap}
\nbbstauthor{Z.~Bern, L.~J.~Dixon, D.~A.~Kosower, R.~Roiban, M.~Spradlin,
  C.~Vergu and A.~Volovich},
\nbbsttitle{The Two-Loop Six-Gluon {MHV} Amplitude in Maximally Supersymmetric
  {Yang}--{Mills} Theory},
\nbbstjournal{\doiref{10.1103/PhysRevD.78.045007}{Phys.~Rev.~D78,~045007~(2008)}},
\nbbsteprint{\arxivref{0803.1465}{arxiv:0803.1465}}.

\bibitem{Drummond:2008aq}
\nbbstauthor{J.~M.~Drummond, J.~Henn, G.~P.~Korchemsky and E.~Sokatchev},
\nbbsttitle{Hexagon {Wilson} loop = six-gluon {MHV} amplitude},
\nbbstjournal{\doiref{10.1016/j.nuclphysb.2009.02.015}{Nucl.~Phys.~B815,~142~(2009)}},
\nbbsteprint{\arxivref{0803.1466}{arxiv:0803.1466}}.

\bibitem{Mason:2010yk}
\nbbstauthor{L.~Mason and D.~Skinner},
\nbbsttitle{The Complete Planar {S}-matrix of {$\mathcal{N}=\mathord{}$4} {SYM}
  as a {Wilson} Loop in Twistor Space},
\nbbstjournal{\doiref{10.1007/JHEP12(2010)018}{JHEP~1012,~018~(2010)}},
\nbbsteprint{\arxivref{1009.2225}{arxiv:1009.2225}}.

\bibitem{Caron-Huot:2010ryg}
\nbbstauthor{S.~Caron-Huot},
\nbbsttitle{Notes on the scattering amplitude / {Wilson} loop duality},
\nbbstjournal{\doiref{10.1007/JHEP07(2011)058}{JHEP~1107,~058~(2011)}},
\nbbsteprint{\arxivref{1010.1167}{arxiv:1010.1167}}.

\bibitem{Eden:2010zz}
\nbbstauthor{B.~Eden, G.~P.~Korchemsky and E.~Sokatchev},
\nbbsttitle{From correlation functions to scattering amplitudes},
\nbbstjournal{\doiref{10.1007/JHEP12(2011)002}{JHEP~1112,~002~(2011)}},
\nbbsteprint{\arxivref{1007.3246}{arxiv:1007.3246}}.

\bibitem{Basso:2015zoa}
\nbbstauthor{B.~Basso, S.~Komatsu and P.~Vieira},
\nbbsttitle{Structure Constants and Integrable Bootstrap in Planar
  {$\mathcal{N}=\mathord{}$4} {SYM} Theory},
\nbbsteprint{\arxivref{1505.06745}{arxiv:1505.06745}}.

\bibitem{Fleury:2016ykk}
\nbbstauthor{T.~Fleury and S.~Komatsu},
\nbbsttitle{Hexagonalization of Correlation Functions},
\nbbstjournal{\doiref{10.1007/JHEP01(2017)130}{JHEP~1701,~130~(2017)}},
\nbbsteprint{\arxivref{1611.05577}{arxiv:1611.05577}}.

\bibitem{Eden:2016xvg}
\nbbstauthor{B.~Eden and A.~Sfondrini},
\nbbsttitle{Tessellating cushions: four-point functions in
  {$\mathcal{N}=\mathord{}$4} {SYM}},
\nbbstjournal{\doiref{10.1007/JHEP10(2017)098}{JHEP~1710,~098~(2017)}},
\nbbsteprint{\arxivref{1611.05436}{arxiv:1611.05436}}.

\bibitem{Fleury:2017eph}
\nbbstauthor{T.~Fleury and S.~Komatsu},
\nbbsttitle{Hexagonalization of Correlation Functions II: Two-Particle
  Contributions},
\nbbstjournal{\doiref{10.1007/JHEP02(2018)177}{JHEP~1802,~177~(2018)}},
\nbbsteprint{\arxivref{1711.05327}{arxiv:1711.05327}}.

\bibitem{Coronado:2018ypq}
\nbbstauthor{F.~Coronado},
\nbbsttitle{Perturbative Four-Point Functions in Planar
  {$\mathcal{N}=\mathord{}$4} {SYM} from Hexagonalization},
\nbbstjournal{\doiref{10.1007/JHEP01(2019)056}{JHEP~1901,~056~(2019)}},
\nbbsteprint{\arxivref{1811.00467}{arxiv:1811.00467}}.

\bibitem{Coronado:2018cxj}
\nbbstauthor{F.~Coronado},
\nbbsttitle{Bootstrapping the simplest correlator in planar
  {$\mathcal{N}=\mathord{}$4} {SYM} at all loops},
\nbbstjournal{\doiref{10.1103/PhysRevLett.124.171601}{Phys.~Rev.~Lett.~124,~171601~(2020)}},
\nbbsteprint{\arxivref{1811.03282}{arxiv:1811.03282}}.

\bibitem{Kostov:2019stn}
\nbbstauthor{I.~Kostov, V.~B.~Petkova and D.~Serban},
\nbbsttitle{Determinant formula for the octagon form factor in
  {$\mathcal{N}=\mathord{}$4} {SYM}},
\nbbstjournal{\doiref{10.1103/PhysRevLett.122.231601}{Phys.~Rev.~Lett.~122,~231601~(2019)}},
\nbbsteprint{\arxivref{1903.05038}{arxiv:1903.05038}}.

\bibitem{Kostov:2019auq}
\nbbstauthor{I.~Kostov, V.~B.~Petkova and D.~Serban},
\nbbsttitle{The Octagon as a Determinant},
\nbbstjournal{\doiref{10.1007/JHEP11(2019)178}{JHEP~1911,~178~(2019)}},
\nbbsteprint{\arxivref{1905.11467}{arxiv:1905.11467}}.

\bibitem{Belitsky:2019fan}
\nbbstauthor{A.~V.~Belitsky and G.~P.~Korchemsky},
\nbbsttitle{Exact null octagon},
\nbbstjournal{\doiref{10.1007/JHEP05(2020)070}{JHEP~2005,~070~(2020)}},
\nbbsteprint{\arxivref{1907.13131}{arxiv:1907.13131}}.

\bibitem{Belitsky:2020qrm}
\nbbstauthor{A.~V.~Belitsky and G.~P.~Korchemsky},
\nbbsttitle{Octagon at finite coupling},
\nbbstjournal{\doiref{10.1007/JHEP07(2020)219}{JHEP~2007,~219~(2020)}},
\nbbsteprint{\arxivref{2003.01121}{arxiv:2003.01121}}.

\bibitem{Fleury:2020ykw}
\nbbstauthor{T.~Fleury and V.~Gon\c{c}alves},
\nbbsttitle{Decagon at Two Loops},
\nbbstjournal{\doiref{10.1007/JHEP07(2020)030}{JHEP~2007,~030~(2020)}},
\nbbsteprint{\arxivref{2004.10867}{arxiv:2004.10867}}.

\bibitem{Caron-Huot:2021usw}
\nbbstauthor{S.~Caron-Huot and F.~Coronado},
\nbbsttitle{Ten dimensional symmetry of {$\mathcal{N}=\mathord{4}$} {SYM}
  correlators},
\nbbstjournal{\doiref{10.1007/JHEP03(2022)151}{JHEP~2203,~151~(2022)}},
\nbbsteprint{\arxivref{2106.03892}{arxiv:2106.03892}}.

\bibitem{Caron-Huot:2023wdh}
\nbbstauthor{S.~Caron-Huot, F.~Coronado and B.~M{\"u}hlmann},
\nbbsttitle{Determinants in self-dual {$\mathcal{N}=\mathord{}$4} {SYM} and
  twistor space},
\nbbstjournal{\doiref{10.1007/JHEP08(2023)008}{JHEP~2308,~008~(2023)}},
\nbbsteprint{\arxivref{2304.12341}{arxiv:2304.12341}}.

\bibitem{Chicherin:2015edu}
\nbbstauthor{D.~Chicherin, J.~Drummond, P.~Heslop and E.~Sokatchev},
\nbbsttitle{All three-loop four-point correlators of half-{BPS} operators in
  planar {$\mathcal{N}=\mathord{}$4} {SYM}},
\nbbstjournal{\doiref{10.1007/JHEP08(2016)053}{JHEP~1608,~053~(2016)}},
\nbbsteprint{\arxivref{1512.02926}{arxiv:1512.02926}}.

\bibitem{Chicherin:2018avq}
\nbbstauthor{D.~Chicherin, A.~Georgoudis, V.~Gon\c{c}alves and R.~Pereira},
\nbbsttitle{All five-loop planar four-point functions of half-{BPS} operators
  in {$\mathcal{N}=\mathord{}$4} {SYM}},
\nbbstjournal{\doiref{10.1007/JHEP11(2018)069}{JHEP~1811,~069~(2018)}},
\nbbsteprint{\arxivref{1809.00551}{arxiv:1809.00551}}.

\bibitem{Caron-Huot:2018kta}
\nbbstauthor{S.~Caron-Huot and A.-K.~Trinh},
\nbbsttitle{All tree-level correlators in {AdS}$_{5}$\texttimes{}S$_{5}$
  supergravity: hidden ten-dimensional conformal symmetry},
\nbbstjournal{\doiref{10.1007/JHEP01(2019)196}{JHEP~1901,~196~(2019)}},
\nbbsteprint{\arxivref{1809.09173}{arxiv:1809.09173}}.

\bibitem{Fernandes:2025eqe}
\nbbstauthor{B.~Fernandes, V.~Gon\c{c}alves, Z.~Huang, Y.~Tang, J.~Vilas~Boas
  and E.~Y.~Yuan},
\nbbsttitle{{AdS}{\texttimes}S Mellin Bootstrap, Hidden 10D Symmetry and
  Five-Point {Kaluza}--{Klein} Functions in {$\mathcal{N}=\mathord{}$4}
  Supersymmetric {Yang}--{Mills} Theory},
\nbbstjournal{\doiref{10.1103/3qyd-n621}{Phys.~Rev.~Lett.~136,~081602~(2026)}},
\nbbsteprint{\arxivref{2507.14124}{arxiv:2507.14124}}.

\bibitem{Du:2024xbd}
\nbbstauthor{X.-E.~Du, Z.~Huang, B.~Wang, E.~Y.~Yuan and X.~Zhou},
\nbbsttitle{Meson correlators in 4d $\mathcal{N}=2$ {SCFT}s and hints for 8d
  structures at weak coupling},
\nbbstjournal{\doiref{10.1007/JHEP04(2025)128}{JHEP~2504,~128~(2025)}},
\nbbsteprint{\arxivref{2412.17260}{arxiv:2412.17260}}.

\bibitem{Bork:2022vat}
\nbbstauthor{L.~V.~Bork, N.~B.~Muzhichkov and E.~S.~Sozinov},
\nbbsttitle{Infrared properties of five-point massive amplitudes in
  {$\mathcal{N}=\mathord{}$4} {SYM} on the Coulomb branch},
\nbbstjournal{\doiref{10.1007/JHEP08(2022)173}{JHEP~2208,~173~(2022)}},
\nbbsteprint{\arxivref{2201.08762}{arxiv:2201.08762}}.

\bibitem{Belitsky:2025bgb}
\nbbstauthor{A.~V.~Belitsky, L.~V.~Bork, R.~N.~Lee, A.~I.~Onishchenko and
  V.~A.~Smirnov},
\nbbsttitle{Five W-boson amplitude is equal to near-null decagon},
\nbbstjournal{\doiref{10.1103/5sf8-fhmt}{Phys.~Rev.~D~113,~086003~(2026)}},
\nbbsteprint{\arxivref{2510.16471}{arxiv:2510.16471}}.

\bibitem{Bargheer:2018jvq}
\nbbstauthor{T.~Bargheer, J.~Caetano, T.~Fleury, S.~Komatsu and P.~Vieira},
\nbbsttitle{Handling handles. Part {II}. Stratification and Data Analysis},
\nbbstjournal{\doiref{10.1007/JHEP11(2018)095}{JHEP~1811,~095~(2018)}},
\nbbsteprint{\arxivref{1809.09145}{arxiv:1809.09145}}.

\bibitem{Olivucci:2021pss}
\nbbstauthor{E.~Olivucci and P.~Vieira},
\nbbsttitle{Stampedes I: fishnet {OPE} and octagon Bootstrap with nonzero
  bridges},
\nbbstjournal{\doiref{10.1007/JHEP07(2022)017}{JHEP~2207,~017~(2022)}},
\nbbsteprint{\arxivref{2111.12131}{arxiv:2111.12131}}.

\bibitem{Olivucci:2022aza}
\nbbstauthor{E.~Olivucci and P.~Vieira},
\nbbsttitle{Null Polygons in Conformal Gauge Theory},
\nbbstjournal{\doiref{10.1103/PhysRevLett.129.221601}{Phys.~Rev.~Lett.~129,~221601~(2022)}},
\nbbsteprint{\arxivref{2205.04476}{arxiv:2205.04476}}.

\bibitem{Eden:2010ce}
\nbbstauthor{B.~Eden, G.~P.~Korchemsky and E.~Sokatchev},
\nbbsttitle{More on the duality correlators/amplitudes},
\nbbstjournal{\doiref{10.1016/j.physletb.2012.02.014}{Phys.~Lett.~B~709,~247~(2012)}},
\nbbsteprint{\arxivref{1009.2488}{arxiv:1009.2488}}.

\bibitem{Adamo:2011dq}
\nbbstauthor{T.~Adamo, M.~Bullimore, L.~Mason and D.~Skinner},
\nbbsttitle{A Proof of the Supersymmetric Correlation Function / {Wilson} Loop
  Correspondence},
\nbbstjournal{\doiref{10.1007/JHEP08(2011)076}{JHEP~1108,~076~(2011)}},
\nbbsteprint{\arxivref{1103.4119}{arxiv:1103.4119}}.

\bibitem{Eden:2011ku}
\nbbstauthor{B.~Eden, P.~Heslop, G.~P.~Korchemsky and E.~Sokatchev},
\nbbsttitle{The super-correlator/super-amplitude duality: Part {II}},
\nbbstjournal{\doiref{10.1016/j.nuclphysb.2012.12.014}{Nucl.~Phys.~B869,~378~(2013)}},
\nbbsteprint{\arxivref{1103.4353}{arxiv:1103.4353}}.

\bibitem{Eden:2011yp}
\nbbstauthor{B.~Eden, P.~Heslop, G.~P.~Korchemsky and E.~Sokatchev},
\nbbsttitle{The super-correlator/super-amplitude duality: Part I},
\nbbstjournal{\doiref{10.1016/j.nuclphysb.2012.12.015}{Nucl.~Phys.~B869,~329~(2013)}},
\nbbsteprint{\arxivref{1103.3714}{arxiv:1103.3714}}.

\bibitem{Basso:2015eqa}
\nbbstauthor{B.~Basso, V.~Gon\c{c}alves, S.~Komatsu and P.~Vieira},
\nbbsttitle{Gluing Hexagons at Three Loops},
\nbbstjournal{\doiref{10.1016/j.nuclphysb.2016.04.020}{Nucl.~Phys.~B907,~695~(2016)}},
\nbbsteprint{\arxivref{1510.01683}{arxiv:1510.01683}}.

\bibitem{Basso:2013vsa}
\nbbstauthor{B.~Basso, A.~Sever and P.~Vieira},
\nbbsttitle{Spacetime and Flux Tube {S}-Matrices at Finite Coupling for
  {$\mathcal{N}=\mathord{}$4} Supersymmetric {Yang}--{Mills} Theory},
\nbbstjournal{\doiref{10.1103/PhysRevLett.111.091602}{Phys.~Rev.~Lett.~111,~091602~(2013)}},
\nbbsteprint{\arxivref{1303.1396}{arxiv:1303.1396}}.

\bibitem{Basso:2013aha}
\nbbstauthor{B.~Basso, A.~Sever and P.~Vieira},
\nbbsttitle{Space-time {S}-matrix and Flux tube {S}-matrix {II}. Extracting and
  Matching Data},
\nbbstjournal{\doiref{10.1007/JHEP01(2014)008}{JHEP~1401,~008~(2014)}},
\nbbsteprint{\arxivref{1306.2058}{arxiv:1306.2058}}.

\bibitem{Basso:2014koa}
\nbbstauthor{B.~Basso, A.~Sever and P.~Vieira},
\nbbsttitle{Space-time {S}-matrix and Flux-tube {S}-matrix {III}. The
  two-particle contributions},
\nbbstjournal{\doiref{10.1007/JHEP08(2014)085}{JHEP~1408,~085~(2014)}},
\nbbsteprint{\arxivref{1402.3307}{arxiv:1402.3307}}.

\bibitem{Basso:2014nra}
\nbbstauthor{B.~Basso, A.~Sever and P.~Vieira},
\nbbsttitle{Space-time {S}-matrix and Flux-tube {S}-matrix {IV}. Gluons and
  Fusion},
\nbbstjournal{\doiref{10.1007/JHEP09(2014)149}{JHEP~1409,~149~(2014)}},
\nbbsteprint{\arxivref{1407.1736}{arxiv:1407.1736}}.

\bibitem{Bargheer:2019kxb}
\nbbstauthor{T.~Bargheer, F.~Coronado and P.~Vieira},
\nbbsttitle{Octagons I: Combinatorics and Non-Planar Resummations},
\nbbstjournal{\doiref{10.1007/JHEP08(2019)162}{JHEP~1908,~162~(2019)}},
\nbbsteprint{\arxivref{1904.00965}{arxiv:1904.00965}}.

\bibitem{Bargheer:2017nne}
\nbbstauthor{T.~Bargheer, J.~Caetano, T.~Fleury, S.~Komatsu and P.~Vieira},
\nbbsttitle{Handling Handles: Nonplanar Integrability in $\mathcal{N}=4$
  Supersymmetric {Yang}--{Mills} Theory},
\nbbstjournal{\doiref{10.1103/PhysRevLett.121.231602}{Phys.~Rev.~Lett.~121,~231602~(2018)}},
\nbbsteprint{\arxivref{1711.05326}{arxiv:1711.05326}}.

\bibitem{Chicherin:2014uca}
\nbbstauthor{D.~Chicherin, R.~Doobary, B.~Eden, P.~Heslop, G.~P.~Korchemsky,
  L.~Mason and E.~Sokatchev},
\nbbsttitle{Correlation functions of the chiral stress-tensor multiplet in
  {$\mathcal{N}=\mathord{}$4} {SYM}},
\nbbstjournal{\doiref{10.1007/JHEP06(2015)198}{JHEP~1506,~198~(2015)}},
\nbbsteprint{\arxivref{1412.8718}{arxiv:1412.8718}}.

\bibitem{Krasko:2015aa}
\nbbstauthor{E.~Krasko and A.~Omelchenko},
\nbbsttitle{Brown's Theorem and its Application for Enumeration of Dissections
  and Planar Trees},
\nbbstjournal{\doiref{10.37236/4129}{The~Electronic~Journal~of~Combinatorics~22,~1.17~(2015)}}.

\bibitem{Vergu:2009tu}
\nbbstauthor{C.~Vergu},
\nbbsttitle{The Two-loop MHV amplitudes in {$\mathcal{N}=\mathord{}$4}
  supersymmetric {Yang}--{Mills} theory},
\nbbstjournal{\doiref{10.1103/PhysRevD.80.125025}{Phys.~Rev.~D~80,~125025~(2009)}},
\nbbsteprint{\arxivref{0908.2394}{arxiv:0908.2394}}.

\bibitem{Usyukina:1993ch}
\nbbstauthor{N.~I.~Ussyukina and A.~I.~Davydychev},
\nbbsttitle{Exact results for three- and four-point ladder diagrams with an
  arbitrary number of rungs},
\nbbstjournal{\doiref{10.1016/0370-2693(93)91118-7}{Phys.~Lett.~B305,~136~(1993)}}.

\bibitem{Bercini:2020msp}
\nbbstauthor{C.~Bercini, V.~Gon\c{c}alves and P.~Vieira},
\nbbsttitle{Light-Cone Bootstrap of Higher Point Functions and {Wilson} Loop
  Duality},
\nbbstjournal{\doiref{10.1103/PhysRevLett.126.121603}{Phys.~Rev.~Lett.~126,~121603~(2021)}},
\nbbsteprint{\arxivref{2008.10407}{arxiv:2008.10407}}.

\bibitem{Basso:2020xts}
\nbbstauthor{B.~Basso, L.~J.~Dixon and G.~Papathanasiou},
\nbbsttitle{Origin of the Six-Gluon Amplitude in Planar
  {$\mathcal{N}=\mathord{}$4} Supersymmetric {Yang}--{Mills} Theory},
\nbbstjournal{\doiref{10.1103/PhysRevLett.124.161603}{Phys.~Rev.~Lett.~124,~161603~(2020)}},
\nbbsteprint{\arxivref{2001.05460}{arxiv:2001.05460}}.

\bibitem{Basso:2022ruw}
\nbbstauthor{B.~Basso, L.~J.~Dixon, Y.-T.~Liu and G.~Papathanasiou},
\nbbsttitle{All-Orders Quadratic-Logarithmic Behavior for Amplitudes},
\nbbstjournal{\doiref{10.1103/PhysRevLett.130.111602}{Phys.~Rev.~Lett.~130,~111602~(2023)}},
\nbbsteprint{\arxivref{2211.12555}{arxiv:2211.12555}}.

\bibitem{Belitsky:2020qir}
\nbbstauthor{A.~V.~Belitsky and G.~P.~Korchemsky},
\nbbsttitle{Crossing bridges with strong {Szeg\"o} limit theorem},
\nbbstjournal{\doiref{10.1007/JHEP04(2021)257}{JHEP~2104,~257~(2021)}},
\nbbsteprint{\arxivref{2006.01831}{arxiv:2006.01831}}.

\bibitem{CaronHuot:2011ky}
\nbbstauthor{S.~Caron-Huot},
\nbbsttitle{Superconformal symmetry and two-loop amplitudes in planar
  {$\mathcal{N}=\mathord{}4$} super {Yang}--{Mills}},
\nbbstjournal{\doiref{10.1007/JHEP12(2011)066}{JHEP~1112,~066~(2011)}},
\nbbsteprint{\arxivref{1105.5606}{arxiv:1105.5606}}.

\bibitem{Belitsky:2025vfc}
\nbbstauthor{A.~V.~Belitsky},
\nbbsttitle{Towards six {W}-boson amplitude at two loops},
\nbbsteprint{\arxivref{2511.20828}{arxiv:2511.20828}}.

\bibitem{Alday:2007he}
\nbbstauthor{L.~F.~Alday and J.~Maldacena},
\nbbsttitle{Comments on gluon scattering amplitudes via {AdS}/{CFT}},
\nbbstjournal{\doiref{10.1088/1126-6708/2007/11/068}{JHEP~0711,~068~(2007)}},
\nbbsteprint{\arxivref{0710.1060}{arxiv:0710.1060}}.

\bibitem{Drummond:2007aua}
\nbbstauthor{J.~M.~Drummond, G.~P.~Korchemsky and E.~Sokatchev},
\nbbsttitle{Conformal properties of four-gluon planar amplitudes and {Wilson}
  loops},
\nbbstjournal{\doiref{10.1016/j.nuclphysb.2007.11.041}{Nucl.~Phys.~B795,~385~(2008)}},
\nbbsteprint{\arxivref{0707.0243}{arxiv:0707.0243}}.

\bibitem{Brandhuber:2007yx}
\nbbstauthor{A.~Brandhuber, P.~Heslop and G.~Travaglini},
\nbbsttitle{{MHV} Amplitudes in {$\mathcal{N}=\mathord{}$4} Super
  {Yang}--{Mills} and {Wilson} Loops},
\nbbstjournal{\doiref{10.1016/j.nuclphysb.2007.11.002}{Nucl.~Phys.~B794,~231~(2008)}},
\nbbsteprint{\arxivref{0707.1153}{arxiv:0707.1153}}.

\bibitem{Belitsky:2021huz}
\nbbstauthor{A.~V.~Belitsky and V.~A.~Smirnov},
\nbbsttitle{An off-shell {Wilson} loop},
\nbbstjournal{\doiref{10.1007/JHEP04(2023)071}{JHEP~2304,~071~(2023)}},
\nbbsteprint{\arxivref{2110.13206}{arxiv:2110.13206}}.

\bibitem{Berends:1987me}
\nbbstauthor{F.~A.~Berends and W.~T.~Giele},
\nbbsttitle{Recursive Calculations for Processes with n Gluons},
\nbbstjournal{\doiref{10.1016/0550-3213(88)90442-7}{Nucl.~Phys.~B306,~759~(1988)}}.

\bibitem{Britto:2004ap}
\nbbstauthor{R.~Britto, F.~Cachazo and B.~Feng},
\nbbsttitle{New recursion relations for tree amplitudes of gluons},
\nbbstjournal{\doiref{10.1016/j.nuclphysb.2005.02.030}{Nucl.~Phys.~B715,~499~(2005)}},
\nbbsteprint{\arxivref{hep-th/0412308}{hep-th/0412308}}.

\bibitem{Britto:2005fq}
\nbbstauthor{R.~Britto, F.~Cachazo, B.~Feng and E.~Witten},
\nbbsttitle{Direct proof of tree-level recursion relation in {Yang}--{Mills}
  theory},
\nbbstjournal{\doiref{10.1103/PhysRevLett.94.181602}{Phys.~Rev.~Lett.~94,~181602~(2005)}},
\nbbsteprint{\arxivref{hep-th/0501052}{hep-th/0501052}}.

\bibitem{Arkani-Hamed:2013jha}
\nbbstauthor{N.~Arkani-Hamed and J.~Trnka},
\nbbsttitle{The Amplituhedron},
\nbbstjournal{\doiref{10.1007/JHEP10(2014)030}{JHEP~1410,~30~(2014)}},
\nbbsteprint{\arxivref{1312.2007}{arxiv:1312.2007}}.

\bibitem{Eden:2017fow}
\nbbstauthor{B.~Eden, P.~Heslop and L.~Mason},
\nbbsttitle{The Correlahedron},
\nbbstjournal{\doiref{10.1007/JHEP09(2017)156}{JHEP~1709,~156~(2017)}},
\nbbsteprint{\arxivref{1701.00453}{arxiv:1701.00453}}.

\bibitem{Arkani-Hamed:2023epq}
\nbbstauthor{N.~Arkani-Hamed, W.~Flieger, J.~M.~Henn, A.~Schreiber and
  J.~Trnka},
\nbbsttitle{{Coulomb} Branch Amplitudes from a Deformed Amplituhedron
  Geometry},
\nbbstjournal{\doiref{10.1103/PhysRevLett.132.211601}{Phys.~Rev.~Lett.~132,~211601~(2024)}},
\nbbsteprint{\arxivref{2311.10814}{arxiv:2311.10814}}.

\bibitem{Jiang:2016ulr}
\nbbstauthor{Y.~Jiang, S.~Komatsu, I.~Kostov and D.~Serban},
\nbbsttitle{Clustering and the Three-Point Function},
\nbbstjournal{\doiref{10.1088/1751-8113/49/45/454003}{J.~Phys.~A49,~454003~(2016)}},
\nbbsteprint{\arxivref{1604.03575}{arxiv:1604.03575}}.

\bibitem{Bargheer:2019exp}
\nbbstauthor{T.~Bargheer, F.~Coronado and P.~Vieira},
\nbbsttitle{Octagons II: Strong Coupling},
\nbbsteprint{\arxivref{1909.04077}{arxiv:1909.04077}}.

\bibitem{Alday:2010vh}
\nbbstauthor{L.~F.~Alday, J.~Maldacena, A.~Sever and P.~Vieira},
\nbbsttitle{Y-system for Scattering Amplitudes},
\nbbstjournal{\doiref{10.1088/1751-8113/43/48/485401}{J.~Phys.~A43,~485401~(2010)}},
\nbbsteprint{\arxivref{1002.2459}{arxiv:1002.2459}}.

\bibitem{Abreu:2020jxa}
\nbbstauthor{S.~Abreu, H.~Ita, F.~Moriello, B.~Page, W.~Tschernow and M.~Zeng},
\nbbsttitle{Two-Loop Integrals for Planar Five-Point One-Mass Processes},
\nbbstjournal{\doiref{10.1007/JHEP11(2020)117}{JHEP~2011,~117~(2020)}},
\nbbsteprint{\arxivref{2005.04195}{arxiv:2005.04195}}.

\bibitem{Canko:2020ylt}
\nbbstauthor{D.~D.~Canko, C.~G.~Papadopoulos and N.~Syrrakos},
\nbbsttitle{Analytic representation of all planar two-loop five-point Master
  Integrals with one off-shell leg},
\nbbstjournal{\doiref{10.1007/JHEP01(2021)199}{JHEP~2101,~199~(2021)}},
\nbbsteprint{\arxivref{2009.13917}{arxiv:2009.13917}}.

\bibitem{Abreu:2021smk}
\nbbstauthor{S.~Abreu, H.~Ita, B.~Page and W.~Tschernow},
\nbbsttitle{Two-loop hexa-box integrals for non-planar five-point one-mass
  processes},
\nbbstjournal{\doiref{10.1007/JHEP03(2022)182}{JHEP~2203,~182~(2022)}},
\nbbsteprint{\arxivref{2107.14180}{arxiv:2107.14180}}.

\bibitem{Abreu:2023rco}
\nbbstauthor{S.~Abreu, D.~Chicherin, H.~Ita, B.~Page, V.~Sotnikov, W.~Tschernow
  and S.~Zoia},
\nbbsttitle{All Two-Loop {Feynman} Integrals for Five-Point One-Mass
  Scattering},
\nbbstjournal{\doiref{10.1103/PhysRevLett.132.141601}{Phys.~Rev.~Lett.~132,~141601~(2024)}},
\nbbsteprint{\arxivref{2306.15431}{arxiv:2306.15431}}.

\bibitem{Rodrigues:2024znq}
\nbbstauthor{R.~Rodrigues},
\nbbsttitle{Two-loop integrals of half-{BPS} six-point functions on a line},
\nbbstjournal{\doiref{10.1007/JHEP05(2024)007}{JHEP~2405,~007~(2024)}},
\nbbsteprint{\arxivref{2402.08463}{arxiv:2402.08463}}.

\bibitem{Henn:2024ngj}
\nbbstauthor{J.~M.~Henn, A.~Matija{\v{s}}i{\'c}, J.~Miczajka, T.~Peraro, Y.~Xu
  and Y.~Zhang},
\nbbsttitle{A computation of two-loop six-point {Feynman} integrals in
  dimensional regularization},
\nbbstjournal{\doiref{10.1007/JHEP08(2024)027}{JHEP~2408,~027~(2024)}},
\nbbsteprint{\arxivref{2403.19742}{arxiv:2403.19742}}.

\bibitem{Abreu:2024yit}
\nbbstauthor{S.~Abreu, D.~Chicherin, V.~Sotnikov and S.~Zoia},
\nbbsttitle{Two-loop five-point two-mass planar integrals and double
  {Lagrangian} insertions in a {Wilson} loop},
\nbbstjournal{\doiref{10.1007/JHEP10(2024)167}{JHEP~2410,~167~(2024)}},
\nbbsteprint{\arxivref{2408.05201}{arxiv:2408.05201}}.

\bibitem{Abreu:2024fei}
\nbbstauthor{S.~Abreu, P.~F.~Monni, B.~Page and J.~Usovitsch},
\nbbsttitle{Planar six-point {Feynman} integrals for four-dimensional gauge
  theories},
\nbbstjournal{\doiref{10.1007/JHEP06(2025)112}{JHEP~2506,~112~(2025)}},
\nbbsteprint{\arxivref{2412.19884}{arxiv:2412.19884}}.

\bibitem{Becchetti:2025oyb}
\nbbstauthor{M.~Becchetti, C.~Dlapa and S.~Zoia},
\nbbsttitle{Canonical differential equations for the elliptic two-loop
  five-point integral family relevant to tt{\textasciimacron}+jet production at
  leading color},
\nbbstjournal{\doiref{10.1103/zt4w-c1jk}{Phys.~Rev.~D~112,~L031501~(2025)}},
\nbbsteprint{\arxivref{2503.03603}{arxiv:2503.03603}}.

\bibitem{Papathanasiou:2022lan}
\nbbstauthor{G.~Papathanasiou},
\nbbsttitle{The {SAGEX} review on scattering amplitudes. Chapter 5: Analytic
  bootstraps for scattering amplitudes and beyond},
\nbbstjournal{\doiref{10.1088/1751-8121/ac7e8e}{J.~Phys.~A~55,~443006~(2022)}},
\nbbsteprint{\arxivref{2203.13016}{arxiv:2203.13016}}.

\bibitem{He:2022ctv}
\nbbstauthor{S.~He, Z.~Li, R.~Ma, Z.~Wu, Q.~Yang and Y.~Zhang},
\nbbsttitle{A study of {Feynman} integrals with uniform transcendental weights
  and their symbology},
\nbbstjournal{\doiref{10.1007/JHEP10(2022)165}{JHEP~2210,~165~(2022)}},
\nbbsteprint{\arxivref{2206.04609}{arxiv:2206.04609}}.

\end{thebibliography}

\end{document}